\documentstyle[12pt,leqno]{article}
\newtheorem{theorem}{Theorem}[section]
\newtheorem{definition}[theorem]{Definition}
\newtheorem{corollary}[theorem]{Corollary}

\newtheorem{lemma}[theorem]{Lemma}
\newtheorem{sublemma}[theorem]{Sublemma}
\newtheorem{proposition}[theorem]{Proposition}

\catcode`\@=11
\def\@begintheorem#1#2{\it \trivlist \item[\hskip
 \labelsep{\bf #1\ #2.}]}
\def\@opargbegintheorem#1#2#3{\it \trivlist\item[\hskip%
 \labelsep{\bf #1\ #2.\ (#3)}]}
\def\@endtheorem{\endtrivlist}

\def\@listI{\leftmargin\leftmargini \parsep 1pt plus 2.5pt
 minus 1pt\topsep 5pt plus 2pt minus 3pt%
 \itemsep 0pt plus 2.5pt minus 1pt}
\let\@listi\@listI
\@listi

\def\@sect#1#2#3#4#5#6[#7]#8{\ifnum #2>\c@secnumdepth%
 \def \@svsec {}\else \refstepcounter {#1}\edef \@svsec%
 {\csname the#1\endcsname. \hskip .1em }\fi \@tempskipa%
 #5\relax \ifdim \@tempskipa >\z@ \begingroup #6\relax%
 \@hangfrom {\hskip #3\relax \@svsec }{\interlinepenalty%
 \@M #8.\par }\endgroup \csname #1mark\endcsname {#7}%
 \addcontentsline {toc}{#1}{\ifnum #2>\c@secnumdepth%
 \else \protect \numberline {\csname the#1\endcsname. }%
 \fi #7}\else \def \@svsechd {#6\hskip #3\@svsec #8.%
 \csname #1mark\endcsname {#7}\addcontentsline {toc}{#1}%
 {\ifnum #2>\c@secnumdepth \else \protect \numberline%
 {\csname the#1\endcsname. }\fi #7}}\fi \@xsect {#5}}

\def\section{\@startsection {section}{1}{\z@ }%
 {-3.5ex plus -1ex minus -.2ex}{2.3ex plus .2ex}{\bf }}

\def\thebibliography#1{%
 \section *{References.\@mkboth {REFERENCES}{REFERENCES}}%
 \list {[\arabic {enumi}]}{\settowidth \labelwidth {[#1]}%
 \leftmargin \labelwidth \advance \leftmargin \labelsep %
 \usecounter {enumi}} \def \newblock %
 {\hskip .11em plus .33em minus -.07em} \sloppy \clubpenalty 4000%
 \widowpenalty 4000 \sfcode`\.=1000\relax}

\def\@maketitle{%
 \newpage \null \vskip 2em
 \begin{center}{\Large\bf \@title \par }
 \vskip 1.5em
 {\large \lineskip .5em
 \begin {tabular}[t]{c}\@author
 \end{tabular}\par } \vskip .8em {June 21, 1994}
 \end{center}\par \vskip 1.5em}

\def\abstract{%
\if@twocolumn \section *{Abstract}
 \else \small\quotation\noindent{\bf Abstract.}\fi}

\catcode`\@=13

\def\qed{\hspace*{\fill}
\mbox{\hphantom{mm}\rule{0.25cm}{0.25cm}}\\}

\def\bigodo{\mbox{$\bigodot$}}
\def\boldk{{\bf k}} \def\bk{{\bf k}}
\def\tp#1#2{{#1}^{\otimes{#2}}}
\def\ident{1\!\!1} \def\id{\ident}
\def\b{\bullet}
\def\bod{\bigodo}
\def\bodvbv{\bod(V,\bv)}
\def\bv{{\cal B}(V)}
\def\db{d_{\cal B}}
\def\calb{{\cal B}}
\def\domega{{d_C}}
\def\dcobar{\domega}
\def\Der{\mbox{\rm Der}}
\def\der#1{\mbox{\rm Der}^{#1}_V(\bod(V,\bv))}
\def\br#1{{\mbox{\rm Br}_{#1}}}
\def\Br#1#2{\mbox{\rm Br}_{#2}(#1)}
\def\T#1{{\cal T}_{#1}}
\def\ck#1#2{C_{#1}(K_{#2})}
\def\int#1{\{0,\ldots,#1\}}
\def\cdots{\!\cdot\!\cdot\!\cdot\!}
\def\bolde{{\bf e}}
\def\bolds{{\bf s}}
\def\boldt{{\bf t}}
\def\boldu{{\bf u}}
\def\hom#1#2#3#4{\mbox{\rm Hom}_{#1}^{#4}(#2,#3)}
\def\HOM#1#2#3{\mbox{\bf Hom}^{#3}(#1,#2)}

\def\dhoch{d_{\mbox{\scriptsize\rm Hoch}}}
\def\dqcohoch{d_{\mbox{\scriptsize\rm q-Car}}}
\def\dcoh{d_{\mbox{\scriptsize\rm Car}}}
\def\dqcoh{\dqcohoch}

\def\od{\odot}
\def\susp{\uparrow\hskip-1mm}
\def\E#1#2{{E^{#1,#2}(A)}}
\def\Mu{\overline\mu}
\def\MU{\hskip.2mm\cdot\hskip.8mm}
\def\ot{\otimes}
\def\otvm{\bigotimes(V,M)}
\def\Bigotimes{{\mbox{$\bigotimes$}}}
\def\susp{\uparrow\kern -.05em}
\def\desusp{\downarrow\kern -.05em}
\def\X#1#2{{X_{{#1},{#2}}}}
\def\dia{\diamond}
\def\Sum{\mbox{$\sum$\hskip1mm}}
\def\deltacurl#1{{\Delta^{\{#1\}}}}
\def\PP#1#2{{\Psi_{{#1},{#2}}}}
\def\BB#1#2{{\Box_{{#1},{#2}}}}
\def\rada#1#2#3{{#1_{#2},\ldots,#1_{#3}}}
\def\Rada#1#2#3{{#1_{#2}|\ldots|#1_{#3}}}
\def\otbar{\overline \otimes}
\def\ots{\otimes\cdots\otimes}
\def\otsbar{\otbar\cdots\otbar}
\def\DC{\deltacurl}
\def\D{\Delta}
\def\proof{\noindent{\bf Proof.}}
\def\BV#1{{\cal B}_{#1}(V)}
\def\REF#1{~(\ref{#1})}
\def\hoj#1#2#3{{#1}^{{(#2,#3)}}}
\def\hhh#1#2#3#4{{#1}^{{(#2,#3)}}_{{#4}}}
\def\sh#1#2{{\rm Sh}_{{#1},{#2}}}
\def\sgn{{\mbox{\rm sgn}}}
\long\def\comment#1\endcomment{{}}
\def\oV{{\overline V}}
\def\exponent#1#2{
 \def\nnnnn{#1}
 \if\nnnnn n \mbox{$\bigotimes$}^{{#1}}#2\else
 \ifnum #1 = 0 {\bf k}\else
 \ifnum #1=1 #2 \else \mbox{$\bigotimes$}^{{#1}}#2
\fi\fi\fi}
\def\jitka#1#2{\hbox{\rm Hom}_{{\bf k}}%
 (\exponent{#1}{\overline V},\exponent{#2}{V})}
\def\OO{\!\otimes\!}
\def\CCCC{\!\cdot\!}
\def\BC{\mbox{$\bigcirc$}}
\def\hatc#1{{\hat C^{#1}(A)}}
\def\mmm{{\overline \mu}}

\def\BC{{[\!]}}
\def\bcc{{\BC}}
\def\Ker{{\mbox{\rm Ker}}}
\def\ix#1#2#3{M[(K_{#1}\!\times\! K_{#2})_{#3}]}

\def\dd{\Delta}\def\ph{\Phi}
\def\oo{\overline{\otimes}}
\def\mmm{\Sum (\BC_1\ot\cdots\ot\BC_5)}
\def\I{((\id\ot\dd\ot\id)(\ph)\ot1)}
\def\II{(\id\ot(\dd\ot\id)\dd\ot\id)(\ph)}
\def\III{((\id\ot\dd)\dd\ot\id^2)(\ph)}
\def\IV{(\id\ot\dd^2)(\ph)}
\def\V{(1\ot(\dd\ot\id^2)(\ph))}
\def\VI{(\ph\ot1^2)}
\def\VII{(\ph\ot1^2)}
\def\VIII{(1\ot(\id^2\ot\dd)(\ph))}
\def\IX{((\dd\ot\id^2)(\ph)\ot1)}
\def\X{((\id^2\ot\dd)(\ph)\ot1)}
\def\XI{(1\ot\ph\ot1)}
\def\XII{((\dd\ot\id)\dd\ot\id^2)(\ph)}
\def\XIII{(\dd\ot\id\ot\dd)(\ph)}

\def\XV{(\id^2\ot(\id\ot\dd)\dd)(\ph)}
\def\XVI{(\id^2\ot(\dd\ot\id)\dd)(\ph)}
\def\XVII{(1\ot\ph\ot1)}
\def\XVIII{(1\ot(\id\ot\dd\ot\id)(\ph))}
\def\IXX{(1^2\ot\ph)}
\def\XX{(\id\ot(\id\ot\dd)\dd\ot\id)(\ph)}
\def\XXI{(1^2\ot\ph)}
\def\XXII{((\dd^2\ot\id)\ph)}
\def\Bigotimes{\mbox{$\bigotimes$}}
\def\vectorjedna{\vector(1,0){11.9}}
\def\vectordva{\vector(0,-1){7.9}}
\def\vectortri{\vector(-1,-1){1.95}}
\def\vectorctiri{\vector(-1,1){1.95}}
\def\vectorpet{\vector(1,0){4.9}}
\def\vectorsest{\vector(1,0){2.9}}
\def\vectorsedm{\vector(0,1){2.4}}
\def\vectorosm{\vector(0,-1){1.4}}
\def\vectordevet{\vector(2,1){2.9}}
\def\vectordeset{\vector(2,1){1.9}}
\def\vectorjedenact{\vector(1,-1){1.95}}
\def\kulka{\makebox(0,0){$\bullet$}}
\newcommand{\rrr}[1]{\makebox(0,0)[r]{\scriptsize$#1$}}
\newcommand{\lll}[1]{\makebox(0,0)[l]{\scriptsize$#1$}}
\newcommand{\ccc}[1]{\makebox(0,0){\scriptsize$#1$}}
\newcommand{\ctv}[1]{\framebox(0.5,0.5){\scriptsize$#1$}}

\begin{document}
\bibliographystyle{plain}
\baselineskip20pt
\parskip3pt

\title{Drinfel'd algebra deformations,
homotopy comodules and the associahedra}
\author{Martin Markl\thanks{Partially
supported by the National Research Counsel, USA}
\hskip2mm and Steve Shnider\thanks{Partially supported by a grant from
the Israel Science Foundation administered by the Israel Academy of
Sciences and Humanities}}

\maketitle

\begin{abstract}
The aim of this work is to construct a cohomology theory
controlling the deformations of a general Drinfel'd algebra $A$ and
thus finish the program which began in~\cite{MS},
\cite{shnider-sternberg:preprint}. The task is accomplished in three
steps. The first step, which was taken in the aforementioned articles,
is the construction of a modified cobar complex adapted to a
non-coassociative comultiplication. The following two steps
each involve a new, highly non-trivial, construction.
The first construction, essentially combinatorial,
defines a differential graded Lie algebra
structure on the simplicial chain complex of the associahedra.
The second construction, of a more algebraic nature, is the definition
of
map of differential graded Lie algebras from the complex defined above
to the algebra of derivations on the bar
resolution. Using the existence of this map and the acyclicity of
the associahedra we can define a so-called homotopy comodule
structure (Definition~\ref{masinka}
below) on the bar resolution of a general Drinfeld algebra.
This in turn allows us to define the desired cohomology theory
in terms of a complex which consists, roughly speaking, of the
bimodule and
bicomodule maps from the bar resolution to the modified cobar
resolution.
The complex is bigraded but not a bicomplex as in the
Gerstenhaber-Schack
theory for bialgebra deformations. The new components of
the coboundary operator are defined via the constructions
mentioned above. The results of the paper were announced
in~\cite{MS:an}.
\end{abstract}

\baselineskip17pt
\section{Introduction}

Recall that a {\em Drinfel'd algebra\/} (or a
{\em quasi-bialgebra\/} in the
original terminology of~\cite{D}) is an object
$A=(V,\MU,\Delta,\Phi)$, where
$(V,\MU,\Delta)$ is an associative, not necessarily coassociative,
unital and counital $\boldk$-bialgebra, $\Phi$ is an
invertible element of
$\tp V3$,
and the usual coassociativity property is replaced by
the condition which we shall refer to as quasi-coassociativity:
\begin{equation}
\label{qcoass}
(\ident\otimes\Delta)\Delta \cdot \Phi
= \Phi\cdot(\Delta\otimes\ident)\Delta,
\end{equation}
where we use the dot $\MU$ to indicate both the (associative)
multiplication on $V$ and the induced multiplication
on $V^{\otimes
3}$.
Moreover, the validity of the following
``pentagon identity'' is required:
\begin{equation}
\label{PENTAGON}
(\ident^2\otimes\Delta)(\Phi)\cdot(\Delta\otimes\ident^2)
(\Phi) =
(1\otimes\Phi)\cdot(\ident
\otimes\Delta\otimes\ident)(\Phi)\cdot(\Phi\otimes1),
\end{equation}
$1\in V$ being the unit element and $\ident$ , the identity map
on $V$. If $\epsilon :V \to \bk$ ($\bk$ being the ground field) is the
counit of the coalgebra $(V,\D)$ then, by definition, $(\epsilon \ot
\id)\Delta = (\id \ot \epsilon)\D = \id$.
A bialgebra is a Drinfel'd algebra with $\Phi=1$.

We have also a natural
splitting $V = \oV \oplus \bk$, $\oV := \Ker(\epsilon)$. We actually
do not need to assume the existence of the counit in the paper. All
the constructions remain valid, except we would not have the ``full
cohomology'' $H^*(A)$ of Definition~\ref{vlacek},
only the reduced version $\hat H^*(A)$ of
Theorem~\ref{vlacek-bez-jednoho-vagonku}, because the ground field
$\bk$ would not have a natural $(V,\MU)$-bimodule structure.

In a recent paper \cite{GS}, Gerstenhaber and Schack have extended the
original Gerstenhaber deformation theory to a deformation theory of
bialgebras. We are interested in extending their theory to the case of
Drinfel'd algebras. In the past one has always been able to construct
a cohomology theory controlling deformations by identifying the
complex  and extracting the differential from
a careful examination of the linearization of the axioms of
structure. In the geometric theory of deformations, Kodaira and
Spencer realized that  the linearization of the compatibility
equations for the transition functions of a complex structure lead to
cohomology with coefficients in the holomorphic tangent sheaf and used
it to study deformations of complex structure. On the algebraic side, in
the deformation theory of associative algebras, Gerstenhaber realized
that if Hochschild cohomology did not already exist, he would have had to
invent it. Similarly, in the deformation of associative
coalgebras there is a dual theory using
the coboundary operator defined by Cartier in the 1950's \cite{Cart55}.
The delightful discovery of Gerstenhaber and Schack is that the
two theories fit together to form a double complex controlling
the deformation of bialgebras and nothing essentially new is needed.
However when we study Drinfel'd algebra deformations there
are some essentially new ingredients and these are not easily
derived from a direct study of the structure axioms as we shall see.

After a quick review of the cohomological approach to the deformation
theory of bialgebras as defined by Gerstenhaber and Schack we shall
describe the rather straightforward extension to deformations
of bialgebras (considered as Drinfel'd algebras with $\Phi = 1$)
in the category of Drinfel'd algebras. Then
we will explain the complications which arise when we try to
extend the theory to deformations for which the initial structure
is a genuine Drinfel'd algebra with the
comultiplication which is not assumed to
be associative. Basically, nonassociativity significantly complicates
the procedure for defining a suitable differential,
that is, an endomorphism with square
zero, which contains all the basic operations in the structure.
This leads us to the new constructions whose definition and properties
will occupy us for the rest of the paper. We close this introductory
section with an outline of the content of the remaining sections.

\begin{figure}
\setlength{\unitlength}{0.055cm}
\begin{center}
\begin{picture}(200,120)(15,-5)

%prvni rada
\put(48,-2){\makebox(0,0){$\circ$}}
\put(78,-2){\makebox(0,0){$\circ$}}
\put(108,-2){\makebox(0,0){$\circ$}}
\put(138,-2){\makebox(0,0){$\circ$}}
\put(168,-2){\makebox(0,0){$\circ$}}
\put(198,-2){\makebox(0,0){$\circ$}}

\put(48,-2){\makebox(0,0){$\times$}}
\put(78,-2){\makebox(0,0){$\times$}}
\put(108,-2){\makebox(0,0){$\times$}}
\put(138,-2){\makebox(0,0){$\times$}}
\put(168,-2){\makebox(0,0){$\times$}}
\put(198,-2){\makebox(0,0){$\times$}}

%druha rada
\put(48,28){\makebox(0,0){$\circ$}\makebox(0,0){$\times$}}
\put(78,28){\makebox(0,0){$\circ$}}
\put(108,28){\makebox(0,0){$\bullet$}}
\put(98,23){\makebox(0,0){\scriptsize$(2,1)$}}
\put(138,28){\makebox(0,0){$\circ$}}
\put(168,28){\makebox(0,0){$\circ$}}
\put(198,28){\makebox(0,0){$\circ$}}

%treti rada
\put(48,58){\makebox(0,0){$\circ$}\makebox(0,0){$\times$}}
\put(78,58){\makebox(0,0){$\bullet$}}
\put(68,53){\makebox(0,0){\scriptsize$(1,2)$}}
\put(108,58){\makebox(0,0){$\circ$}}
\put(138,58){\makebox(0,0){$\circ$}}
\put(168,58){\makebox(0,0){$\circ$}}
\put(198,58){\makebox(0,0){$\circ$}}

%ctvrta rada
\put(48,88){\makebox(0,0){$\circ$}\makebox(0,0){$\times$}}
\put(78,88){\makebox(0,0){$\circ$}}
\put(108,88){\makebox(0,0){$\circ$}}
\put(138,88){\makebox(0,0){$\circ$}}
\put(168,88){\makebox(0,0){$\circ$}}
\put(198,88){\makebox(0,0){$\circ$}}

%pata rada
\put(48,118){\makebox(0,0){$\circ$}\makebox(0,0){$\times$}}
\put(78,118){\makebox(0,0){$\circ$}}
\put(108,118){\makebox(0,0){$\circ$}}
\put(138,118){\makebox(0,0){$\circ$}}
\put(168,118){\makebox(0,0){$\circ$}}
\put(198,118){\makebox(0,0){$\circ$}}

\put(78,58){\vector(0,1){28}\vector(1,0){28}}
\put(108,28){\vector(0,1){28}\vector(1,0){28}}

\end{picture}
\end{center}

\caption{\label{fig1}
The Gerstenhaber-Schack bialgebra deformation complex.
The bottom row and
left hand column are deleted (deletion indicated by $\times$).
The infinitesimal deformation of the
multiplication is at position (1,2) and of the comultiplication is at
position (2,1). Vertical arrows represent the Hochschild differential
and horizontal arrows represent the Cartier (coalgebra) differential.}
\end{figure}
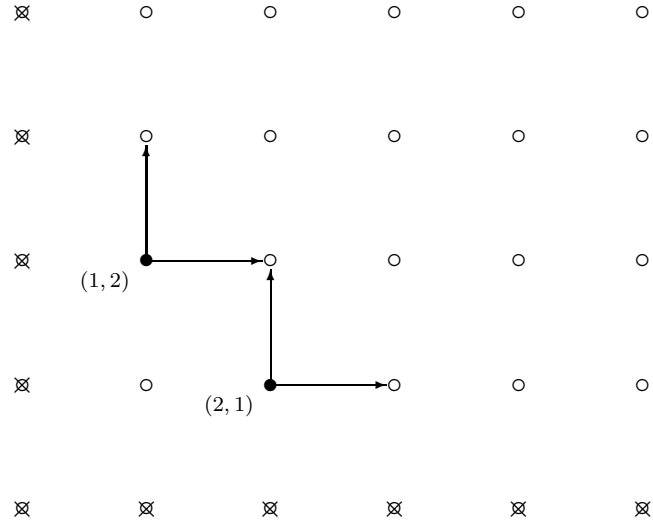

Our basic framework is the theory of formal deformations of algebraic
structures as formulated by Gerstenhaber \cite{gerst-Ann}. For an
introductory presentation, see \cite{shnider-sternberg:book}.
If we deform the multiplication the infinitesimal will lie in
Hom$_{\bk}(V\otimes V, V)$, the space of $\bk$-linear maps
$V\otimes V \to V$. The infinitesimal deformation of the
comultiplication lies in
Hom$_{\bk}(V,V\otimes V)$. So we introduce a
``double complex'' and follow the convention
of Gerstenhaber and Schack by considering Hom$_{\bk}(V^{\otimes q},
V^{\otimes p})$
as having total degree $p+q-1$ and placing it at position $(p,q)$,
column $p$, row $q$ as in Figure~\ref{fig1}.
The infinitesimal deformation of the multiplication is located at
position
$(1,2)$, and the fact that the deformed multiplication is
 associative implies that it is sent
to zero by the Hochschild differential represented by the
vertical arrow. This is just the old Gerstenhaber deformation theory
for
associative algebras, \cite {gerst-Ann}.
Dually, the infinitesimal comultiplication is located at
$(2,1)$ and the (co)associativity of the deformed comultiplication
says
that it is sent to zero by the coalgebra differential, introduced by
Cartier in the 50's, \cite{Cart55}, called sometimes also the
coHochschild differential, which is
represented by the horizontal arrow.

The delightful discovery of Gerstenhaber and Schack is that
compatibility of multiplication and comultiplication implies that two
contributions at $(2,2)$ (from $(1,2)$ horizontally and
 from $(2,1)$ vertically) mutually cancel. Thus
the sum of the infinitesimal multiplication and comultiplication give
a
cocycle for the total complex.

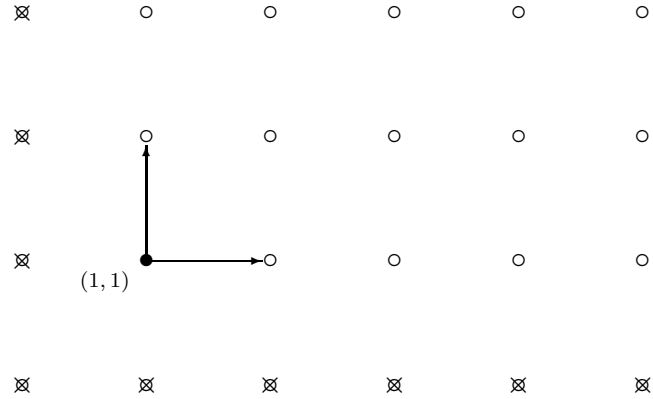
\begin{figure}
\setlength{\unitlength}{0.055cm}
\begin{center}
\begin{picture}(200,90)(15,-5)

%prvni rada
\put(48,-2){\makebox(0,0){$\circ$}}
\put(78,-2){\makebox(0,0){$\circ$}}
\put(108,-2){\makebox(0,0){$\circ$}}
\put(138,-2){\makebox(0,0){$\circ$}}
\put(168,-2){\makebox(0,0){$\circ$}}
\put(198,-2){\makebox(0,0){$\circ$}}

\put(48,-2){\makebox(0,0){$\times$}}
\put(78,-2){\makebox(0,0){$\times$}}
\put(108,-2){\makebox(0,0){$\times$}}
\put(138,-2){\makebox(0,0){$\times$}}
\put(168,-2){\makebox(0,0){$\times$}}
\put(198,-2){\makebox(0,0){$\times$}}

%druha rada
\put(48,28){\makebox(0,0){$\circ$}\makebox(0,0){$\times$}}
\put(78,28){\makebox(0,0){$\bullet$}}
\put(68,23){\makebox(0,0){\scriptsize$(1,1)$}}
\put(108,28){\makebox(0,0){$\circ$}}
\put(138,28){\makebox(0,0){$\circ$}}
\put(168,28){\makebox(0,0){$\circ$}}
\put(198,28){\makebox(0,0){$\circ$}}

%treti rada
\put(48,58){\makebox(0,0){$\circ$}\makebox(0,0){$\times$}}
\put(78,58){\makebox(0,0){$\circ$}}
\put(108,58){\makebox(0,0){$\circ$}}
\put(138,58){\makebox(0,0){$\circ$}}
\put(168,58){\makebox(0,0){$\circ$}}
\put(198,58){\makebox(0,0){$\circ$}}

%ctvrta rada
\put(48,88){\makebox(0,0){$\circ$}\makebox(0,0){$\times$}}
\put(78,88){\makebox(0,0){$\circ$}}
\put(108,88){\makebox(0,0){$\circ$}}
\put(138,88){\makebox(0,0){$\circ$}}
\put(168,88){\makebox(0,0){$\circ$}}
\put(198,88){\makebox(0,0){$\circ$}}

\put(78,28){\vector(0,1){28}\vector(1,0){28}}

\end{picture}
\end{center}
\caption{\label{fig2}
Two infinitesimal deformations are equivalent if they differ
by a coboundary in the restricted complex.}
\end{figure}

{}From here the theory ``follows the book''. In the (associative and
coassociative) bialgebra
deformation theory we must delete the bottom row and leftmost column
as in
Figure~\ref{fig1}. Then
an infinitesimal bialgebra deformation, as we have just seen, is a
cocycle of
total degree two = $p+q-1$ with two components corresponding
to multiplication and comultiplication.
Two infinitesimal deformations differ by coboundary in this deleted
complex if and only if they are related by an
infinitesimal ``change of coordinates'', and hence should be
considered
equivalent. See Figure~\ref{fig2}.

\begin{figure}
\setlength{\unitlength}{0.055cm}
\begin{center}
\begin{picture}(200,120)(15,-5)

%prvni rada
\put(48,-2){\makebox(0,0){$\circ$}}
\put(78,-2){\makebox(0,0){$\circ$}}
\put(108,-2){\makebox(0,0){$\circ$}}
\put(138,-2){\makebox(0,0){$\circ$}}
\put(168,-2){\makebox(0,0){$\circ$}}
\put(198,-2){\makebox(0,0){$\circ$}}

\put(48,-2){\makebox(0,0){$\times$}}
\put(78,-2){\makebox(0,0){$\times$}}
\put(108,-2){\makebox(0,0){$\times$}}
\put(138,-2){\makebox(0,0){$\times$}}
\put(168,-2){\makebox(0,0){$\times$}}
\put(198,-2){\makebox(0,0){$\times$}}

%druha rada
\put(48,28){\makebox(0,0){$\circ$}\makebox(0,0){$\times$}}
\put(78,28){\makebox(0,0){$\circ$}}
\put(108,28){\makebox(0,0){$\circ$}}
\put(138,28){\makebox(0,0){$\circ$}}
\put(168,28){\makebox(0,0){$\circ$}}
\put(198,28){\makebox(0,0){$\circ$}}

%treti rada
\put(48,58){\makebox(0,0){$\circ$}\makebox(0,0){$\times$}}
\put(78,58){\makebox(0,0){$\circ$}}
\put(108,58){\makebox(0,0){$\circ$}}
\put(138,58){\makebox(0,0){$\circ$}}
\put(168,58){\makebox(0,0){$\circ$}}
\put(198,58){\makebox(0,0){$\circ$}}

%ctvrta rada
\put(48,88){\makebox(0,0){$\circ$}\makebox(0,0){$\times$}}
\put(78,88){\makebox(0,0){$\circ$}}
\put(108,88){\makebox(0,0){$\circ$}}
\put(138,88){\makebox(0,0){$\circ$}}
\put(168,88){\makebox(0,0){$\circ$}}
\put(198,88){\makebox(0,0){$\circ$}}

%pata rada
\put(48,118){\makebox(0,0){$\circ$}\makebox(0,0){$\times$}}
\put(78,118){\makebox(0,0){$\circ$}}
\put(108,118){\makebox(0,0){$\circ$}}
\put(138,118){\makebox(0,0){$\circ$}}
\put(168,118){\makebox(0,0){$\circ$}}
\put(198,118){\makebox(0,0){$\circ$}}

\put(138,28){\put(-10,-5){\makebox(0,0)%
 {\scriptsize$(3,1)$}}\makebox(0,0){$\bullet$}\vector(0,1){28}%
 \vector(1,0){28}}
\put(108,58){\put(-10,-5){\makebox(0,0)%
 {\scriptsize$(2,2)$}}\makebox(0,0){$\bullet$}\vector(0,1){28}%
 \vector(1,0){28}}
\put(78,88){\put(-10,-5){\makebox(0,0)%
 {\scriptsize$(1,3)$}}\makebox(0,0){$\bullet$}\vector(0,1){28}%
 \vector(1,0){28}}

\end{picture}
\end{center}
\caption{\label{fig3}
The obstruction to extending the infinitesimal one higher power in
$t$ is a cocycle of total degree three as drawn.
If this cocycle is a coboundary the
extension is possible.}
\end{figure}
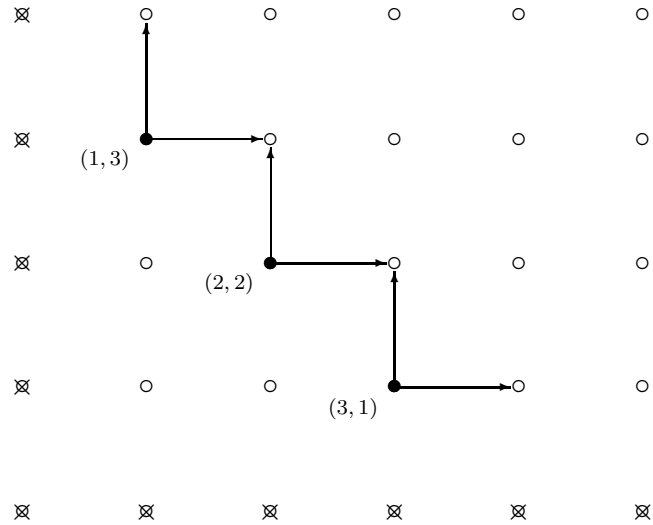

Substituting the putative infinitesimal deformation, i.e. a cocycle
of total degree two, in the three structure equations,
associativity, coassociativity, and compatibility generates
a cocycle of total degree three with three components. See
Figure~\ref{fig3}. The
deformation can be extended to one higher
degree in the deformation parameter
$t$ if and only if this
so called ``obstruction cocycle'' is a coboundary.

\begin{figure}
\setlength{\unitlength}{0.055cm}
\begin{center}
\begin{picture}(200,120)(15,-5)

%prvni rada
\put(48,-2){\makebox(0,0){$\circ$}\makebox(0,0){$\times$}}
\put(78,-2){\makebox(0,0){$\circ$}}
\put(108,-2){\makebox(0,0){$\circ$}}
\put(138,-2){\makebox(0,0){$\circ$}}
\put(168,-2){\makebox(0,0){$\circ$}}
\put(198,-2){\makebox(0,0){$\circ$}}

%druha rada
\put(48,28){\makebox(0,0){$\circ$}\makebox(0,0){$\times$}}
\put(78,28){\makebox(0,0){$\circ$}}
\put(108,28){\makebox(0,0){$\circ$}}
\put(138,28){\makebox(0,0){$\circ$}}
\put(168,28){\makebox(0,0){$\circ$}}
\put(198,28){\makebox(0,0){$\circ$}}

%treti rada
\put(48,58){\makebox(0,0){$\circ$}\makebox(0,0){$\times$}}
\put(78,58){\makebox(0,0){$\circ$}}
\put(108,58){\makebox(0,0){$\circ$}}
\put(138,58){\makebox(0,0){$\circ$}}
\put(168,58){\makebox(0,0){$\circ$}}
\put(198,58){\makebox(0,0){$\circ$}}

%ctvrta rada
\put(48,88){\makebox(0,0){$\circ$}\makebox(0,0){$\times$}}
\put(78,88){\makebox(0,0){$\circ$}}
\put(108,88){\makebox(0,0){$\circ$}}
\put(138,88){\makebox(0,0){$\circ$}}
\put(168,88){\makebox(0,0){$\circ$}}
\put(198,88){\makebox(0,0){$\circ$}}

%pata rada
\put(48,118){\makebox(0,0){$\circ$}\makebox(0,0){$\times$}}
\put(78,118){\makebox(0,0){$\circ$}}
\put(108,118){\makebox(0,0){$\circ$}}
\put(138,118){\makebox(0,0){$\circ$}}
\put(168,118){\makebox(0,0){$\circ$}}
\put(198,118){\makebox(0,0){$\circ$}}

\put(78,58){\put(-10,-5){\makebox(0,0)%
 {\scriptsize$(1,2)$}}\makebox(0,0){$\bullet$}\vector(0,1){28}%
 \vector(1,0){28}}
\put(108,28){\put(-10,-5){\makebox(0,0)%
 {\scriptsize$(2,1)$}}\makebox(0,0){$\bullet$}\vector(0,1){28}%
 \vector(1,0){28}}
\put(138,-2){\put(-10,-5){\makebox(0,0)%
 {\scriptsize$(3,0)$}}\makebox(0,0){$\bullet$}\vector(0,1){28}%
 \vector(1,0){28}}

\end{picture}
\end{center}
\caption{\label{fig4}
Bialgebra to Drinfel'd algebra deformation. The bottom row is
enabled
but not the left column. The deformed comultiplication need not be
coassociative, so the horizontal arrow at (2,1) need not give zero.
The
infinitesimal of $\Phi$ lies at position (3,0).}
\end{figure}
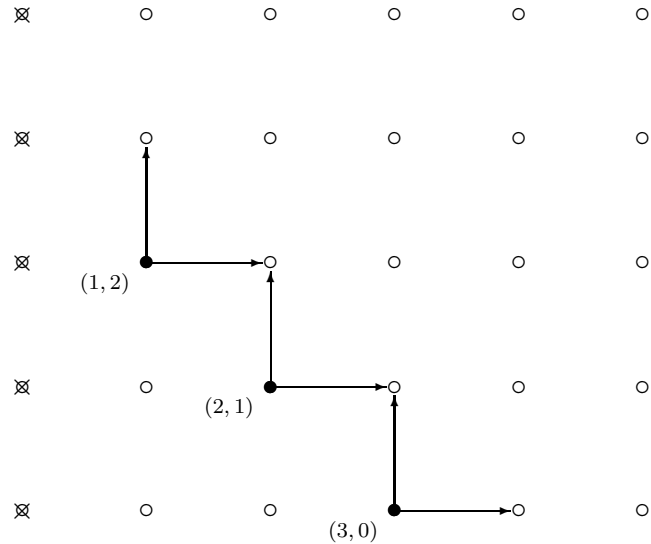

Now what happens if we start with an associative and coassociative
bialgebra,
but allow the deformed object to be a Drinfel'd algebra? Amazingly,
all that has to be done is to include the bottom row!
Indeed, the horizontal arrow from position $(2,1)$ may no longer
send the infinitesimal comultiplication to zero, since the
comultiplication need not be coassociative.
The infinitesimal deformation now includes
the infinitesimal of $\Phi$ which lies in $V^{\otimes 3} =
\hbox{Hom}_{\bk}(\bk,V^{\otimes 3})$ and so is located at $(3,0)$.
The fact that the
comultiplication is coassociative up to conjugation by
$\Phi$ translates, at the infinitesimal level, into the
assertion that the contributions as $(3,1)$ coming
horizontally from $(2,1)$ and vertically from $(3,0)$ cancel. The
pentagon
identity on $\Phi$ considered as a fourth structure equation
introduces
a fourth component to the obstruction cocycle.
At the infinitesimal level the pentagon identity implies that the
image of the infinitesimal of $\Phi$ in position $(4,0)$
 under the coalgebra differential
vanishes. The situation is as described by Figure~\ref{fig4}.

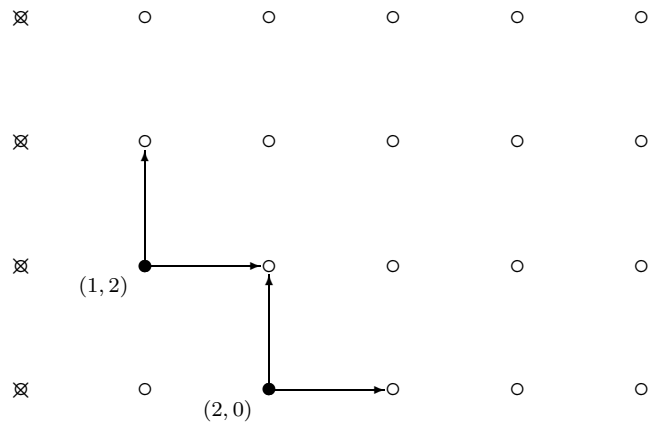
\begin{figure}
\setlength{\unitlength}{0.055cm}
\begin{center}
\begin{picture}(200,90)(10,-5)

%prvni rada
\put(48,-2){\makebox(0,0){$\circ$}\makebox(0,0){$\times$}}
\put(78,-2){\makebox(0,0){$\circ$}}
\put(108,-2){\makebox(0,0){$\circ$}}
\put(138,-2){\makebox(0,0){$\circ$}}
\put(168,-2){\makebox(0,0){$\circ$}}
\put(198,-2){\makebox(0,0){$\circ$}}

%druha rada
\put(48,28){\makebox(0,0){$\circ$}\makebox(0,0){$\times$}}
\put(78,28){\makebox(0,0){$\circ$}}
\put(108,28){\makebox(0,0){$\circ$}}
\put(138,28){\makebox(0,0){$\circ$}}
\put(168,28){\makebox(0,0){$\circ$}}
\put(198,28){\makebox(0,0){$\circ$}}

%treti rada
\put(48,58){\makebox(0,0){$\circ$}\makebox(0,0){$\times$}}
\put(78,58){\makebox(0,0){$\circ$}}
\put(108,58){\makebox(0,0){$\circ$}}
\put(138,58){\makebox(0,0){$\circ$}}
\put(168,58){\makebox(0,0){$\circ$}}
\put(198,58){\makebox(0,0){$\circ$}}

%ctvrta rada
\put(48,88){\makebox(0,0){$\circ$}\makebox(0,0){$\times$}}
\put(78,88){\makebox(0,0){$\circ$}}
\put(108,88){\makebox(0,0){$\circ$}}
\put(138,88){\makebox(0,0){$\circ$}}
\put(168,88){\makebox(0,0){$\circ$}}
\put(198,88){\makebox(0,0){$\circ$}}

\put(78,28){\put(-10,-5){\makebox(0,0)%
 {\scriptsize$(1,2)$}}\makebox(0,0){$\bullet$}\vector(0,1){28}%
 \vector(1,0){28}}
\put(108,-2){\put(-10,-5){\makebox(0,0)%
 {\scriptsize$(2,0)$}}\makebox(0,0){$\bullet$}\vector(0,1){28}%
 \vector(1,0){28}}

\end{picture}
\end{center}
\caption{\label{fig5}
Equivalence in bialgebra to Drinfel'd algebra deformations. The
infinitesimal
of the twist lies at position (2,0).}
\end{figure}

For Drinfel'd algebras there is a broader definition of
equivalence. One must
not only allow for change of coordinates, but also for twisting (see
the original paper~\cite{D} for the definition).
The infinitesimal of the twisting is located at position $(2,0)$.
So allowing for this modified infinitesimal
equivalence, two three cocycles of the total complex give equivalent
deformations if
they differ by a total coboundary as indicated in Figure~\ref{fig5}.

\begin{figure}
\setlength{\unitlength}{0.055cm}
\begin{center}
\begin{picture}(200,120)(15,-5)

%prvni rada
\put(48,-2){\makebox(0,0){$\circ$}\makebox(0,0){$\times$}}
\put(78,-2){\makebox(0,0){$\circ$}}
\put(108,-2){\makebox(0,0){$\circ$}}
\put(138,-2){\makebox(0,0){$\circ$}}
\put(168,-2){\makebox(0,0){$\circ$}}
\put(198,-2){\makebox(0,0){$\circ$}}

%druha rada
\put(48,28){\makebox(0,0){$\circ$}\makebox(0,0){$\times$}}
\put(78,28){\makebox(0,0){$\circ$}}
\put(108,28){\makebox(0,0){$\circ$}}
\put(138,28){\makebox(0,0){$\circ$}}
\put(168,28){\makebox(0,0){$\circ$}}
\put(198,28){\makebox(0,0){$\circ$}}

%treti rada
\put(48,58){\makebox(0,0){$\circ$}\makebox(0,0){$\times$}}
\put(78,58){\makebox(0,0){$\circ$}}
\put(108,58){\makebox(0,0){$\circ$}}
\put(138,58){\makebox(0,0){$\circ$}}
\put(168,58){\makebox(0,0){$\circ$}}
\put(198,58){\makebox(0,0){$\circ$}}

%ctvrta rada
\put(48,88){\makebox(0,0){$\circ$}\makebox(0,0){$\times$}}
\put(78,88){\makebox(0,0){$\circ$}}
\put(108,88){\makebox(0,0){$\circ$}}
\put(138,88){\makebox(0,0){$\circ$}}
\put(168,88){\makebox(0,0){$\circ$}}
\put(198,88){\makebox(0,0){$\circ$}}

%pata rada
\put(48,118){\makebox(0,0){$\circ$}\makebox(0,0){$\times$}}
\put(78,118){\makebox(0,0){$\circ$}}
\put(108,118){\makebox(0,0){$\circ$}}
\put(138,118){\makebox(0,0){$\circ$}}
\put(168,118){\makebox(0,0){$\circ$}}
\put(198,118){\makebox(0,0){$\circ$}}

\put(78,88){\put(-10,-5){\makebox(0,0)%
 {\scriptsize$(1,3)$}}\makebox(0,0){$\bullet$}\vector(0,1){28}%
 \vector(1,0){28}}
\put(108,58){\put(-10,-5){\makebox(0,0)%
 {\scriptsize$(2,2)$}}\makebox(0,0){$\bullet$}\vector(0,1){28}%
 \vector(1,0){28}}
\put(138,28){\put(-10,-5){\makebox(0,0)%
 {\scriptsize$(3,1)$}}\makebox(0,0){$\bullet$}\vector(0,1){28}%
 \vector(1,0){28}}
\put(168,-2){\put(-10,-5){\makebox(0,0)%
 {\scriptsize$(4,0)$}}\makebox(0,0){$\bullet$}\vector(0,1){28}%
 \vector(1,0){28}}

\end{picture}
\end{center}
\caption{\label{fig6}
The obstruction to extending the infinitesimal one higher
power in $t$ is a cocycle of total degree three.}
\end{figure}
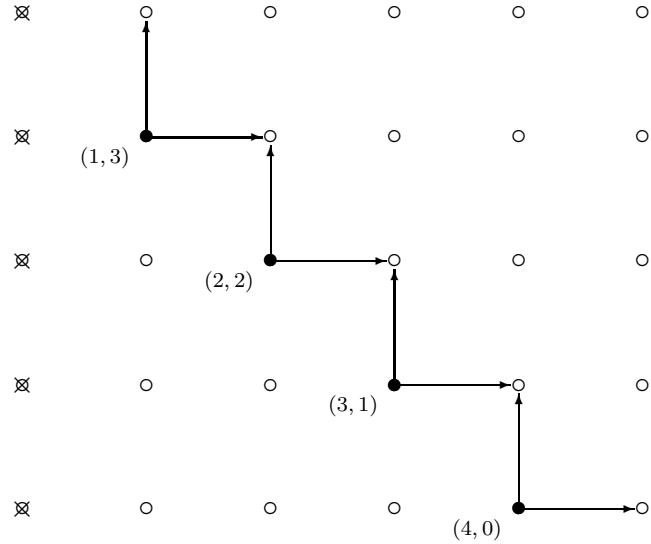

Once again, there is a quadratic expression in a putative
infinitesimal
deformation which is a cocycle of total degree three. The
vanishing of the corresponding total cohomology
class determines whether or not the infinitesimal deformation can be
extended to a power series of one higher order in $t$. See
Figure~\ref{fig6}.

This, in outline, is the cohomological approach to the theory of
deformations
of bialgebras, when the deformed object is a Drinfel'd algebra. What
happens
when the undeformed structure is that of a Drinfel'd algebra and the
comultiplication is not necessarily associative? The first problem is
that the standard construction of the cobar differential does not have
square zero. We must consider the maps $(\Delta\otimes \id)\Delta$ and
$(\id\otimes \Delta)\Delta$ as taking values in different
three-fold tensor products defined by different bracketings.
After distinguishing between different bracketings we use
the quasi-coassociativity condition~(\ref{qcoass}) to define an
equivalence between them.
This leads to a modified cobar resolution whose construction is
reviewed in
Section 2.

There is no need to change the definition of the bar resolution since
a Drinfel'd algebra is, in particular, an associative algebra.
However the complex which captures both the algebra and coalgebra
structures
consists of the
$A$-bimodule, $A$-bicomodule homomorphisms from the bar resolution to
the
cobar resolution and this requires the definition of an $A$-bicomodule
structure on the bar resolution and an $A$-bimodule structure on the
modified cobar resolution. Once again there are some surprises.
The $A$-bimodule structure on the modified cobar resolution presents
no difficulties, but the $A$-bicomodule structure on the
bar resolution does. In general the definition of comodules over a
Drinfel'd
algebra requires special
attention since the usual compatibility
conditions for comodules are based on
the assumption that the comultiplication is coassociative. There is a
natural extension of the definition of an $A$-(bi)comodule which
involves
using $\Phi$ to relate the different placement of parentheses. The
new ingredient which complicates (we think ``enriches'') our theory
is that the obvious definition of a comodule structure on the bar
complex,
the one which reduces to the standard one for a bialgebra, does not
satisfy the required compatibility condition involving $\Phi$. Thus we
are led to introduce the notion of a homotopy comodule structure
in Section 3.

This definition involves a hierarchy of operators
generalizing the standard comodule structure operators. The first new
operator can be defined directly but the complications involved
in defining the higher operators becomes unmanageable very quickly
without some auxiliary construction. The key here is the
definition of a differential graded Lie algebra structure
on the simplicial chain
complex of the associahedra as given in Section 4.
 This allows us to keep track of the
intricate combinatorics related to nonassociativity.

In Section 5 we define a homomorphism of differential graded Lie
algebras from the complex defined in Section 4 to the derivations of the
bar complex. In Section 6 we describe the relation of all the
preceding
constructions to the deformation theory of Drinfel'd algebras. As an
example we consider the Drinfel'd quantization of $U({\cal G})$ for
a simple Lie algebra ${\cal G}$. The
last section contains the proofs of some technical lemmas.

\section{Algebraic preliminaries}

All algebraic objects in the paper will be considered over a fixed
field $\bk$ of characteristic zero. We will systematically use the
Koszul sign convention meaning that whenever we commute two
``things''
of degrees $p$ and $q$, respectively,
we multiply by the sign $(-1)^{pq}$. For a graded vector space $M$,
let $\susp M$ (resp. $\desusp M$) be
the suspension (resp. the desuspension) of $M$, i.e. the graded
vector space defined by $(\susp M)_p: = M_{p-1}$ (resp. $(\desusp M)_p
:= M_{p+1}$).
We have the obvious
natural isomorphisms $\susp : M \to \susp M$ and $\desusp : M\to
\desusp M$.

For a $(V,\MU)$-module $N$, define the following generalization of
the $M$-{\em construction\/} of \cite[par.~3]{MS}.
Let $F^*=\bigoplus_{n\geq 0}F^n$ be
the free unitary nonassociative $\boldk$-algebra generated by
$N$, graded by the length of words. The space $F^n$
is the direct sum of
copies of $N^{\ot n}$ over the set $\br n$ of full bracketings of $n$
symbols, $F^n = \bigoplus_{u\in {\rm Br}_n}N^{\ot n}_u$.
For example, $F^0=\bk$, $F^1 = N$, $F^2 =
N^{\ot 2}$, $F^3 = N^{\ot 3}_{(\b\b)\b} \oplus
N^{\ot 3}_{\b(\b\b)}$,~etc.
The algebra $F^*$ admits a
natural left action,
$(a,f)\mapsto a\b f$, of $(V,\MU)$ given by the rules:
\begin{itemize}
\item[(i)]
on $F^0=\bk$, the action is given by the augmentation $\epsilon$,
\item[(ii)]
on $F^1=N$, the action is given by the action of $V$ on $N$
and
\item[(iii)]
$a\b (f\star g)= \sum(\Delta'(a)\b f)\star(\Delta''(a)\b g)$,
\end{itemize}
where $\star$ stands for the multiplication in $F^*$ and we use the
notation $\Delta(a)= \sum \Delta'(a)\otimes \Delta''(a)$. The right
action $(f,b)\mapsto f\b b$ is defined by similar rules. These
actions define on $F^*$ the structure of a $(V,\MU)$-bimodule.

Let $\sim$ be the relation on $F^*$ $\star$-multiplicatively generated
by the expressions of the form
\begin{equation}
\label{relation}
\Sum\left((\Phi_1\b x)\star\left((\Phi_2\b
y)\rule{0mm}{4mm}\star(\Phi_3\b
z)\right)\right)
\sim\Sum\left(\left((x\b\Phi_1)\star\rule{0mm}{4mm}
(y\b\Phi_2)\right)\star
(z\b\Phi_3)\right),
\end{equation}
where $\Phi = \sum\Phi_1\otimes \Phi_2\otimes\Phi_3$ and $x,y,z\in
F^*$.
Put $\bigodo(N):= F/\sim$.
Just as in~\cite[Proposition~3.2]{MS}
one proves that the $\b$-action induces on $\bigodo(N)$
the structure of
a $(V,\MU)$-bimodule (denoted again by~$\b$) and that
$\star$ induces
on $\bigodo(N)$ a nonassociative multiplication denoted by
$\odot$. The operations are related by
\begin{equation}
\label{opulka}
a\b (f\od g)= \Sum(\Delta'(a)\b f)\od(\Delta''(a)\b g)
\mbox{ and }
(f\od g)\b b = \Sum(f\b \Delta'(b))\od (g \b \Delta''(b)),
\end{equation}
for $a,b\in V$ and $f,g\in \bigodo(N)$.
The multiplication $\odot$ is quasi-associative in the sense
\begin{equation}
\label{quasi-ass}
\Sum(\Phi_1\b x)\odot\left((\Phi_2\b
y)\rule{0mm}{4mm}\odot(\Phi_3\b
z)\right)
=\Sum\left((x\b\Phi_1)\odot\rule{0mm}{4mm}
(y\b\Phi_2)\right)\odot
(z\b\Phi_3)
\end{equation}
 Since the defining
relations~(\ref{relation}) are homogeneous with respect to length,
the grading of $F^*$ induces on $\bod(N)$ the grading $\bod^*(N)=
\bigoplus_{i\geq 0}\bod^i(N)$. If $N$ itself is a graded vector
space, we have also the obvious second grading, $\bod(N) =
\bigoplus_{j}\bod(N)^j$, which coincides with the first grading if
$N$ is concentrated in degree~1.

Let $\mbox{\rm Der}^n_V(\bod N)$ denote the set of
$(V,\MU)$-linear derivations of degree $n$ (relative to the
second grading) of the (nonassociative) graded algebra
$\bod(N)^*$. One sees immediately that there is an one-to-one
correspondence between the elements $\theta \in\mbox{\rm
Der}^n_V(\bod N)$
and $(V,\MU)$-linear homogeneous degree~$n$ maps $f:N^*\to
\bod(N)^*$.

For the Drinfel'd algebra itself, $\Delta:V\rightarrow V\otimes V=
V\odot V$ determines an element
$d_C\in \mbox{\rm Der}^1_V(\bod V).$ By the quasi-coassociativity
relation~(\ref{qcoass}) we have $(\id\ot \Delta)\Delta(v) =
\Phi\cdot(\Delta
\ot \id)\Delta(v) \cdot \Phi^{-1}$ and the defining relation of the
$M$-construction~(\ref{relation}) thus give
$(\Delta \odot \id)\Delta(v) =
(\id\odot \Delta )\Delta(v)$ for $v\in V$,
therefore $d_C^2=0$ follows from the
same line of argument as in the coassociative case.
Then $(\bod V, d_C)$ is the {\em modified cobar resolution}. For more
details see \cite{MS,shnider-sternberg:preprint}.

If $N=X\oplus Y$, then $\bod(X\oplus Y)$ is naturally
bigraded, $\bod^{*,*}(X\oplus Y)= \bigoplus_{i,j\geq
0}\bod^{i,j}(X\oplus Y)$, this bigrading being defined by saying that
a monomial $w$ belongs to $\bod^{i,j}(X\oplus Y)$ if there are exactly
$i$
(resp.~$j$) occurrences of the elements of $X$ (resp.~$Y$) in $w$.
If $X,Y$ are graded vector spaces then there is a second bigrading
just as above.

Let $(\bv,\db)$ be the (two-sided) normalized
bar resolution of the algebra
$(V,\MU)$ (see~\cite[Chapter~X]{McL}),
but considered with the opposite grading.
This means that $\bv$ is
the graded space, $\bv = \bigoplus_{n\leq
1}\calb_n(V)$, where $\calb_1(V)=V$, $\calb_0(V)=V\otimes V$, and
for $n\leq- 1$, $\calb_n(V)$ is the free $(V,\MU)$-bimodule on
$\oV^{\otimes( -n)}$,
i.e.~the vector space $V\ot \oV^{\otimes(-n)}\ot V$
with the action of ($V,\MU)$ given by
\[
u\cdot (a_0\otimes\cdots\otimes a_{-n+1}):= (u\cdot
a_0\otimes\cdots\otimes a_{-n+1})\
\mbox{ and }\
(a_0\otimes\cdots\otimes a_{-n+1})\cdot w := (a_0\otimes\cdots\otimes
a_{-n+1}\cdot w)
\]
for $u,v,a_0,a_{-n+1}\in V$ and $a_1,\ldots,a_{-n}\in \oV$.
If we use the more compact
notation (though a nonstandard one), writing $(a_0|\cdots|a_{-n+1})$
instead of $a_0\otimes\cdots\otimes a_{-n+1}$, the differential $\db
: \calb_n(V)
\to \calb_{n+1}(V)$ is defined as
\[
\db(a_0|\cdots|a_{-n+1}):=
\sum_{0\leq i\leq -n}(-1)^{i}(a_0|\cdots|a_i\cdot a_{i+1}|
\cdots|a_{-n+1}).
\]
Here, as is usual in this context, we make no distinction between
the elements of $V/\bk \cdot 1$ and their representatives in $\oV$. We
use the same convention throughout all the paper.

Notice that the differential $\db$ is a $(V,\MU)$-linear map. We
have two more $(V,\MU)$-linear maps, namely the `coactions' $\lambda
:\bv \to V\odot \bv \mbox{ and } \rho : \bv \to \bv\odot V$ given by
\begin{eqnarray*}
\lambda(a_0|\cdots|a_{-n+1})&:=&
\sum\Delta'(a_0)\cdots\Delta'(a_{-n+1})\od
(\Delta''(a_0)|\cdots|\Delta''(a_{-n+1})),\mbox{ and}
\\
\rho(a_0|\cdots|a_{-n+1})&:=&\sum
(\Delta'(a_0)|\cdots|\Delta'(a_{-n+1}))\od\Delta''(a_0)\cdots
\Delta''(a_{-n+1}).
\end{eqnarray*}
A simple computation shows that these ``coactions'' are compatible
with the differential in the sense that
\begin{equation}
\label{compatibility}
(\id \od \db)(\lambda) =\lambda(\db)\mbox{ and }
(\db \od \id)(\rho) = \rho(\db).
\end{equation}

Most of the constructions presented here would work also when we
use the un-normalized bar resolution instead of the normalized one,
but in the normalized case we have better control over the objects
of Sublemma~\ref{slune}.

\section{Homotopy comodule structure on the bar resolution}

Recall that a (differential) {\em bicomodule\/} over a coassociative
coalgebra $B = (V,\Delta)$ is a (differential) graded vector space
$(M,d_M)$, $\deg(d_M) = 1$, together with two homogeneous degree zero
linear maps $\lambda : M \to V\ot M$ and $\rho : M \to M\ot V$ (the
{\em coactions\/}) such that
\begin{eqnarray*}
& (\id \ot d_M)(\lambda) = \lambda (d_M),\
(d_M \ot \id)(\rho) = \rho( d_M),
\\
&(\Delta\ot \id)(\lambda) = (\id\ot\lambda)(\lambda),\
 (\lambda\ot \id)(\rho) = (\id\ot\rho)(\lambda)\
\mbox{ and }
(\id \ot \Delta)(\rho)= (\rho\ot\id)(\rho).&
\end{eqnarray*}
Here, of course, $V$ is considered as a graded vector space
concentrated in degree zero.

We formulate the following lemma as a motivation for the definition of
the homotopy comodule structure given in Definition~\ref{masinka}.
The proof is an easy exercise and we leave it to the reader.

\begin{lemma}
\label{oslicek}
\hskip-3pt
Let $(M,d_M)$ be a differential graded vector space and let $\otvm\!
:= \! \mbox{$\bigotimes(\susp\! V\!\bigoplus\! \susp\! M)$}$.
Let $(\lambda,\rho)$ be a
$(V,\Delta)$-bicomodule structure on $(M,d_M)$ and define
$D \in \Der^1(\otvm)$ as $D := D_{-1} + D_0$, where
\begin{eqnarray*}
&D_{-1}|_{\susp M}:= \susp d_M \desusp,\ D_{-1}|_{\susp V}:= 0,&
\\
&
D_0|_{\susp M}:= (\susp\ot\susp)(\lambda+\rho)(\desusp),
\mbox{ and }
D_0|_{\susp V}:= (\susp\ot\susp)(\Delta)(\desusp).&
\end{eqnarray*}
Then $D^2=0$.

The above formula defines a one-to-one correspondence
between $(V,\Delta)$-bicomodule structures on $(M,d_M)$ and
differentials $D=D_{-1}+D_0 \in \Der^1(\otvm)$ where the
components $D_i$ are of
degree $i+1$ relative to the first grading in $V$
(the number of factors of $V$) and satisfy the conditions
$$D_{-1}|_{\susp V}= 0,\ \mbox{ and }\ D_0|_{\susp V}=
(\susp\ot\susp)(\Delta)(\desusp).$$
\end{lemma}

We think that some comments on how to interpret the formulas of the
lemma are in order. Let
$m\in M$ be a homogeneous element, $\deg(m) = p$. We write as usual
$\lambda(m) = \sum \lambda'(m)\ot \lambda''(m)$ and $\rho(m) = \sum
\rho'(m)\ot \rho''(m)$ with $\lambda'(m),\rho''(m)\in V$ and
$\lambda''(m),\rho'(m)\in M_p$. Then
\begin{eqnarray*}
D_0(\susp m) &=&
(\susp \ot \susp)(\lambda+\rho)(m) = (\susp \ot \susp)(\lambda'(m)\ot
\lambda''(m)) + (\susp \ot \susp)(\rho'(m)\ot\rho''(m))
\\
&=& \susp \lambda'(m)\ot \susp \lambda''(m)+
(-1)^p (\susp \rho'(m)\ot \susp \rho''(m)).
\end{eqnarray*}
Here the sign $(-1)^p$ is due to the commuting of
$\rho'(m)$ (a ``thing'' of
degree $p$) and the map $\susp$ (a ``thing'' of degree $1$).
It is useful sometimes to identify an element with its image under
the map $\susp$ (or $\desusp$). Under this identification, the above
equation can be written as
\[
D_0(m) = \lambda(m) + (-1)^{(|m|+1)}\cdot \rho(m).
\]
Notice the unexpected sign $(-1)^{(|m|+1)}$ which is due to the fact
that $m$ here represents $\susp m$, and $|\susp m| = p+1$. We see
that we must be extremely careful when using this shorthand,
especially as far as the sign issue is concerned.

Suppose for the moment that $A=(V,\MU,\Delta,\Phi)$ is a Drinfel'd
algebra with $\Phi = 1$. In this case $(V,\MU,\Delta)$ is an ordinary
(associative and coassociative)
bialgebra and $\odot$ can be replaced by $\otimes$
 everywhere in the paragraph defining the coactions on $\bv$.
One easily verifies that
$\lambda: \bv \to V \ot \bv$ and $\rho: \bv \to \bv \ot V$
describe a $(V,\Delta)$-bicomodule structure on the
differential space $(\bv,\db)$. Lemma~\ref{oslicek} then gives a
differential $D = D_{-1} + D_0 \in \Der^1(\bigotimes(V,\bv)$.

In the general case with $\Phi \not= 1$ all the above definitions make
sense, but we need not have $D^2=0$. Let us discuss this situation
more carefully. Put $\bodvbv := \bod(\susp V \oplus \susp \bv)$ and
define the derivations $D_{-1}, D_0\in \Der^1(\bodvbv)$ by
\begin{eqnarray*}
&D_{-1}|_{\susp \bv}:= \susp \db \desusp,\ D_{-1}|_{\susp V}:= 0,&
\\
& D_0|_{\susp \bv}:= (\susp\bod\susp)(\lambda+\rho)(\desusp),
\mbox{ and }
D_0|_{\susp V}:= (\susp\bod\susp)(\Delta)(\desusp).&
\end{eqnarray*}
Because of the $(V,\MU)$-linearity of the maps $\db$,
$\lambda$ and $\rho$,
the derivations $D_{-1}$ and $D_0$ are each
$(V,\MU)$-linear, $D_{-1},D_0
\in \der 1$ and clearly
\[
D_{-1}(\bod(V,{\cal B}(V))^{i,j})
\subset \bod(V,{\cal B}(V))^{i,j+1}\mbox{ and }
D_0(\bod(V,{\cal B}(V))^{i,j})
\subset \bod(V,{\cal B}(V))^{i+1,j}
\]
for any $n\leq -1$ and $i,j \geq 0$. The desired condition $D^2=0$
requires
\begin{equation}
\label{kralicek}
D_{-1}^2=0,\ D_{-1} D_0 + D_0 D_{-1} = 0\mbox{ and }
D_0^2 = 0.
\end{equation}

\begin{lemma}
\label{kozulka}
In the situation above, we have two of the three conditions for
$D^2=0$,
namely,
\[
D_{-1}^2=0\mbox{ and the graded commutator } D_{-1} D_0+D_0 D_{-1} =
0.
\]
\end{lemma}

\noindent{\bf Proof.}
It is enough to verify the conditions on the
generators $\susp V$ and $\susp \bv$ of $\bodvbv$. As for the first
condition, $D_{-1}^2|_{\susp V}$ is obviously trivial while
$D_{-1}^2|_{\susp \bv} = \susp \db^2 \desusp =0$.

The second equation is obviously trivial on $\susp V$ while on $\susp
\bv$ we obtain
\begin{eqnarray*}
 D_{-1} D_0 + D_0 D_{-1}&=&
D_{-1} (\susp \odot \susp)(\lambda+\rho)(\desusp) + D_0 \susp \db
\desusp
\\
&=& -(\susp \odot (\susp\db))(\lambda)( \desusp) + ((\susp \db) \odot
\susp)(\rho) (\desusp) + (\susp \odot \susp)(\lambda+\rho)(\db)(
\desusp)
\\
&=&
(\susp \odot \susp)[-(\id \odot \db)(\lambda) - (\db \odot \id)(\rho)
+ (\lambda+\rho)(\db)] (\desusp)
\end{eqnarray*}
which follows from~(\ref{compatibility}).
The sign introduced in the second line comes from moving $D_{-1}$ past
$\susp$ and the sign in the third line comes from moving $\susp$
past $\db$.\qed

Let us discuss the condition $D_0^2 =0$. On $\susp V$ we have
\begin{eqnarray*}
D_0^2 &=& D_0(\susp \odot \susp)(\Delta)(\desusp) =
[((\susp \odot \susp)(\Delta)\odot \susp) - (\susp \odot(\susp \odot
\susp)(\Delta))](\Delta)(\desusp)
\\
&=&
(\susp \odot \susp \odot \susp)[(\Delta \odot \id)\Delta - (\id \odot
\Delta)\Delta](\desusp).
\end{eqnarray*}
Notice that, for $v\in V$,
$(\Delta \odot \id)\Delta(v) \in \bod^3(V)$ is represented by
$(\Delta \ot \id)\Delta(v) \in V^{\ot 3}$, while
$(\id\odot \Delta )\Delta(v)
\in \bod^3(V)$ is represented by
$(\id\ot \Delta)\Delta(v) \in V^{\ot 3}$. The condition
$(\Delta \odot \id)\Delta(v) =
(\id\odot \Delta )\Delta(v)$ implies that $D_0^2|_{\susp V}=0$
just as it implies $d_C^2=0$.
On $\susp \bv$ we have
\[
D_0^2 = (\susp \odot \susp \odot \susp)[ (\Delta\odot \id)\lambda -
(\id \odot \lambda)\lambda -(\id \odot\rho)\lambda
+ (\lambda \odot \id)\rho + (\rho \odot \id)\rho
-(\id \odot \Delta)\rho ](\desusp)
\]
which is equivalent to the following three equations
\begin{equation}
\label{comodule}
(\Delta\odot \id)\lambda =
(\id \odot \lambda)\lambda,\
(\lambda \odot \id)\rho = (\id \odot
\rho)\lambda \mbox{ and } (\id \odot \Delta)\rho =
(\rho \odot \id)\rho
\end{equation}
which are obvious analogs of the last three
conditions from the definition
of a bicomodule. These conditions may be violated
already on ${\cal B}_{-2}(V)$ as we explain after the
next definition.

Lemma~\ref{kozulka} suggests the possibility of interpreting
the derivation $D_0$ as
an infinitesimal deformation of $D_{-1}$ and we
may try to integrate
this infinitesimal deformation.
The formal definition is the following.

\begin{definition}
\label{masinka}
By a homotopy $(V,\Delta)$-bicomodule structure on
$(\bv,\db)$ we mean
a sequence $\{D_k \in \der 1;\ k\geq 1\}$ such
that (relative to the first bigrading)
\[
D_k(\bod^{i,j}(V,{\cal B}(V)))\subset \bod^{i+k+1,j}(V,{\cal
B}(V))
\]
for $i,j \geq 0$, $k\geq 1$, and
that $D:= D_{-1} + D_0 + \sum_{k\geq 1}D_k$ has square zero,
$D^2=0$.
\end{definition}

One of the central results
of the paper (Corollary~\ref{central}) is that such a
homotopy bicomodule structure exists.

We now come back to the failure of ${\cal B}_{-2}(V)$
to satisfy the comodule compatibility conditions.
Take for example the first equation of~(\ref{comodule}).
Using generalized Sweedler notation, we set
\[
(\Delta\otimes \id)\Delta(a)=\Sum a'_{(1)}\otimes a'_{(2)}
\otimes a'_{(3)}\quad ( \id\otimes \Delta)\Delta(a)=
\Sum a''_{(1)}\otimes a''_{(2)}
\otimes a''_{(3)}.
\]
For $(a|b)\in V\otimes V = {\cal B}_{-2}(V)$ we get
\begin{eqnarray*}
\lefteqn{
[(\Delta\odot \id)\lambda-(\id\odot
\lambda)\lambda](a|b)=}
\\
&&\!\!=\!\Sum \{(a'_{(1)}b'_{(1)}\odot a'_{(2)}b'_{(2)})\odot
(a'_{(3)}| b'_{(3)})- a''_{(1)}b''_{(1)}\odot
(a''_{(2)}b''_{(2)}\odot(a''_{(3)}| b''_{(3)}))\}
\\
&&\!\!=\!\Phi^{-1}\b\Sum\left\{ (\Phi_1 a'_{(1)}b'_{(1)}\odot
\Phi_2 a'_{(2)}b'_{(2)})\odot(\Phi_3 a'_{(3)}| b'_{(3)})
- (a''_{(1)}b''_{(1)}\Phi_1\odot a''_{(2)}b''_{(2)}\Phi_2)
\odot (a''_{(3)}|b''_{(3)}\Phi_3)\right\}
\\
&&\!\!=\!\Phi^{-1}\b\Sum\left\{ (a''_{(1)}\Phi_1 b'_{(1)}\odot
a''_{(2)}\Phi_2 b'_{(2)})\odot
(a''_{(3)}\Phi_3| b'_{(3)})
- (a''_{(1)}\Phi_1 b'_{(1)}\odot
a''_{(2)}\Phi_2b'_{(2)})\odot (a''_{(3)}|\Phi_3b'_{(3)})\right\}
\\
&&\!\!=\!\Phi^{-1}\b\Sum\left\{ (a''_{(1)}\Phi_1b'_{(1)}\odot
a''_{(2)}\Phi_2 b'_{(2)})\odot
[(a''_{(3)}\Phi_3|b'_{(3)})-
(a''_{(3)}|\Phi_3b'_{(3)})]\right\}
\end{eqnarray*}
The expression
$(a''_{(3)}\Phi_3| b'_{(3)})- (a''_{(3)}|\Phi_3b'_{(3)})$
inside the right brackets in the last line is exactly
$D_{-1}(a''_{(3)}| \Phi_3|b'_{(3)})$
therefore the deviation is, up to the
overall factor, $\Phi^{-1}$ equal to $(\id\odot \id\odot D_{-1})$
acting on
\begin{equation}
\Sum (a''_{(1)}\Phi_1b'_{(1)}\odot a''_{(2)}\Phi_2b'_{(2)})\odot
(a''_{(3)}|\Phi_3 | b'_{(3)}).\label{13.5.1}
\end{equation}
This suggests the form for the operator $D_1$.
Consider the diagram for the Stasheff associahedron $K_3$ (see the
following paragraph for the definition)
\begin{equation}
(\b\b)\b\stackrel{\Phi}{\longrightarrow}\b(\b\b )\label{13.5.2}
\end{equation}
and the comultiplications associated to the two vertices
$$
\Delta^{(\b\b)\b}(a)=\Sum a'_{(1)}\otimes a'_{(2)}\otimes
a'_{(3)}\quad\quad \Delta^{\b(\b\b)}(a)=
\Sum a''_{(1)}\otimes a''_{(2)}\otimes a''_{(3)}.
$$
Then (\ref{13.5.1}) is a kind of product of three terms
$$\Delta^{\b(\b\b)}(a)\quad\Phi\quad \Delta^{(\b\b)\b}(b)$$
corresponding to (\ref{13.5.2}) but reversing the order
because in the correspondence between functional notation
and arrow notation,
\[
a\quad \stackrel{T}{\longrightarrow}\quad b\quad\hbox{becomes}\quad
 b\quad=\quad T(a).
\]

All the operators $D_{n-2}$ will correspond to the geometry of the
associahedron $K_n$
in precisely this way.
For any $k$ simplex $\sigma=[v_0, v_1,\dots,v_k]$
spanned by $k+1$ of the vertices
of $K_n$ we will define an operator $M_n[\sigma]$
which is a product of higher order comultiplications and
coassociativity
operators.
The details will be given in the next two sections.

\section{Some properties of the Stasheff associahedra}

Let $K_n$ be, for $n\geq 2$, the Stasheff
associahedron~\cite{stasheff:TAMS63}.
It is an $(n-2)$-dimensional cellular complex whose $i$-dimensional
cells are indexed by the set $\Br in$ of all (meaningful) insertions
of $(n-i-2)$ pairs of brackets between $n$ symbols, with suitably
defined incidence maps. There is, for any $a,b \geq 2$, $0\leq i \leq
a-2$, $0\leq j \leq b-2$ and $1\leq t \leq b$, a map
\[
(-,-)_t : \Br ia \times \Br jb \to \Br{i+j}{a+b-1},\ u\times v \mapsto
(u,v)_t,
\]
where $(u,v)_t$ is given by the insertion of $(u)$ at the $t$-th
place in $v$.
This map defines, for $a,b \geq 2$ and $1\leq t\leq b$, the inclusions
$\iota_t :K_a\times K_b \hookrightarrow \partial K_{a+b-1}$ and it is
known that
\begin{equation}
\label{boundary}
\partial K_n = \textstyle\bigcup_J \iota_t (K_a \times K_b),
\end{equation}
where $J=\{(a,b,t);\ a+b=n+1, a,b\geq 2 \mbox{ and }1\leq t\leq
b\}$.

Let $\prec$ be the partial order on the set $\br n:= \Br 0n$ of the
vertices of $K_n$ defined by saying that
$u\prec v$ if and only if $v$ is obtained from $u$ by the
substitution $(w_1,w_2)w_3\mapsto w_1(w_2,w_3)$ with some
$w_i\in \br{n_i}$, $i=1,2,3$, $n_1+n_2+n_3\leq n$. Let $\xi_n$ be
the (unique) minimal element of $\br n$.

The following proposition, as well as Proposition~\ref{laurinka}, is
of a purely combinatorial nature and can be proved by the induction.
We
leave the proofs of both propositions to the reader.

\begin{proposition}
\label{observation}
\[
K_n = \mbox{\rm Cone}
(\textstyle\bigcup_{J'}\iota_t (K_a \times K_b);\xi_n),
\]
where $J'=\{(a,b,t) \in J;\ t\geq 2\}$.
\end{proposition}

The cone structure above induces on $K_n$ inductively the
triangulation $\T n$ as follows:
\begin{enumerate}
\item[(i)]
$\T2$ is the only possible triangulation of $K_2=\{\b\b\}$ and
\item[(ii)]
for $n\geq 2$, $\T n$ is induced by the cone structure
of Proposition~\ref{observation} from
the triangulations $\T a\times \T b$ of $K_a\times K_b$,
$a+b=n+1$, $a,b\geq 2$, which are already defined, by the induction
assumption.
\end{enumerate}

\begin{proposition}
\label{laurinka}
The triangulation $\T n$ has the property that $\prec$
induces a total order on the vertices of any simplex
of $\T n$.
\end{proposition}

\begin{figure}
\setlength{\unitlength}{.05cm}
\begin{center}
\thicklines
\begin{picture}(100,100)(-50,-12.5)
\def\id{1\!\! 1}
\put(-25,0){\vector(1,0){48}}
\put(25,0){\vector(1,2){24}}
\put(0,75){\vector(2,-1){48}}
\put(-50,50){\vector(2,1){48}}
\put(-50,50){\vector(1,-2){24}}
\put(-15,-10){\mbox{\scriptsize$(\id\otimes \Delta\otimes \id)\Phi$}}
\put(-25,55){\mbox{\scriptsize
$(\id^2\otimes \Delta)\Phi(\Delta\otimes\id^2)\Phi$}}
\put(-30,20){\mbox{\scriptsize
$[(\id\otimes\Delta\otimes\id)\Phi] (\Phi\otimes 1)$}}
\put(-60,25){\mbox{\scriptsize$\Phi\otimes 1$}}
\put(-50,65){\mbox{\scriptsize$(\Delta\otimes \id^2)\Phi$}}
\put(25,65){\mbox{\scriptsize$\id^2\otimes \Delta\Phi$}}
\put(45,30){\mbox{\scriptsize$1\otimes\Phi$}}

%koule v rozich
\put(25,0){\makebox(0,0){$\bullet$}}
\put(50,50){\makebox(0,0){$\bullet$}}
\put(0,75){\makebox(0,0){$\bullet$}}
\put(-50,50){\makebox(0,0){$\bullet$}}
\put(-25,0){\makebox(0,0){$\bullet$}}
\def\b{\bullet}

%zavorkovani ve vrcholech
\put(30,0){\scriptsize{$\b((\b\b)\b)$}}
\put(55,50){\scriptsize{$\b(\b(\b\b))$}}
\put(-10,85){\scriptsize{$(\b\b)(\b\b)$}}
\put(-75,50){\scriptsize{$((\b\b)\b)\b$}}
\put(-50,0){\scriptsize{$(\b(\b\b))\b$}}

\put(-50,50){\line(3,-2){30}}
\put(10,10){\vector(3,-2){15}}
\put(-50,50){\vector(1,0){99}}

\end{picture}
\end{center}
\caption{\label{fig7}
The distinguished triangulation of $K_4$.}
\end{figure}
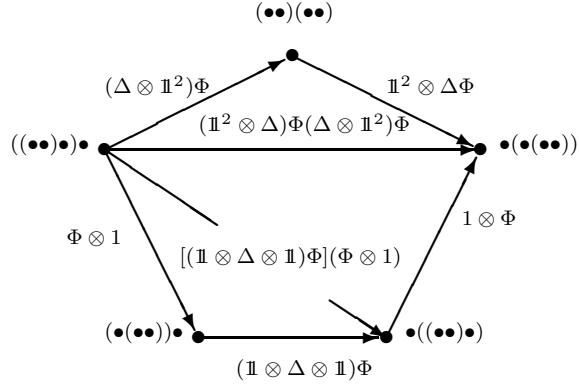

We will consider $\T n$ as an oriented triangulation with
the orientation induced by $\prec$. It will also be necessary
to consider the reversed orientation which we will denote
$\tilde\T n$.
Figure~\ref{fig7} shows the triangulation $\T n$ for the pentagon
$K_4$. This figure shows also the associativity operators which we
introduce later.

The following definition is a differential and bigraded analog of the
``pre-Lie system'' introduced in~\cite{G} (compare also the notion of
a nonunitary differential operad of~\cite{M:operads}).

\begin{definition}
\label{comp}
By a (right) differential comp algebra we mean a bigraded differential
space $(\X**,d_X)$, $d_X(\X **)\subset \X{*-1}*$, together
with a system of bilinear operations
\[
\dia_t : \X pa \otimes \X qb \to \X{p+q}{a+b-1}
\]
given for any $1\leq t \leq b$ such that, for $\bolds \in \X pa$,
$\boldt \in \X qb$ and $\boldu \in \X rc$,
\begin{equation}
\label{111}
\bolds\dia_i(\boldt\dia_j\boldu) =
\left\{
\begin{array}{ll}
(-1)^{p\cdot q}\cdot \boldt\dia_{j+a-1}(\bolds\dia_i\boldu),&
\mbox{ for }1\leq i\leq j-1,
\\
(\bolds \dia_{i-j+1}\boldt)\dia_j\boldu, &
\mbox{ for }j\leq i \leq b+j-1, \mbox{ and}
\\
(-1)^{p\cdot q}\cdot\boldt\dia_j(\bolds\dia_{i-b+1}\boldu),&
\mbox{ for }i\geq j+b.
\end{array}
\right.
\end{equation}
and, for any $\bolds \in \X pa$, and $\boldt \in \X qb$,
$1\leq t \leq b$,
\begin{equation}
\label{222}
d_X(\bolds\dia_t \boldt) = d_X(\bolds)\dia_t \boldt + (-1)^p\cdot
\bolds\dia_t d_X(\boldt).
\end{equation}
\end{definition}

Let $\ck in$ denote the set of $i$-dimensional oriented
simplicial chains of
$\T n$ with coefficients in $\boldk$ and let $d_S
:\ck in\to \ck {i-1}n$ be the simplicial differential.
For $\bolds\in \ck pa$ and $\boldt \in \ck qb$, $p,q \geq 0$,
$a,b\geq 2$ and $1\leq t\leq b$,
let $\bolds \times\boldt \in C_{p+q}(K_a\times K_b)$
denote the simplicial
homology cross product (see~\cite[VIII.8]{McL}) and put
\begin{equation}
\label{ccc}
\bolds\diamond_t\boldt:=
({\iota_t})_*
(\bolds\times\boldt)\in \ck {p+q}{a+b-1}.
\end{equation}

\begin{proposition}
\label{3.4}
The simplicial chain complex $(\ck **,d_S)$ together with
operations $\dia_i$
introduced above is a
differential comp algebra in the sense of Definition~\ref{comp}.
\end{proposition}

\noindent{\bf Proof.}
Elementary combinatorial arguments give that, for
$\alpha\in \Br pa$, $\beta
\in \Br qb$ and $\gamma \in \Br rc$,
\[
(\alpha,(\beta,\gamma)_j)_i =
\left\{
\begin{array}{ll}
(\beta,(\alpha,\gamma)_i)_{j+a-1},&
\mbox{ for }1\leq i\leq j-1,
\\
((\alpha,\beta)_{i-j+1},\gamma)_j,&
\mbox{ for }j\leq i \leq b+j-1, \mbox{ and}
\\
(\beta,(\alpha,\gamma)_{i-b+1})_j,&
\mbox{ for }i\geq j+b.
\end{array}
\right.
\]
This implies that, for $x \in K_a$, $y\in K_b$ and $z\in K_c$,
\[
\iota_i(x\times \iota_j(y\times z)) =
\left\{
\begin{array}{ll}
\iota_{j+a-1}(y\times \iota_i(x\times z)),&
\mbox{ for }1\leq i\leq j-1,
\\
\iota_j(\iota_{i-j+1}(x\times y)\times z),&
\mbox{ for }j\leq i \leq b+j-1, \mbox{ and}
\\
\iota_j(y\times \iota_{i-b+1}(x\times z)),&
\mbox{ for }i\geq j+b.
\end{array}
\right.
\]
Let $1\leq i \leq j-1$, $\bolds \in \ck pa$, $\boldt \in \ck qb$ and
$\boldu \in \ck rc$. Then, by definition,
\def\iotastar#1{{\iota_{#1}}_*}
\begin{eqnarray*}
\bolds\dia_i(\boldt\dia_j\boldu) &=& \iotastar i
(\bolds \times \iotastar j
(\boldt \times \boldu)) =
\\
&=& [\iota_i(\id \times \iota_j)]_*(\bolds \times
\boldt\times \boldu)
= [\iota_{j+a-1}(\id \times \iota_i)(S_{21}\times \id)]_*
(\bolds \times
\boldt\times \boldu)=
\\
&=&
(-1)^{pq}\cdot [\iota_{j+a-1}(\id \times \iota_i)]_*
(\boldt \times \bolds
\times \boldu) = (-1)^{pq}\cdot \boldt\dia_{j+a-1}
(\bolds\dia_i\boldu),
\end{eqnarray*}
which proves the first case of~(\ref{111}). Here $S_{21}$
denotes the transposition of factors in the cartesian product,
 $S_{21}(x\times y) = y\times x$, and we
used the associativity
and (graded) commutativity property of the cross product,
see~\cite{McL}. The
remaining cases can be discussed similarly.
As for the condition~(\ref{222}), we have
\begin{eqnarray*}
d_S(\bolds\dia_t \boldt)& =& d_S \iotastar t
(\bolds\times \boldt) = \iotastar
t(d_S(\bolds\times \boldt))=
\\
&=&
\iotastar t(d_S(\bolds)\times \boldt+(-1)^p\cdot\bolds
\times d_S(\boldt)) =
d_S(\bolds)\dia_t \boldt + (-1)^p \cdot\bolds\dia_t
d_S(\boldt).
\end{eqnarray*}
Here we used the derivation property of the cross product~\cite{McL}.
\qed

We have also the following analog of~\cite[Theorem~2]{G}.

\begin{proposition}
\label{3.5}
Let $(\X**,d_X,\dia_*)$ be a differential comp algebra as in
Definition~\ref{comp}. Put, for $\bolds \in \X pa$
and $\boldt \in \X qb$,
\[
\bolds\diamond\boldt:=
\sum_{1\leq t\leq b}(-1)^{(a+1)(t+q+1)}\cdot \bolds\dia_t \boldt.
\]
Then $\dia$ induces on $X^* := \bigoplus_{a-p-1=*}\X pa$
the structure of a
differential graded pre-Lie algebra (compare the nondifferential
analog of this construction in~{\rm \cite[par.~2]{G}}, that is
\begin{eqnarray*}
\bolds\diamond(\boldt\diamond
\boldu)-(\bolds\diamond\boldt)\diamond\boldu &=&
(-1)^{|\bolds|\cdot|\boldt|}(\boldt\diamond(\bolds\diamond
\boldu)-(\boldt\diamond\bolds)\diamond\boldu )
\mbox{ and }
\\
d_X(\bolds \diamond \boldt) &=& d_X \bolds \diamond
\boldt+(-1)^{|\bolds|}\cdot
\bolds \diamond d_X \boldt
\end{eqnarray*}
for any $\bolds \in X^{|\bolds|}$, $\boldt \in X^{|\boldt|}$
and $\boldu \in
X^{|\boldu|}$.
Consequently, the formula
\[
[\bolds,\boldt]:= \bolds \diamond\boldt -
(-1)^{|\bolds|\cdot|\boldt|}\boldt\diamond\bolds
\]
defines on $(X^*,d_X)$ the structure of a differential
graded Lie algebra.
\end{proposition}

\noindent
{\bf Proof.} A straightforward verification.
\qed

We have the obvious corollary of Proposition~\ref{3.4} and
Proposition~\ref{3.5}.

\begin{corollary}
\label{central}
The comp structure of~(\ref{ccc}) induces on
$C(K)^*:= \bigoplus_{n-p-1=*}\ck pn$
the structure of a differential graded Lie algebra
$(C(K)^*,[-,-],d_S)$.
\end{corollary}

We end this section with the following proposition

\begin{proposition}
There exists a sequence $\bolde_n\in \ck n{n+2}\subset C(K)^1$, $n\geq
0$
of ``fundamental classes'' such that
\begin{equation}
\label{fundamental}
\bolde_0=1\cdot (\b\b) \mbox{ and }
\mbox{$d_S(\bolde_n)+\frac12 \sum_{i+j=n-1}[\bolde_i,\bolde_j]=0$}.
\end{equation}
\end{proposition}

\noindent
{\bf Proof.}
The proof is based on the acyclicity of the associahedra
(see~\cite{shnider-sternberg:book,stasheff:TAMS63}). Suppose
inductively that
$\bolde_k$ have been
constructed for any $k < n$. Then, since all the $\bolde_i$ have
degree $1$,
\begin{eqnarray*}
d_S(\mbox{$\frac12$} \Sum_{i+j=n-1}[\bolde_i,\bolde_j]) &=&
\mbox{$\frac12$} \Sum_{i+j=n-1}([d_S(\bolde_i),\bolde_j]-
[\bolde_i,d_S(\bolde_j)])=
\\
&=&-\mbox{$\frac14$} \Sum_{\alpha+\beta+\gamma=n-2}
([[\bolde_\alpha,\bolde_\beta], \bolde_\gamma]-
[\bolde_\gamma,[\bolde_\alpha,\bolde_\beta]])=\\
&=&\mbox{$\frac12$} \Sum_{\alpha+\beta+\gamma=n-2}
[\bolde_\gamma,[\bolde_\alpha,\bolde_\beta]])=0,
\end{eqnarray*}
by the Jacobi identity.
Therefore $\frac12 \sum_{i+j=n-1}[\bolde_i,\bolde_j]$
is a cycle in $\ck
{n-1}{n+2}$ and the existence of $\bolde_n$ follows from the acyclicity of
$K_{n+2}$.
\qed

Let us give an explicit description of the $\dia_i$-product.
By an $(r,s)$-{\em shuffle\/} we mean, for $r,s \geq
1$, a partition of the set
$\{1,\ldots,r+s\}$ into two disjoint subsets
$\{i_1<\cdots < i_r\}$ and $\{j_1
< \cdots <j_s\}$. Denote by $\sh r s$ the set of all
$(r,s)$-shuffles. For
$\phi \in \sh rs$, let $\sgn(\phi)$ denote the signum
of the permutation
$(1,\ldots,r+s)\mapsto (i_1,\ldots,i_r,j_1,\ldots,j_s)$.
Any $\phi \in \sh
rs$ defines a sequence
$\{(\phi'(l),\phi''(l));\ 0\leq l\leq r+s\}
\subset (0,\ldots,r)\times (0,\ldots,s)$ by
\[
\phi'(l) := \mbox{Card}\{k;\ i_k \leq l\}
\mbox{ and }
\phi''(l) := \mbox{Card}\{k;\ j_k \leq l\},\ 0\leq l \leq r+s.
\]
For $\bolds = [\rada v0r]\in c_r(K_a)$
and $\boldt = [\rada w0r]\in C_s(K_b)$,
the simplicial cross product is given by
\[
\bolds \times \boldt :=
\sum_{\phi \in \sh rs}
\sgn(\phi)[(v_{\phi'(0)},w_{\phi''(0)}),
\ldots,(v_{\phi'(r+s)},w_{\phi''(r+s)})] \in C_{r+s}(K_a \times K_b),
\]
see~\cite{McL}. This gives the formula
\[
\bolds \dia_t \boldt =
\sum_{\phi \in \sh rs}
\sgn(\phi)[(v_{\phi'(0)},w_{\phi''(0)})_t,
\ldots,(v_{\phi'(r+s)},w_{\phi''(r+s)})_t] \in C_{r+s}(K_{a+b-1}),\
1\leq
t\leq b.
\]

Before concluding this section we would like to note a further
analogy between $C(K)^*$ and the algebraic structures investigated
by Gerstenhaber in [6]. There is a second product $*$
on $\susp C(K)^*$ which is compatible with the $\diamond$-product in
the sense that,
for any $\rho\in C_r(K_a)$, the operation $\rho\diamond(-)$
is a $*$-derivation of degree $a-r-1$.

To begin with we have an imbedding
$$
\phi:K_a\times K_b\rightarrow K_{a+b}
$$
given by concatenation of bracketings, $\phi(u,v):=(u)(v).$
For $\sigma\in C_s(K_b)$ and $\tau\in C_t(K_c)$
define
$$\sigma * \tau:= (-1)^{bt}\phi_*(\sigma\times \tau)\in
C_{s+t}(K_{b+c}),$$
where $\times$ is the simplicial homology cross product. Since
$\phi(\phi\times \id)\neq \phi(\id\times \phi)$ this product is
not associative. Nor is it graded commutative.
A straightforward calculation shows that, for $\rho\in C_r(K_a)$,
$$
\rho\diamond (\sigma *\tau )=(\rho\diamond \sigma)*\tau
+(-1)^{(a-r-1)(s+b)}\sigma *(\rho\diamond \tau).
$$
Furthermore,
$$
d_S(\sigma *\tau)= (d_S\sigma)*\tau + (-1)^{s+b}\sigma *(d_S \tau).
$$
What we have just described is presented for its intrinsic
interest. We do not as yet have an application for this product in
deformation theory.

\section{Main theorem}

Let $D_{-1},D_0 \in \der 1$ be as in Lemma~\ref{kozulka}.
Observe that $\der*$ is a differential graded
Lie algebra, the bracket being given by the (graded) commutator of
derivations and the differential defined by
$\nabla(\theta):= [D_{-1}, \theta]$.
Let $(C(K)^*,[-,-],d_S)$ be the graded
Lie algebra from Corollary~\ref{central}.

\begin{theorem}
\label{existence_of_homomorphism}
There exists a homomorphism
\[
m :(C(K)^*,[-,-],d_S)\to (\der*,[-,-],\nabla)
\]
of differential graded Lie algebras such that
$m(\bolde_0)=D_0$. The homomorphism $m$ has the properties
that
\begin{eqnarray*}
m(\ck kn)(\bod
(V,\calb(V))^{i,j})&\subset&\bod(V,\calb(V))^{i+n-1,j+k},\\
m(\ck kn)(\bod^{i,j}
(V,\calb(V)))&\subset&\bod^{i+n-1,j}(V,\calb(V)),
i,j,k\geq 0,\ n\geq 2,
\end{eqnarray*}
for $i,j,k\geq 0,\ n\geq 2$.
\end{theorem}

As a corollary, we get the existence of a homotopy $
(V,\Delta)$-bicomodule
structure on the bar resolution $(\bv,\db)$.

\begin{corollary}
\label{andulka}
The derivations $D_k:= m(\bolde_k)$, $ k\geq 1$, define a
homotopy $(V,\Delta)$-bicomodule
structure on the bar resolution $(\bv,\db)$.
\end{corollary}

\noindent
{\bf Proof.} Let $D = D_{-1}+D_0+ \sum_{k\geq 1}D_k$.
The only thing which
has to be verified is the condition $D^2 =0$.
Looking at the homogeneous components
we see that $D^2=0$ is equivalent to
\[
[D_{-1},D_k]+
\mbox{$\frac12$}\Sum_{i+j=k-1}
[D_i, D_j] = 0
\]
which is, by the definition of the
derivations $D_k$ and Theorem~\ref{existence_of_homomorphism}, the
same as
\[
m(d_S(\bolde_{k})+\mbox{$\frac12$}\Sum_{i+j=k-1}
[\bolde_{i},\bolde_{j}]) = 0,
\]
which follows from~(\ref{fundamental}).
\qed

As in~\cite{MS}, define inductively, for any $u\in \br n$,
the iterated comultiplication
$\deltacurl u : V \to V^{\ot n}$ by
\[
\deltacurl \b = \id,\ \deltacurl {\b\b} = \Delta \mbox{ and }
\deltacurl {(u,v)_t} = (\id^{\ot (t-1)} \ot
\deltacurl u \ot \id^{\ot (n-t)})
\deltacurl v,
\]
$u \in \br m$, $v\in \br n$ and $1\leq t\leq n$.
For example, $\deltacurl
{\b(\b\b)} = (\id \ot \Delta)\Delta$, $\deltacurl
{(\b\b)(\b\b)} = (\Delta\ot
\Delta)\Delta$, etc. It is also a good exercise to prove that,
for $x\in \br a$,
$y\in \br b$ and $z\in \br c$,
\begin{eqnarray}
\label{mysicka}
\deltacurl{x(yz)}& =& (\deltacurl x \ot
\deltacurl y \ot \deltacurl z)(\id \ot
\Delta)\Delta\ \ \mbox{ and }
\\
\deltacurl{(xy)z}& =& (\deltacurl x \ot \deltacurl y
\ot \deltacurl z)(\Delta
\ot \id)\Delta.
\nonumber
\end{eqnarray}
Sometimes we use the ``generalized Sweedler notation'',
meaning that for
$a\in V$ and $u\in \br n$ we write
\begin{equation}
\label{Swe}
\DC u (a) = \Sum\hoj au1 \ots \hoj aun:=\Sum\Bigotimes_{i=1}^n
\hoj aun \in V^{\ot n}.
\end{equation}
For any $u,v\in \br n$ consider the {\em associativity
operator\/} $\PP uv \in V^{\ot n}$ having the following
two properties.
\begin{itemize}
\item[(i)]
$\PP uu = 1^{\ot n}$ and $\PP uv \cdot \PP vw = \PP uw$, for any
$u,v,w \in \br n$ (the quasigroup property).
\item[(ii)]
Let $u = (x(yz),w)_t$ and $v = ((xy)z,w)_t$,
for some $x\in \br a$, $y\in \br
b$, $z\in \br c$, $w\in \br m$, with $a+b+c+m = n+1$.
In plain words this means
that $v$ was obtained by the substitution $x(yz)
\mapsto (xy)z$ at the $t$-th
place in $u$. Then
\[
\PP uv = 1^{\ot (t-1)}\ot (\deltacurl x \ot \deltacurl y\ot
\deltacurl
z)(\Phi)\ot 1^{\ot (m-t)}.
\]
\end{itemize}
These two conditions really define $\PP uv$ for
all $u,v \in \br n$.
Indeed, for any $u,v\in \br n$ there exists a
sequence of ``elementary moves'' from $u$ to $v$ as in (ii),
and we may define
$\PP uv$ using the quasigroup
property (i). The coherence then says that this
definition does not depend on
the particular choice of elementary moves. The
following lemma gives a
relation between the associativity operators and
iterated comultiplications.
\begin{lemma}
For any $u,v \in \br n$ and $a\in V$,
\begin{equation}
\label{angre}
\PP uv \cdot \DC v(a) = \DC u(a)\cdot \PP uv.
\end{equation}
Using the notation
of\REF{Swe}, we may rewrite\REF{angre}\ symbolically as
\begin{equation}
\label{angrevar}
\Sum\Bigotimes_{i=1}^n \Psi_{u,v}^i \cdot \hoj avi =
\Sum\Bigotimes_{i=1}^n \hoj aui\cdot \Psi_{u,v}^i,
\end{equation}
with $\PP uv = \sum \Psi_{u,v}^1\ots \Psi_{u,v}^n$.
\end{lemma}

\noindent
{\bf Proof.}
First, let $u=x(yz)$ and $v=(xy)z$. Then we have,
by the definition of the
associativity operator and\REF{mysicka},
\begin{eqnarray*}
\PP uv \cdot \DC v (a) &=&
(\DC x\ot \DC y\ot \DC z)(\Phi)\cdot (\DC x\ot \DC y\ot \DC
z)(\Delta\ot
\id)\Delta (a) =
\\
&=& (\DC x\ot \DC y\ot \DC z)[\Phi \cdot (\Delta\ot\id)\Delta(a)] =
\\
&=&
(\DC x\ot \DC y\ot \DC z)[(\id \ot \Delta)\Delta(a)\cdot \Phi] =
\\
&=&
(\DC x\ot \DC y\ot \DC z)(\id \ot \Delta)\Delta(a)\cdot
(\DC x\ot \DC y\ot \DC
z)(\Phi) = \DC u(a)\PP uv.
\end{eqnarray*}
{}From this we prove~(\ref{angre}) easily
for any ``elementary move'' as in~(ii)
above. The statement for a general couple $u$,
$v$ then follows from the fact
that there exists a sequence of elementary moves from
$u$ to $v$ and the resulting operator $\Psi_{u,v}$ is independent
of the choice of particular sequence.
\qed

Let $u\in \br n$ and $\rada x1n \in \bod(N)$.
Denote by $\bod_u(\rada x1n)$
the product of the elements $\rada x1n$ in the
nonassociative algebra
$\bod(N)$ with the bracketing $u$. For example,
$\bod_{\b(\b\b)}(x_1,x_2,x_3) = x_1\odot(x_2\odot x_3)$, etc.

Let $\rada x1n$ be elements from $\bod (N)$
and let $\phi = \sum \phi_1\ot
\cdots \ot \phi_n \in V^{\ot n}$. By $\phi \b (\rada x1n)$ we
denote the
element $\sum (\phi_1\b x_1,\ldots,\phi_n\b x_n)$.
In other words, we
denote by the same symbol the left action of the
algebra $(V,\MU)$ on
$\bod(N)$ and the induced action of $(V^{\ot n},\MU)$
on $\bod^n(N)$.
The same notation with $\phi$ on the right will be used also for the
right action.
The following lemma gives a characterization of the
associativity operators
up to a central element.

\begin{lemma}
\label{45}
Let $u,v\in \br n$ and $\rada x1n \in \bod(N)$. Then
\begin{equation}
\label{hodinky}
\odot_u(\PP uv \b (\rada x1n)) = \odot_v((\rada x1n)\b \PP uv).
\end{equation}
\end{lemma}

\noindent
{\bf Proof.}
The lemma follows
immediately from the definition of the associativity operator and
the defining relation~(\ref{relation}).
\qed

Notice that, for $u = \b(\b\b)$ and $v = (\b\b)\b$, $\PP uv= \Phi$
and~(\ref{hodinky}) says that
\[
\Phi \b [x_1\odot (x_2\odot x_3)] = [(x_1\odot x_2)\odot x_2]\b \Phi,
\]
which is exactly~(\ref{relation}).

For $u,v\in \br n$ and $\rada x1n \in \bod(N)$, let
$\BB uv(\rada x1n) :=
\odot_u(\PP uv \b (\rada x1n))$. We have the following lemma.

\begin{lemma}
\label{susenka}
For $u,v,w \in \br n$, $\rada x1n \in \bod(N)$ and $a,b\in V$ we have
\begin{eqnarray*}
a\b \BB uv(\rada xn1)& =& \BB uv(\deltacurl v(a)\b (\rada x1n)),
\\
\BB uv(\rada xn1) \b b& =& \BB uv((\rada x1n)\b \deltacurl u(b)),
\\
\BB uw(\PP wv \b (\rada x1n))&=&\BB uv(\rada x1n)
\mbox{ and }
\\
\BB uw((\rada x1n)\b \PP vu)& =& \BB vw(\rada x1n).
\end{eqnarray*}
\end{lemma}

\noindent
{\bf Proof.}
The first two equations follow immediately from~\cite[Lemma~3.1]{MS}.
As for the third one, we have
\begin{eqnarray*}
\BB uw(\PP wv \b (\rada x1n))&=& \odot_u(\PP uw \b
(\PP wv \b (\rada x1n))) =
\\
&=&
\odot_u((\PP uw \cdot \PP wv) \b (\rada x1n))=
\\
&=& \odot_u(\PP uv \b (\rada x1n)) = \BB uv(\rada x1n).
\end{eqnarray*}
Here we use that $\PP uw \cdot \PP wv =\PP uv$
which follows from the quasigroup
property of the associativity operator. The last
equation can be proved
similarly.
\qed

\begin{lemma}
\label{EPS}
Let $u,v\in \br a$, $z,w\in \br b$, $\rada x1a,\rada y1b \in
\bod(N)$ and
$1\leq t \leq b$. Then
\[
\BB zw(\rada y1{t-1},\BB uv(\rada x1a),\rada y{t+1}b)\! =\!
\BB {(u,z)_t}{(v,w)_t}(\rada y1{t-1},\rada x1a,\rada y{t+1}b).
\]
\end{lemma}

\proof\
An easy exercise on the definitions which we leave for the reader.
\qed

The map $m$ of Theorem~\ref{existence_of_homomorphism}
will be defined using
another map
\[
M[\sigma] :\bv \to \bod^{n-1,1}(V,\bv),
\]
given for any $\tilde \T n$-oriented
$k$-simplex $\sigma= [\rada v0k]$ of $K_n$, $n\geq 2$,
$k\geq 1$ and $\rada
v0k\in \br n$, $v_k \prec v_{k-1}\prec \cdots \prec v_0$ (recall that
$\tilde \T n$ is the orientation opposite to the one induced by the
partial order of vertices of the associahedron $K_n$).
For any such $\sigma$, $m(\sigma)$ will be a derivation given by
\begin{equation}
\label{fear}
m(\sigma) |_{\susp V} := 0
\mbox{ and }
m(\sigma)|_{\susp \bv} := (\susp^{\odot n})\circ
M[\sigma]\circ\desusp.
\end{equation}
Let $P(q,k)$ be,
for $k\geq 0$ and $q\geq 1$, the set of all
subdivisions of the segment of
$(q+2)$ elements into $(k+1)$ subsegments
such that the first and the last
segment of the subdivision is nonempty.
The elements of $P(q,k)$ can be
encoded by nondecreasing maps $\pi : (1\ldots,k)\to
(1,\ldots,q+1).$ The corresponding
subdivision will be then
\[
(0,\ldots,\pi(1)-1)(\pi(1),\ldots,\pi(2)-1)
\cdots(\pi(k),\ldots,q+1).
\]
Let $\hat a = a_0\ot\cdots \ot a_{q+1}\in
{\cal B}_{-q}(V)$. For $0\leq j\leq
k$ write
\[
\pi_j(\hat a) := a_{\pi(j)}\ots a_{\pi(j+1)-1}.
\]
Denote also, for $\pi \in P(q,k)$,
\begin{equation}
o(\pi)=\sum_{i=1}^k \pi(i).
\label{opi}
\end{equation}

Let $\otbar$ denote the multiplication in the tensor product of tensor
algebras. In particular,
for $\rada x1m \in (\Bigotimes V)^{\ot d}$ with
$x_j = x_j^1\ot \cdots \ot x_j^d$ and $x^i_j\in \Bigotimes V$
for $1\leq j\leq m$, we have
\[
x_1\otbar \cdots\otbar x_m :=
(x^1_1\ots x^1_m)\ot(x^2_1\ots x^2_m)\ots
(x^d_1\ots x^d_m)\in (\Bigotimes V)^{\otimes d}
\]
Using this notation, let $\Psi_\pi[\sigma](\hat a) \in \bigotimes^n
V^{\ot(q+k+2)}$ be defined, for any $\pi \in P(q,k)$, as
\[
\Psi_\pi[\sigma](\hat a):=
\deltacurl{v_0}(\pi_0(\hat a))\otbar \PP
{v_0}{v_1}\otbar \deltacurl{v_1}(\pi_1(\hat a))\otsbar \PP
{v_{k-1}}{v_k}\otbar \deltacurl{v_k}(\pi_k(\hat a)).
\]
Here we used the shorthand $\deltacurl{v_j}(\pi_j(\hat a))$ for
\[
\deltacurl{v_j}(a_{\pi(j)})\otsbar\deltacurl{v_j}(a_{(\pi(j+1)-1)}).
\]
Note that there is an order preserving bijection between the factors
of this product and the elements (vertices and edges) of the directed
 graph determined by the one skeleton of $\sigma$.

Since the multiplication $\MU$ on $V$ is associative,
we can iterate it to a
map $\Mu : V^{\ot m} \to V$, for any $m\geq 1$. Writing
$\Psi_\pi[\sigma](\hat a)$ as
$\sum \phi_1\ots \phi_n$, with $\phi_i \in
V^{\ot(q+k+2)}$, $1\leq i\leq n$,
let $\Mu_r(\Psi_\pi[\sigma](\hat a))
\in V^{\ot(r-1)}\ot {\cal B}_{-q-k}(V) \ot
V^{\ot(n-r)}$ be defined as
\[
\Mu_r(\Psi_\pi[\sigma](\hat a)) :=
\Sum \Mu(\phi_1)\ots \Mu(\phi_{r-1}) \ot \phi_r \ot
\Mu(\phi_{r+1}) \ots \Mu(\phi_n).
\]
For any $1\leq r\leq n$, define
\[
M^r_\pi[\sigma](\hat a):=
\BB {v_k}{v_0}(\Mu_r(\Psi_\pi[\sigma](\hat a)).
\]
Here, as usual, we identify the multilinear map $\BB {v_k}{v_0}$
with a linear map on the tensor product.
Finally we introduce a sign factor:
$$\sgn(\pi,\sigma) := (-1)^{o(\pi)+kn},$$
where $\sigma\in C_k(K_n)$
and define
\[
M[\sigma] := \sum_{1\leq r\leq n}M^r[\sigma](\hat a)\
\mbox{with}\
M^r[\sigma] := \sum_{\pi \in P(q,k)}
\sgn(\pi,\sigma)\cdot
M^r_\pi[\sigma]
\]

\begin{lemma}
\label{Poo}\label{Puu}
For any $k$-simplex $\sigma = [\rada v0k]$ of $K_n$
\begin{equation}
\label{Oo}
m(d_S\sigma) = \nabla(m(\sigma))
\end{equation}
\end{lemma}

\proof\
Let us express first the terms of~(\ref{Oo})
via the defining map $M[-]$. We have
\[
m(d_S \sigma) = \susp^{\odot n} M[d_S\sigma]\desusp
\]
while
\begin{eqnarray*}
\nabla(m(\sigma))&=&
[D_{-1}\circ m(\sigma)]
+(-1)^{k+n}\cdot [m(\sigma)\circ D_{-1}]=
\\
&=&
D_{-1}(\susp^{\odot n}M[\sigma])\desusp
+(-1)^{k+n}\cdot m(\sigma)(\susp \db\desusp)=
\\
&=&\sum_{1\leq r\leq n}(-1)^{n-1}\cdot (\susp^{\odot n})(\id^{\odot
(r-1)}\odot\db \odot \id^{\odot(n-r)})M^r[\sigma]\desusp+
\\
&&+(-1)^{k+n}\cdot (\susp^{\odot n})M[\sigma]\db\desusp
\end{eqnarray*}
The sign $(-1)^{n-1}$ is explained as follows:
\begin{eqnarray*}
\lefteqn{
D_{-1}(\susp^{\odot n}M^r[\sigma])=}\\
&=&\sum_{1\leq r\leq n}(\id^{\odot(r-1)}\odot \susp \db \desusp \odot
\id^{\odot(n-r)})( \susp^{\odot n}M^r[\sigma]) =
\mbox{/transpose $\susp^{\odot(r-1)}$ , $\susp \db
\desusp$/}
\\
&=&\sum_{1\leq r\leq n}(-1)^{(r-1)}\cdot (\susp^{\odot(r-1)}\odot
\susp \db \odot
\susp^{\odot(n-r)})M[\sigma]=
\mbox{/transpose $\db$\, , $\susp^{\odot(n-r)}$/}
\\
&=&\sum_{1\leq r\leq n}(-1)^{(r-1) + (n-r)}\cdot
(\susp^{\odot(r-1)}\odot \susp \odot
\susp^{\odot(n-r)})(\id^{\odot(r-1)}\odot \db \odot \id^{\odot(n-r)})
M[\sigma] =
\\
&=&\sum_{1\leq r\leq n}(-1)^{n-1}\cdot (\susp^{\odot n})(\id^{\odot
(r-1)}\odot\db \odot \id^{\odot(n-r)})M^r[\sigma]
\end{eqnarray*}
which gives the sign as claimed.

Looking at the component (in $\bod^{n-1,1}(V)$)
having an element of $\bv$ at the $r$-th place we see
that~(\ref{Oo})
is equivalent to
\begin{equation}
\label{oO}
M^r[d_S\sigma](\hat a) = (-1)^{n-1}\cdot(\id^{\odot
(r-1)}\odot\db \odot \id^{\odot(n-r)})M^r[\sigma](\hat a)+
(-1)^{n+k}\cdot M^r[\sigma](\db(\hat a)),
\end{equation}
which has to be satisfied for any $1\leq r\leq n$.

Let us discuss the first term on the right hand side of~(\ref{oO}).
Let $\hat a\in \BV{-q}$ so that $M^r[\sigma](\hat a)\in\BV{-q-k}.$
The differential $\db$ applied on an element from $\BV{-q-k}$ may be
decomposed as $\db = \sum_{0\leq j\leq q+k}(-1)^j \db^j$, where
\[
\db^j(\Rada b0{k+q+1}) := (\Rada b0{j-1}|b_j\cdot b_{j+1}|\Rada
b{j+2}{k+q+1}),
\]
for any $(\Rada b0{k+q+1})\in\BV{-q-k}$. We may then write
\begin{eqnarray*}
\lefteqn{
(-1)^{n-1}\cdot(\id^{\odot
(r-1)}\odot\db \odot \id^{\odot(n-r)})M^r[\sigma](\hat a)=}
\\
&&=
 \Sum_S (-1)^{n+j-1}\cdot\sgn(\pi,\sigma)\cdot(\id^{\odot
(r-1)}\odot\db^j \odot \id^{\odot(n-r)})M^r_\pi[\sigma](\hat a),
\end{eqnarray*}
where $S := \{(j,\pi);\ 0\leq j\leq k+q,\ \pi \in P(q,k)\}$.

Our proof of (19) involves distinguishing the various kinds of
products
which appear on the right side: products of two $a$ terms, products of
an $a$ term and a $\Psi$ term in either order, and products of
two $\Psi$ terms. We have the following disjoint decomposition of $S$
where the subscripts indicate the type of product associated to the
pairs in a given set.
\[
S = S'_{a\Psi}\cup S''_{a\Psi} \cup S'_{\Psi a}\cup S''_{\Psi a} \cup
S^1_{\Psi\Psi} \cup \cdots \cup S^{k-1}_{\Psi\Psi}\cup S_{aa}
\]
where
\begin{eqnarray*}
S'_{a\Psi} &:=& \{(j,\pi)\in S;\ j=0, \pi(1)=1\},
\\
S'_{\Psi a} &:=& \{(j,\pi)\in S;\ j=k+q, \pi(k)=q+1\},
\\
S''_{a\Psi} &:=& \{(j,\pi)\in S;\ \mbox{there exists $l$,
$1\leq l \leq k$,
s.t.~$\pi(l)>\pi(l-1)$ and $j=\pi(l)+l-2$}\}\setminus S'_{a\Psi},
\\
S''_{\Psi a} &:=& \{(j,\pi)\in S;\ \mbox{there exists $l$,
$1\leq l \leq k$,
s.t.~$\pi(l) < \pi(l+1)$ and $j=\pi(l)+l-1$}\}\setminus S'_{\Psi a},
\\
S^l_{\Psi\Psi} &:=& \{(j,\pi)\in S;\ \pi(l)=\pi(l+1),\
j = \pi(l)+l-1\},\
1\leq l\leq k-1,\mbox{ and}
\\
S_{aa} &:=& \{(j,\pi)\in S;\ \mbox{there exists $l$, $1\leq l <k$,
such that $\pi(l)+l -1< j < \pi(l+1)+l-1$}\}.
\end{eqnarray*}

Continuing this approach consider the second term on the right hand
side
of~(\ref{oO}). The
differential $\db$ is applied to an element of $\BV{-q}$,
thus we may write
\[
M^r[\sigma](\db(\hat a)) = \Sum_T(-1)^j\cdot
\sgn(\pi,\sigma)\cdot M^r_\pi[\sigma]
(\db^j(\hat a)),
\]
where $T := \{(j,\pi);\ 0\leq j\leq q,\ \pi \in P(q-1,k)\}$.
As for the term on
the left hand side of~(\ref{oO}), we again have a decomposition
$d_S = \sum_{0\leq j\leq
k}(-1)^j d^j_S$, where
\[
d_S^j[\rada v0k] := [\rada v0{j-1},\rada v{j+1}k],\
\mbox{for $0\leq j \leq k$}.
\]
We may write
\[
M^r[d_S\sigma](\hat a) = \Sum_R(-1)^j \cdot
\sgn(\pi,d^j_S(\sigma))\cdot
M^r_\pi[d^j_S\sigma](\hat a)
\]
where $R := \{(j,\pi);\ 0\leq j\leq k,\ \pi \in P(q,k-1)\}$.
We can decompose
$R = R^0\cup R^1 \cup \cdots \cup R^k$ with $R^l :=
\{(j,\pi)\in R;\ j= l\}$.

The terms in (20) cancel in pairs arising from the
isomorphisms (bijections) between the various sets of
indices as described in the following two sublemmas.
\begin{sublemma}
\label{cancelpair}
There are isomorphisms $\alpha : S''_{a\Psi}\cong
S''_{\Psi a}$, $\beta :
T\cong S_{aa}$, $\gamma : R^0 \cong S'_{a \Psi}$,
$\delta : R^k \cong
S'_{\Psi a}$ and $\phi^l : R^l \cong S^l_{\Psi\Psi}$,
$1\leq l\leq k-1$.
\end{sublemma}

We give an explicit description of the isomorphisms.
The isomorphism $\alpha : S''_{a\Psi}\cong
S''_{\Psi a}$ is given by
$\alpha(j,\pi) := (j,\alpha(\pi))$, where
\[
\alpha(\pi)(s)=
\left\{
\begin{array}{ll}
\pi(s),& \mbox{for } i\not= l, \mbox{ and}
\\
\pi(s)-1,& \mbox{for } s=l.
\end{array}
\right.
\]
The isomorphism $\beta : T\cong S_{aa}$, $\beta(j,\pi) :=
(\beta(j),\beta(\pi))$ is given in the following way.
For any $j$, $0\leq
j\leq q$, there exists exactly one $l$ such that
$\pi(l)\leq j\leq
\pi(l+1)-1$. Define then $\beta(j) := j+l$ and
\[
\beta(\pi)(s)=
\left\{
\begin{array}{ll}
\pi(s),& \mbox{for } s\leq l, \mbox{ and}
\\
\pi(s)+1,& \mbox{for } s>l.
\end{array}
\right.
\]
The isomorphism $\gamma : R^0 \cong S'_{a \Psi}$ is defined as
$\gamma(0,\pi) := (0,\gamma(\pi))$, where
\[
\gamma(\pi)(s)=
\left\{
\begin{array}{ll}
1,& \mbox{for } s=1, \mbox{ and}
\\
\pi(s-1),& \mbox{for } s\geq2.
\end{array}
\right.
\]
The isomorphism $\delta : R^k \cong S'_{\Psi a}$ is defined as
$\delta(k,\pi):= (k+q, \delta(\pi))$, where
\[
\delta(\pi)(s)=
\left\{
\begin{array}{ll}
\pi(s),& \mbox{for } s\leq k-1, \mbox{ and}
\\
q+1,& \mbox{for } s=k.
\end{array}
\right.
\]
Finally, the isomorphism $\phi^l : R^l \cong S^l_{\Psi\Psi}$ is,
for $1\leq
l\leq k-1$, defined by $\phi^l(l,\pi):=
(\pi(l)+l-1,\phi^l(\pi))$, where
\[
\phi^l(\pi)(s)=
\left\{
\begin{array}{ll}
\pi(s),& \mbox{for } s\leq l, \mbox{ and}
\\
\pi(s-1),& \mbox{for } s>l.
\end{array}
\right.
\]
Lemma~\ref{Puu}\ will obviously follow from the next sublemma.

\begin{sublemma}
\label{sblm}
The following equations hold.
\begin{eqnarray}
\label{A}
\lefteqn{\sgn(\pi,\sigma)\cdot
(\id^{\odot(r-1)}\odot \db^j \odot \id^{\odot(n-r)})
M^r_\pi[\sigma](\hat a)+}
\\
\nonumber
&& + \sgn(\alpha(\pi),\sigma)\cdot
(\id^{\odot(r-1)}\odot \db^{j} \odot
\id^{\odot(n-r)})M^r_{\alpha(\pi)}[\sigma](\hat a) = 0,
\end{eqnarray}
for $(j,\pi)\in S''_{a\Psi}$,
\begin{eqnarray}
\label{B}
\lefteqn{(-1)^{n+\beta(j)-1}\cdot \sgn(\beta(\pi),\sigma)\cdot
(\id^{\odot(r-1)}\odot \db^{\beta(j)} \odot
\id^{\odot(n-r)})M^r_{\beta(\pi)}[\sigma]
(\hat a)+\hskip3cm}
\\
&&
\nonumber\hskip3.5cm
+(-1)^{k+n+j}\cdot\sgn(\pi,\sigma)\cdot
M^r_\pi[\sigma](\db^j(\hat a)) =0,
\end{eqnarray}
for $(j,\pi)\in T$,
\begin{eqnarray}
\label{C}
\lefteqn{\sgn(\pi,d_S^0(\sigma))\cdot
M^r_\pi[d_S^0(\sigma)](\hat a)=}
\\
&&=
\nonumber
(-1)^{n-1}\cdot\sgn(\gamma(\pi),\sigma)\cdot
(\id^{\odot(r-1)}\odot \db^{0} \odot
\id^{\odot(n-r)})M^r_{\gamma(\pi)}[\sigma](\hat a),
\end{eqnarray}
for $(0,\pi)\in R^0$,
\begin{eqnarray}
\label{D}
\lefteqn{(-1)^k\cdot\sgn(\pi,d_S^k(\sigma))\cdot
M^r_\pi[d_S^k(\sigma)](\hat a)=}
\\
\nonumber
&&=
(-1)^{n+k+q-1}\cdot\sgn(\delta(\pi),\sigma)\cdot
(\id^{\odot(r-1)}\odot \db^{k+q} \odot
\id^{\odot(n-r)})M^r_{\delta(\pi)}[\sigma]
(\hat a),
\end{eqnarray}
for $(k,\pi)\in R^k,\ \mbox{and}$
\begin{eqnarray}
\label{E}
\lefteqn{(-1)^l\cdot \sgn(\pi,d_S^l(\sigma))\cdot
M^r_\pi[d_S^l(\sigma)](\hat a)=}
\\
\nonumber
&&=
(-1)^{n+\pi(l)+l}\cdot\sgn(\phi^l(\pi),\sigma,\hat a)\cdot
(\id^{\odot(r-1)}\odot \db^{\pi(l)+l-1} \odot
\id^{\odot(n-r)}M^r_{\phi^l(\pi)}[\sigma](\hat a),
\end{eqnarray}
for $(l,\pi)\in R^l, 1\leq l\leq k-1$.
\end{sublemma}

\noindent
The proof of the sublemma is straightforward but rather technical
and will be postponed to the last section.

The next lemma is the key to proving that $m$ is a homomorphism of
Lie algebras.

\begin{lemma}
\label{kava}
Let $\sigma = [\rada v0r]\in C_r(K_a)$, $\tau =
[\rada w0s]\in C_s(K_b)$
and $\hat a = a_0\ots a_{q+1} \in \BV{-q}$. Then
\begin{eqnarray}
\label{Y}
\lefteqn{M^l[\sigma \dia_t \tau](\hat a)=}
\\
\nonumber
&&=
\left\{
\begin{array}{ll}
(-1)^{(a+1)s+(b+1)r}\cdot
(\id^{\odot( t-1)}\odot M^{l-t+1}
[\sigma]\odot \id^{\odot( b-t)})M^t[\tau] (\hat
a ),& l\in [t,t+a-1],
\\
0,& \mbox{otherwise}.
\end{array}
\right.
\end{eqnarray}
\end{lemma}
We show first that this lemma would imply that
\begin{equation}
\label{X}
m([\sigma,\tau])(\susp \hat a) = [m(\sigma),m(\tau)](\susp \hat a),
\end{equation}
for any $\sigma, \tau$ and $\hat a$ as above,
which would obviously finish
the proof of Theorem~\ref{existence_of_homomorphism}.
On the left hand side of\REF{X}\ we have
\begin{eqnarray*}
m([\sigma,\tau])(\susp \hat a) &=&
\sum_{1\leq t\leq b}(-1)^{(a+1)(t+s+1)}\cdot
 m(\sigma \dia_t \tau)(\susp \hat a) -
\\
&&-(-1)^{(a+r+1)(b+s+1)}\cdot
\sum_{1\leq t\leq a}(-1)^{(b+1)(t+r+1)}\cdot
m(\tau \dia_t \sigma)(\susp \hat a) =
\\
&=& \susp^{\odot(a+b-1)}\{\sum_{1\leq t\leq b}
(-1)^{(a+1)(t+s+1)}\cdot M[\sigma \dia_t \tau](\hat a) -
\\
&&\hphantom{\susp^{\odot(a+b-1)}x}
-(-1)^{(a+r+1)(b+s+1)}\cdot
\sum_{1\leq t\leq a}(-1)^{(b+1)(t+r+1)}
\cdot M[\tau \dia_t \sigma](\hat a)\}.
\end{eqnarray*}
On the other hand
\begin{eqnarray*}
\lefteqn{
[m(\sigma),m(\tau)](\susp \hat a) =}
\\
&&=
[m(\sigma)\circ m(\tau)](\susp \hat a) -(-1)^{(a+r+1)(b+s+1)}\cdot
[m(\tau)\circ m(\sigma)](\susp \hat a) =
\\
&&= m(\sigma)(\susp^{\odot b} M[\tau](\hat a)) -
(-1)^{(a+r+1)(b+s+1)}\cdot
m(\tau)(\susp^{\odot a} M[\sigma](\hat a))=
\\
&&= \susp^{\odot(a+b-1)} \{\sum_{1\leq t\leq b}
(-1)^{(r+a+1)(t+1)+r(b+t)}
\cdot (\id^{\odot( t-1)}\odot
M[\sigma] \odot \id^{\odot( b-t)})M^t[\tau](\hat a) -
\\
&&\hphantom{\susp^{\odot(a+b-1)}\{x}
-\sum_{1\leq t\leq a}(-1)^{(a+r+t)(b+s+1)+s(a+t)}
\cdot (\id^{\odot t-1}\odot
M[\sigma] \odot \id^{\odot a-t})M^t[\sigma](\hat a)\}.
\end{eqnarray*}

The signs which appear in the last two lines arise as follows.
\begin{eqnarray*}
\lefteqn{
m(\sigma)(\susp^{\odot b}M[\tau])=}
\\
&=&\sum_{1\leq t\leq b}(\id^{\odot (t-1)}\odot \susp^{\odot
a}M[\sigma]\desusp \odot \id^{\odot(b-t)})\susp ^{\odot b} M[\tau] =
\mbox{ /switch $\susp^{\odot
a}M[\sigma]\desusp$ and $\susp^{\odot (t-1)}$/}
\\
&=&\sum_{1\leq t\leq b} (-1)^{(t+1)(r+a+1)}\cdot
(\susp^{\odot (t-1)}\odot \susp^{\odot
a}M[\sigma] \odot \susp^{\odot(b-t)})M[\tau]=
\mbox{ /switch $M[\sigma]$ and $ \susp^{\odot(b-t)}$/}
\\
&=& \sum_{1\leq t\leq b}
(-1)^{(r+a+1)(t+1)+r(b+t)}\cdot(\susp^{\odot( t-1)}\odot\susp^{\odot
a}
 \odot \susp^{\odot( b-t)})
(\id^{\odot( t-1)}\odot
M[\sigma] \odot \id^{\odot( b-t)})M^t[\tau] =
\\
&=& \susp^{\odot(a+b-1)} \sum_{1\leq t\leq b}
(-1)^{(r+a+1)(t+1)+r(b+t)}\cdot
(\id^{\odot( t-1)}\odot
M[\sigma] \odot \id^{\odot( b-t)})M^t[\tau](\hat a).
\end{eqnarray*}
This explains the factor $(-1)^{r(b+t)}$ in the formula, the
explanation of the second factor $(-1)^{s(a+t)}$ is the
same.

Summing\REF{Y} over $l$ we get
\[
\sum_{1\leq t\leq b}
M[\sigma \dia_t \tau](\hat a)=(-1)^{(a+1)s+(b+1)r}\cdot
\sum_{1\leq t\leq b}
(\id^{\odot (t-1)}\odot
M[\sigma] \odot \id^{\odot( b-t)})M^t[\tau](\hat a)
\]
while the same formula with the r\^oles of
$\sigma$ and $\tau$ interchanged
gives
\[
\sum_{1\leq t\leq a}
M[\tau \dia_t \sigma](\hat a) = (-1)^{(a+1)s+(b+1)r}\cdot
\sum_{1\leq t\leq a}(\id^{\odot( t-1)}\odot
M[\tau] \odot \id^{\odot(a-t)})M^t[\sigma](\hat a)
\]
and\REF{X}\ follows. In fact lemma~\ref{kava}
implies a bit more, it says that
the map $m$ preserves suitably defined comp structures.

Let us discuss the term on the left
hand side of\REF{Y}. For $\phi\in \sh rs$ denote
(see the end of par.~3 for
the notation)
\[
\sigma\dia_{t,\phi}\tau :=[(v_{\phi'(0)},w_{\phi''(0)})_t,
\ldots,(v_{\phi'(r+s)},w_{\phi''(r+s)})_t] \in C_{r+s}(K_{a+b-1}).
\]
Then
\[
M^l[\sigma\dia_t \tau](\hat a) = \Sum_U
\sgn(\pi,\sigma\dia_{t,\phi}\tau)\cdot\sgn(\phi)\cdot
M^l_\pi[\sigma\dia_{t,\phi}\tau](\hat a),
\]
where $U := \{ (\phi,\pi);\ \phi \in \sh rs,\ \pi \in P(q,r+s)\}$. On
the
other hand,
\begin{eqnarray*}
\lefteqn{
(\id^{\odot t-1}\odot M^{l-t+1}[\sigma]
\odot \id^{\odot b-t})M^t[\tau] (\hat
a ) =}
\\
&&= \Sum_V\sgn(\pi',\sigma,)\cdot
\sgn(\pi'',\tau)\cdot
(\id^{\odot t-1}\odot M_{\pi'}^{l-t+1}
[\sigma]\odot \id^{\odot b-t})M_{\pi''}^t[\tau] (\hat a ),
\end{eqnarray*}
with $V:= \{(\pi',\pi'');\ \pi' \in P(q+s,r),\ \pi'' \in P(q,s)\}$.
Let us
construct an isomorphism $\chi :V \cong U$ as follows.
Let $\pi'' \in P(q,s)$ and let,
just for the moment, $\rada a0{q+1}$ be a
sequence of independent symbols.
Consider the sequence
\[
(\rada b0{s+q+1}):=
(\rada a0{\pi''(1)-1},\Psi_1,\rada a{\pi''(1)}
{\pi''(s)-1},\Psi_{s},\rada
a{\pi''(s)}{q+1}),
\]
where $\rada \Psi1{s}$ are other
independent symbols. Let $\pi' \in
P(q+s,r)$ and define
\[
(\rada c0{r+s+q+1}):=
(\rada b0{\pi'(1)-1},\overline \Psi_1,\rada b{\pi'(1)}{\pi'(r)-1},
\overline\Psi_{r},\rada
b{\pi'(s)}{s+q+1}),
\]
for a sequence $\rada{\overline \Psi}1{s}$ of
independent symbols. We may
then write
\[
(\rada c0{r+s+q+1}) = (\rada a0{\pi(1)-1},X_1,
\rada a{\pi(1)}{\pi(r+s)-1},
X_{r+s},\rada a{\pi(r+s)}{q+1})
\]
for some $\pi \in P(q,r+s)$ and $X_h \in \{\rada \Psi1{s},\rada
{\overline\Psi}1{r}\}$, $1\leq h \leq r+s$. Define then
\[
\{i_1< \cdots< i_r\}:= \{ h;\ X_h = \overline\Psi_h\}
\mbox{ and }
\{j_1< \cdots< j_s\}:= \{ h;\ X_h = \Psi_h\}.
\]
These data determine a shuffle $\phi \in \sh rs$.
Put $\chi(\pi',\pi''):= (\phi,\pi)$.
The description of $\pi$ in terms of $\pi'$ and $\pi''$ is
as follows. Let
$$\bar i_k=\mbox{\rm max}\{i|\ \pi''(i)+i\leq \pi'(k)\},$$
then the sequence $\pi(i),\, 1\leq i\leq r+s$ is given by
\[
\pi''(1),\ldots, \pi''(\bar i_1),\pi'(1)-\bar i_1,\ldots,
\pi''(\bar i_r),\pi'(r)-\bar i_r,
\left\{ \begin{array}{lcr}
\ldots\pi''(s)
&\mbox{if}&\bar i_r<s\\
\ldots\pi'(r)-s &\mbox{if}&\bar i_r=s
\end{array}
\right.
\]
The corresponding shuffle $\phi$ is given by those indices
for which the related associativity operator
has the form $1^{\otimes c}\otimes \Psi_{v,v'}\otimes 1^{\otimes c'}.$
These are the indices $ i$ where $\pi(i)=\pi'(a)-\bar i_a,i_a=\bar i_a
+a.$
Then
$$\sgn(\phi)=(-1)^{\sum i_a -a}=(-1)^{\sum \bar i_a}.$$

\ Lemma~\ref{kava} would then
follow from the following sublemma.

\begin{sublemma}
\label{slune}
$M^l_\pi[\sigma \dia_{t,\phi} \tau](\hat a) = 0$
for $l \not\in [t,t+a-1]$, $\phi \in \sh rs$ and $\pi
\in P(q,r+s)$, while
\begin{eqnarray*}
\lefteqn{
\sgn(\pi,\sigma\dia_{t,\phi}\tau)\cdot\sgn(\phi)\cdot
M^l_\pi[\sigma \dia_{t,\phi} \tau](\hat a) =}
\\
&&=
(-1)^{(a+1)s+(b+1)r}\cdot
\sgn(\pi',\sigma)\cdot
\sgn(\pi'',\tau)\cdot
(\id^{\odot( t-1)}\odot M_{\pi'}^{l-t+1}[\sigma]
\odot \id^{\odot( b-t)})M^t_{\pi''}[\tau] (\hat a)
\end{eqnarray*}
for $(\phi,\pi)= \chi (\pi',\pi'')$, $(\pi',\pi'') \in V$
and $l\in [t,t+a-1]$.
\end{sublemma}

\noindent
The proof, technical although straightforward, will be given
in the last section.

\section{Cohomology and the deformations}

As we have already observed, both $\bod(V,\bv)^*$ ($=\bod(\susp
V\oplus \susp \bv)$) and $\bod^*(V)$
($=\bod(\susp V)^*$) are graded $(V,\mu)$-modules (via the
$\bullet$-action). The $\odot$-multiplication yields on both
$\bod(V,\bv)^*$ and $\bod^*(V)$ the structures of graded
$\bod^*(V)$-bimodules and both structures are related
by~(\ref{opulka}). Let $\bod^{\prime}(V,\bv)=\bod^{*,1}(V,\bv)$ be
the submodule with precisely one factor of $\bv$.
Denote by $\HOM{\bod^\prime(V,\bv)^*}{\bod^*(V)}n$
the set of all degree $n$
homogeneous maps $f:\bod^\prime(V,\bv)^*\to\bod^*(V)$
which are both $\bod^*(V)$ and $(V,\mu)$-linear.
For such a map put
$d(f):= f\circ D+(-1)^n \dcobar \circ f$, where $D$ is as in
Corollary~\ref{andulka} and $\dcobar$ is the degree
one derivation on $\bod^*(V)$ defined by
$\dcobar|_{\susp V}:= (\susp \odot
\susp)\Delta\desusp$.

\begin{definition}
\label{vlacek}
The cohomology of a Drinfel'd algebra $A =
(V,\cdot,\Delta,\Phi)$ is the graded vector space
\[
H^*(A) := H^*(C^*(A),d),
\]
where $C^*(A):= \HOM{\bod^\prime(V,\bv)^*}{\bod^*(V)}*$ and the
differential $d$ is as above.
\end{definition}

Let us try to understand better the structure of the cohomology above.
Using the fact that $\bod^\prime(V,\bv)^*$ is, in a sense, the free
$\bod^*(V)$-bimodule on the $(V,\mu)$-bimodule $\susp\BV*$,
we may write
\[
\mbox{$\HOM{\bod^\prime(V,\bv)^*}{\bod^*(V)}n \cong
\hom V{\BV*}{\bod^*(V)}{n+1},$}
\]
where $\hom V{\BV*}{\bod^*(V)}{n+1}$ denotes the set of
$(V,\mu)$-linear maps $f:\BV* \to \bod^*(V)$ of degree $(n+1)$.
Using the fact that $\BV*$ is the free $(V,\mu)$-module on
$\bigotimes^{-*}(\oV)$ (recall that $\oV = \Ker(\epsilon)$),
we get the identification
\[
\mbox{$\hom V{\BV*}{\bod^*(V)}{n+1}\cong
\hom{\boldk}{\bigotimes^{-*}(\oV)}{\bod^*(V)}{n+1},$}
\]
where $\hom{\boldk}{\bigotimes^{-*}(\oV)}{\bod^*(V)}{n+1}$ denotes
the set of degree $(n+1)$ $\boldk$-linear homogeneous maps $g:
\bigotimes^{-*}(\oV)\to \bod^*(V)$. Finally, invoking the canonical
isomorphism $J:\bod^*(V)\cong \bigotimes^*(V)$ of graded vector
spaces constructed in~\cite[Proposition~3.3]{MS}
we get
\[
\mbox{$\hom{\boldk}{\bigotimes^{-*}(\oV)}%
{\bod^*(V)}{n+1}\cong\hom{\boldk}%
{\bigotimes^{-*}(\oV)}{\bigotimes^*(V)}{n+1},$}
\]
Summing up the above remarks, we get the canonical identification
\[
\mbox{$C^n(A)
\cong\hom{\boldk}{\bigotimes^{-*}(\oV)}%
{\bigotimes^*(V)}{n+1},\ n\geq 0.$}
\]
This is the underlying vector space of
the Gerstenhaber-Schack complex \cite{GS}
which controls the deformation of coassociative bialgebras.
In fact, our complex can be considered as a kind of deformation of the
Gerstenhaber-Schack complex as follows.

Let us simplify the notation by setting
$\E ij:=
\hom{\boldk}{\bigotimes^j(\oV)}{\bigotimes^i(V)}{}$.
Then it is
immediate to see that $C^n(A) = \bigoplus_{i+j=n+1}\E ij$ and that
the differential $d$ decomposes as $d=\dhoch +\dqcohoch
+\sum_{k=1}^{\infty}d_k$, where
\begin{eqnarray}
f\mapsto f\circ D_{-1}\quad\mbox{ induces}\quad\dhoch:\E **
\rightarrow\E *{*+1},\nonumber\\
f\mapsto f\circ D_0 +(-1)^{|f|}\cdot d_C\circ f\quad\mbox{
induces}\quad
 \dqcohoch:\E **\rightarrow \E {*+1}*\label{zebrulka}\\
\mbox{ and } \quad f\mapsto f\circ D_k\quad \mbox{ induces}
\quad d_k:\E **\rightarrow\E{*+k+1}{*-k}.\ k\geq 1\nonumber
\end{eqnarray}
Here $\dhoch$ is the differential of the Hochschild complex of the
associative algebra $(V,\MU)$ with coefficients in the bimodule
$\bigotimes^*(V)$, $\dqcohoch$ is the analog of the Cartier
differential (see \cite{Cart55})
for the (noncoassociative) coalgebra $(V,\Delta)$,
and the maps $d_k$ correspond to the
derivations $D_k$ of Corollary~\ref{andulka}, for
$k\geq1$.
Bigraded differential complexes where the differential
decomposes in as in~(\ref{zebrulka}) are sometimes called
"multicompexes" (although the first author prefers to call them
monsters).

In general, when $\Phi\neq 1$, we do
not have $\dqcohoch^2=0$. However when $\Phi=1$, $\dqcohoch^2=0$ and
$\dhoch +\dqcohoch$ is the differential defined in \cite{GS}.
Moreover,
in this case the operators $D_k$ vanish, which easily follows from the
fact that we work with {\em normalized\/} bar resolution.
So the cohomology of our complex is the same as that of
the Gerstenhaber-Schack complex for the bialgebra cohomology. Notice
also that the complex $(\E *0,\dqcoh)$ is exactly the
$M$-construction $(M^*,d_{q\Omega})$ of~\cite[par.~3]{MS}.

Now we arrive at our goal of defining a cohomology theory
controlling the deformations
of Drinfel'd algebras. The meaning of a ``cohomology theory
controlling
the deformations'' has been explained in the introduction. For more
details we refer the reader to the classical
exposition~\cite{gerst-Ann} or more recent
treatments~\cite{fox:JPAA93,M:tang,shnider-sternberg:book}.

\begin{theorem}
\label{vlacek-bez-jednoho-vagonku}
Deformations of a Drinfel'd algebra $A = (V,\cdot,\D,\Phi)$ are
controlled by the cohomology
\[
\hat H^*(A) := H^*(\hat C(A),\hat d),
\]
where $\hat C^n(A):= \bigoplus\{ \E ij;\ i+j= n+1,\ i\geq 1\}$ and
$\hat d : = d|_{\hat C(A)}$.
\end{theorem}

\noindent
{\bf Proof.} Observe first that
\begin{eqnarray*}
\hatc 1 &=& \jitka11 \oplus V^2,
\\
\hatc 2 &=& \jitka 21 \oplus \jitka 12 \oplus V^3,\
\mbox{and}
\\
\hatc 3 &=& \jitka 13 \oplus \jitka 22 \oplus \jitka 13 \oplus V^4.
\end{eqnarray*}

Let $h = (\chi,D,\Psi) \in \hatc 2$ be an infinitesimal deformation.
 By definition, $\hat d(h)=0$
is equivalent to
\begin{eqnarray}
\label{a}
\dhoch(\chi)&=& 0\ \mbox{ in } \jitka31,
\\
\label{b}
\dqcoh(\chi)+\dhoch(D)&=& 0 \ \mbox{ in } \jitka22,
\\
\label{c}
d_1(\chi)+\dqcoh(D)+\dhoch(\Psi)&=& 0\ \mbox{ in } \jitka13,
\mbox{ and }
\\
\label{d}
d_2(\chi)+d_1(D)+\dqcoh(\Psi)&=&0\ \mbox{ in } \jitka04.
\end{eqnarray}
We shall see that these are exactly the infinitesimal forms of
the structure equations,
\REF{a}\ is the infinitesimal form of the associativity of
$\mu$, \REF{b}\ is the infinitesimal form of the compatibility
between $\mu$ and $\D$, \REF{c} is the infinitesimal form of the
quasi-coassociativity\REF{qcoass}\ and\REF{d}\ is the infinitesimal
form of the pentagon identity\REF{PENTAGON}.
{}From this it will follow that $\hat d(h) = 0$
if and only if $(\mu + t\chi, \Delta + tD, \Phi +t\Psi)$ is an
infinitesimal deformation of the algebra $A = (V,\mu,\D,\Phi)$,
$t$ being an independent variable.

Suppose $g = (\phi,F)\in \hatc1$. By definition, $\hat d(g) = h$ means
that
\begin{eqnarray}
\dhoch(\phi)&=&\chi,\label{e}\\
\dqcoh(\phi)+\dhoch(F) &=& D\label{f}\\
d_1(\phi)+\dqcoh(F)&=& \Psi.\label{g}
\end{eqnarray}
We shall show that the above equations describe the infinitesimal
action of the infinitesimal automorphism $\phi$ and the infinitesimal
twisting $F$ on the infinitesimal deformation $(\chi,D,\Psi)$,
hence $\hat H^2(A)$ describes the infinitesimal deformations modulo
infinitesimal automorphisms and twistings.
A first order deformation is equivalent to the trivial first order
 deformation if and only if the infinitesimal cocycle cobounds and
two extensions (which both cobound the same obstruction cocycle) are
equivalent if and only if their difference cobounds.

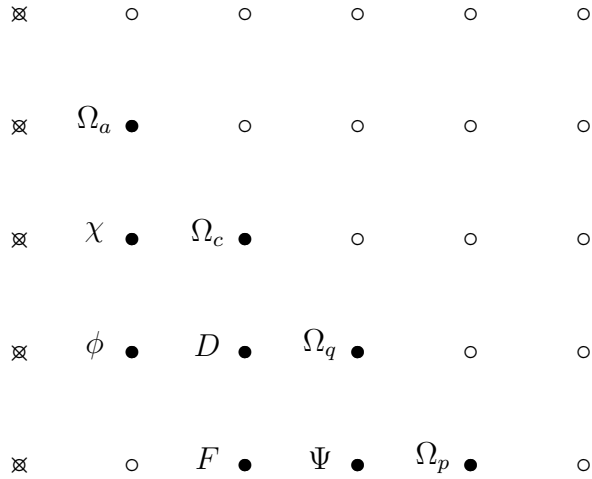
\begin{figure}

\setlength{\unitlength}{.05cm}
\begin{center}
\begin{picture}(200,120)(15,-5)

%prvni rada
\put(48,-2){\makebox(0,0){$\circ$}\makebox(0,0){$\times$}}
\put(78,-2){\makebox(0,0){$\circ$}}
\put(108,-2){\makebox(0,0){$\circ$}}
\put(138,-2){\makebox(0,0){$\circ$}}
\put(168,-2){\makebox(0,0){$\circ$}}
\put(198,-2){\makebox(0,0){$\circ$}}

%druha rada
\put(48,28){\makebox(0,0){$\circ$}\makebox(0,0){$\times$}}
\put(78,28){\makebox(0,0){$\circ$}}
\put(108,28){\makebox(0,0){$\circ$}}
\put(138,28){\makebox(0,0){$\circ$}}
\put(168,28){\makebox(0,0){$\circ$}}
\put(198,28){\makebox(0,0){$\circ$}}

%treti rada
\put(48,58){\makebox(0,0){$\circ$}\makebox(0,0){$\times$}}
\put(78,58){\makebox(0,0){$\circ$}}
\put(108,58){\makebox(0,0){$\circ$}}
\put(138,58){\makebox(0,0){$\circ$}}
\put(168,58){\makebox(0,0){$\circ$}}
\put(198,58){\makebox(0,0){$\circ$}}

%ctvrta rada
\put(48,88){\makebox(0,0){$\circ$}\makebox(0,0){$\times$}}
\put(78,88){\makebox(0,0){$\circ$}}
\put(108,88){\makebox(0,0){$\circ$}}
\put(138,88){\makebox(0,0){$\circ$}}
\put(168,88){\makebox(0,0){$\circ$}}
\put(198,88){\makebox(0,0){$\circ$}}

%pata rada
\put(48,118){\makebox(0,0){$\circ$}\makebox(0,0){$\times$}}
\put(78,118){\makebox(0,0){$\circ$}}
\put(108,118){\makebox(0,0){$\circ$}}
\put(138,118){\makebox(0,0){$\circ$}}
\put(168,118){\makebox(0,0){$\circ$}}
\put(198,118){\makebox(0,0){$\circ$}}

\put(108,-2){\makebox(0,0){$\bullet$}}
\put(98,0){\makebox(0,0){$F$}}
\put(138,-2){\makebox(0,0){$\bullet$}}
\put(128,0){\makebox(0,0){$\Psi$}}
\put(168,-2){\makebox(0,0){$\bullet$}}
\put(158,0){\makebox(0,0){$\Omega_p$}}
\put(78,28){\makebox(0,0){$\bullet$}}
\put(68,30){\makebox(0,0){$\phi$}}
\put(108,28){\makebox(0,0){$\bullet$}}
\put(98,30){\makebox(0,0){$D$}}
\put(138,28){\makebox(0,0){$\bullet$}}
\put(128,30){\makebox(0,0){$\Omega_q$}}
\put(78,58){\makebox(0,0){$\bullet$}}
\put(68,60){\makebox(0,0){$\chi$}}
\put(108,58){\makebox(0,0){$\bullet$}}
\put(98,60){\makebox(0,0){$\Omega_c$}}
\put(78,88){\makebox(0,0){$\bullet$}}
\put(68,90){\makebox(0,0){$\Omega_a$}}

\end{picture}
\caption{\label{fig8}
Positions in $E^{**}$ of infinitesimal automorphism $\phi$,
twisting $F$, infinitesimal deformations
$\chi,D,\Psi$ of multiplication, comultiplication, and
$\Phi$, respectively, and obstruction classes $\Omega_a$, $\Omega_c$,
$\Omega_q$ and $\Omega_p$
of the structure equations, associativity, compatibility,
quasi-associativity and pentagon identities, respectively.}
\end{center}
\end{figure}

We now proceed to demonstrate the statements made above by giving
the explicit formulae for the terms of (\ref{a}) to (\ref{g}),
separating the terms according to the type of operator,
see Figure~\ref{fig8} for the positions of relevant elements in the
bicomplex.

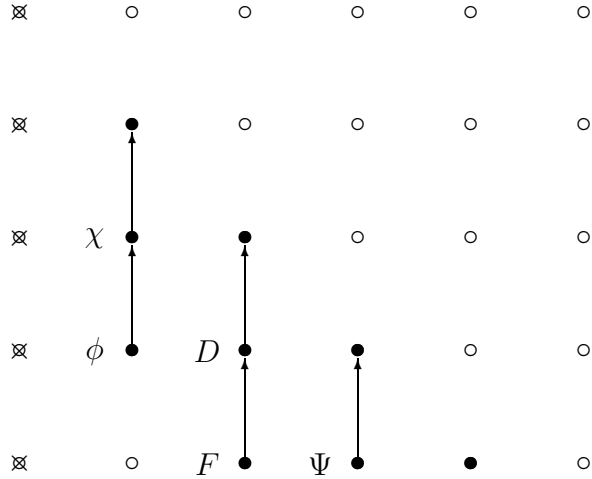
\begin{figure}
\setlength{\unitlength}{.05cm}
\begin{center}
\begin{picture}(200,120)(15,-5)

%prvni rada
\put(48,-2){\makebox(0,0){$\circ$}\makebox(0,0){$\times$}}
\put(78,-2){\makebox(0,0){$\circ$}}
\put(108,-2){\makebox(0,0){$\circ$}}
\put(138,-2){\makebox(0,0){$\circ$}}
\put(168,-2){\makebox(0,0){$\circ$}}
\put(198,-2){\makebox(0,0){$\circ$}}

%druha rada
\put(48,28){\makebox(0,0){$\circ$}\makebox(0,0){$\times$}}
\put(78,28){\makebox(0,0){$\circ$}}
\put(108,28){\makebox(0,0){$\circ$}}
\put(138,28){\makebox(0,0){$\circ$}}
\put(168,28){\makebox(0,0){$\circ$}}
\put(198,28){\makebox(0,0){$\circ$}}

%treti rada
\put(48,58){\makebox(0,0){$\circ$}\makebox(0,0){$\times$}}
\put(78,58){\makebox(0,0){$\circ$}}
\put(108,58){\makebox(0,0){$\circ$}}
\put(138,58){\makebox(0,0){$\circ$}}
\put(168,58){\makebox(0,0){$\circ$}}
\put(198,58){\makebox(0,0){$\circ$}}

%ctvrta rada
\put(48,88){\makebox(0,0){$\circ$}\makebox(0,0){$\times$}}
\put(78,88){\makebox(0,0){$\circ$}}
\put(108,88){\makebox(0,0){$\circ$}}
\put(138,88){\makebox(0,0){$\circ$}}
\put(168,88){\makebox(0,0){$\circ$}}
\put(198,88){\makebox(0,0){$\circ$}}

%pata rada
\put(48,118){\makebox(0,0){$\circ$}\makebox(0,0){$\times$}}
\put(78,118){\makebox(0,0){$\circ$}}
\put(108,118){\makebox(0,0){$\circ$}}
\put(138,118){\makebox(0,0){$\circ$}}
\put(168,118){\makebox(0,0){$\circ$}}
\put(198,118){\makebox(0,0){$\circ$}}

\put(108,-2){\makebox(0,0){$\bullet$}}
\put(138,-2){\makebox(0,0){$\bullet$}}
\put(168,-2){\makebox(0,0){$\bullet$}}
\put(78,28){\makebox(0,0){$\bullet$}}
\put(108,28){\makebox(0,0){$\bullet$}}
\put(138,28){\makebox(0,0){$\bullet$}}
\put(78,58){\makebox(0,0){$\bullet$}}
\put(108,58){\makebox(0,0){$\bullet$}}
\put(78,88){\makebox(0,0){$\bullet$}}

\put(78,58){\line(0,1){25}}
\put(78,78){\vector(0,1){8}}
\put(68,58){\makebox(0,0){$\chi$}}
\put(78,28){\line(0,1){25}}
\put(78,48){\vector(0,1){8}}
\put(68,28){\makebox(0,0){$\phi$}}
\put(108,28){\line(0,1){25}}
\put(108,48){\vector(0,1){8}}
\put(98,28){\makebox(0,0){$D$}}

\put(108,-2){\line(0,1){25}}
\put(108,18){\vector(0,1){8}}
\put(98,-2){\makebox(0,0){$F$}}
\put(138,-2){\line(0,1){25}}
\put(138,18){\vector(0,1){8}}
\put(128,-2){\makebox(0,0){$\Psi$}}
\end{picture}

\caption{\label{fig9}
The Hochschild differentials of representative elements.}
\end{center}
\end{figure}

We use the same convention as in the
introduction, i.e.~we represent our bigraded complex
as a non-negative integer lattice in the plane
with the position $i^{\rm th}$ column $j^{\rm th}$ row (counting
upwards) occupied by $\E ij$.
First we describe the Hochschild differentials represented in
Figure~\ref{fig9}.

The differential
$\dhoch : \jitka21 \to \jitka31$ is given by
\[
\dhoch(\chi)(v_1,v_2,v_3) = v_1\cdot \chi(v_2,v_3) -
\chi(v_1\cdot v_2,v_3)+\chi(v_1, v_2\cdot v_3)
-\chi(v_1,v_2)\cdot v_3,
\]
for $\chi \in \jitka21,\ v_1,v_2,v_3 \in \oV$.
This is the variational derivative of the associativity expression
$(v_1v_2) v_3 - v_1(v_2v_3)$ for the deformed multiplication
relative to an infinitesimal change in the multiplication.
It appears in (\ref{a}) and its vanishing is the usual cocycle
condition
for infinitesimal deformation of an associative multiplication.

The differential
$\dhoch : \jitka12 \to \jitka22$ is given by
\begin{eqnarray*}
\dhoch(D)& =&\Sum D'(v_1)\cdot \D'(v_2)\ot D''(v_1)\cdot\D''(v_2)-
\\
&&-
D(v_1\cdot v_2)+\Sum
\D'(v_1)\cdot D'(v_2)\ot \D''(v_1)\cdot D''(v_2),
\end{eqnarray*}
for $D =\Sum D'\ot D'' \in \jitka12$, $\D = \Sum \D' \ot \D''$ (the
Sweedler notation) and $v_1,v_2\in \oV$.
This is the variational derivative of the condition for compatibility
of multiplication and comultiplication with respect to an
infinitesimal change in the comultiplication, $D$.
It is one of the summands in (\ref{b}). We complete the discussion
of this and the remaining equations once we have described all the
summands in a particular equation.

 The differential
$\dhoch : \jitka03 \to \jitka13$ is given by
\[
\dhoch(\Psi)(v) = (\id\ot \D)\D(v)\cdot \Psi - \Psi \cdot (\D\ot
\id)\D(v),
\]
for $\Psi \in \jitka03$ and $v\in \oV$. This is the variational
derivative of the quasi-coassociativity condition relative to an
infinitesimal
variation, $\Psi,$ in the (co)associativity operator
$\Phi$ of the Drinfel'd algebra.

 The differential
$\dhoch : \jitka11 \to \jitka21$ is given by
\[
\dhoch(\phi)(v_1,v_2) = v_1\cdot \phi(v_2) - \phi(v_1\cdot v_2)+
\phi(v_1)\cdot v_2,
\]
for $\phi \in \jitka11,\ v_1,v_2 \in \oV$.
This is the variational derivative of multiplication relative to an
infinitesimal automorphism. Just as in the
classical theory of deformations of associative algebras, if
the infinitesimal deformation of multiplication is the
Hochschild coboundary of $\phi$ then the automorphism
$id + t\phi$ transforms it to zero.

The differential
$\dhoch: \jitka02 \to\jitka12$ is given by
\[
\dhoch (F)(a) = F \cdot \Delta(a) - \Delta(a)\cdot F
\]
for $F \in \jitka02 =\bigotimes^2 V$ and $a\in \oV$.
This expresses the variational derivative of comultiplication with
respect to an infinitesimal twisting, $F$, which
appears in (\ref{f}) and will be discussed below.

\begin{figure}
\setlength{\unitlength}{.05cm}
\begin{center}
\begin{picture}(200,120)(15,-5)

%prvni rada
\put(48,-2){\makebox(0,0){$\circ$}\makebox(0,0){$\times$}}
\put(78,-2){\makebox(0,0){$\circ$}}
\put(108,-2){\makebox(0,0){$\circ$}}
\put(138,-2){\makebox(0,0){$\circ$}}
\put(168,-2){\makebox(0,0){$\circ$}}
\put(198,-2){\makebox(0,0){$\circ$}}

%druha rada
\put(48,28){\makebox(0,0){$\circ$}\makebox(0,0){$\times$}}
\put(78,28){\makebox(0,0){$\circ$}}
\put(108,28){\makebox(0,0){$\circ$}}
\put(138,28){\makebox(0,0){$\circ$}}
\put(168,28){\makebox(0,0){$\circ$}}
\put(198,28){\makebox(0,0){$\circ$}}

%treti rada
\put(48,58){\makebox(0,0){$\circ$}\makebox(0,0){$\times$}}
\put(78,58){\makebox(0,0){$\circ$}}
\put(108,58){\makebox(0,0){$\circ$}}
\put(138,58){\makebox(0,0){$\circ$}}
\put(168,58){\makebox(0,0){$\circ$}}
\put(198,58){\makebox(0,0){$\circ$}}

%ctvrta rada
\put(48,88){\makebox(0,0){$\circ$}\makebox(0,0){$\times$}}
\put(78,88){\makebox(0,0){$\circ$}}
\put(108,88){\makebox(0,0){$\circ$}}
\put(138,88){\makebox(0,0){$\circ$}}
\put(168,88){\makebox(0,0){$\circ$}}
\put(198,88){\makebox(0,0){$\circ$}}

%pata rada
\put(48,118){\makebox(0,0){$\circ$}\makebox(0,0){$\times$}}
\put(78,118){\makebox(0,0){$\circ$}}
\put(108,118){\makebox(0,0){$\circ$}}
\put(138,118){\makebox(0,0){$\circ$}}
\put(168,118){\makebox(0,0){$\circ$}}
\put(198,118){\makebox(0,0){$\circ$}}

\put(108,-2){\makebox(0,0){$\bullet$}}
\put(138,-2){\makebox(0,0){$\bullet$}}
\put(168,-2){\makebox(0,0){$\bullet$}}
\put(78,28){\makebox(0,0){$\bullet$}}
\put(108,28){\makebox(0,0){$\bullet$}}
\put(138,28){\makebox(0,0){$\bullet$}}
\put(78,58){\makebox(0,0){$\bullet$}}
\put(108,58){\makebox(0,0){$\bullet$}}
\put(78,88){\makebox(0,0){$\bullet$}}

\put(78,58){\line(1,0){25}}
\put(98,58){\vector(1,0){8}}
\put(68,58){\makebox(0,0){$\chi$}}
\put(78,58){\line(1,0){25}}
\put(98,58){\vector(1,0){8}}
\put(68,28){\makebox(0,0){$\phi$}}
\put(78,28){\vector(1,0){28}}
\put(108,-2){\line(1,0){25}}
\put(128,-2){\vector(1,0){8}}
\put(98,-2){\makebox(0,0){$F$}}

\put(108,28){\line(1,0){25}}
\put(128,28){\vector(1,0){8}}
\put(108,38){\makebox(0,0){$D$}}

\put(138,-2){\line(1,0){25}}
\put(158,-2){\vector(1,0){8}}
\put(138,8){\makebox(0,0){$\Psi$}}
\end{picture}
\end{center}
\caption{\label{fig10}
The modified Cartier operator on representative elements.}
\end{figure}
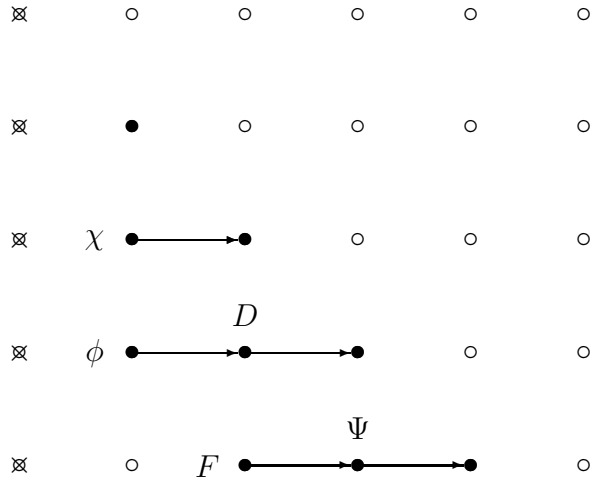

Next consider the operator $\dqcoh$ on the same elements,
see Figure~\ref{fig10}. The operator
$\dqcoh : \jitka21 \to \jitka22$ is given by
\begin{eqnarray*}
\dqcoh(\chi)(v_1,v_2) &=& \Sum\D'(v_1)\cdot \D'(v_2)\ot
\chi(\D''(v_1),\D''(v_2)) -\D \chi(v_1,v_2)+
\\
&&+\Sum \chi(\D'(v_1),\D'(v_2))\ot
\D''(v_1)\cdot\D''(v_2),
\end{eqnarray*}
for $\chi \in \jitka21$, $\D = \Sum \D' \ot \D''$
and $v_1,v_2 \in \oV$.
This gives the variational derivative of the compatibility condition
for multiplication and comultiplication relative to an infinitesimal
variation, $\chi,$ in multiplication. Equation (\ref{b}) says that
the total variational derivative of the compatibility condition
relative to a combined infinitesimal variation of multiplication
and comultiplication is zero. This is clearly a necessary and
sufficient
condition for the infinitesimal preservation of compatibility.

The operator $\dqcoh : \jitka12 \to \jitka13$ is given by
\[
\dqcoh(D)(v)= (\id \ot D)\D(v)\cdot \Phi - \Phi \cdot (\D\ot
\id)D(v)+(\id \ot \D)D(v)\cdot \Phi - \Phi \cdot (D\ot
\id)\D(v),
\]
for $D \in \jitka 12$, $v\in \oV$, which
is the variational derivative of the quasi-coassociativity
relative to an infinitesimal change in comultiplication.
It appears in (\ref{c}) and we need to consider yet one more
term before concluding the discussion of this equation.

\def\de{\D}
The operator $\dqcoh : \jitka0{3} \to \jitka04$ is given by
\begin{eqnarray*}
\dqcoh(\Psi)&\!\!=\!\!&
(1\OO \Psi)\CCCC(\id\OO\de\OO\id)(\Phi)\CCCC(\Phi\OO1)-
(\id^2\OO\de)(\Phi)\CCCC(\de\OO\id^2)(\Psi)+
\\
&&+(1\OO\Phi)\CCCC(\id\OO\de\OO\id)(\Psi)\CCCC(\Phi\OO1)-
(\id^2\OO\de)(\Psi)\CCCC(\de\OO\id^2)(\Phi)+
\\
&&+(1\OO\Phi)\CCCC(\id\OO\de\OO\id)(\Phi)\CCCC(\Psi\OO1).
\end{eqnarray*}
for $\Psi \in \jitka03 = \bigotimes^3V$.
It represents the variational derivative of the pentagon identity
relative to an infinitesimal variation, $\Psi$, in $\Phi$.

The operator $\dqcoh : \jitka11 \to \jitka12$
for $\phi \in \jitka11$ and $v\in \oV$ is given by
\[
\dqcoh(\phi)(v) = (\id\ot \phi)\D(v)-\D\phi(v)+(\phi\ot\id)\D,
\]
which expresses the variational derivative of the comultiplication
relative to an infinitesimal automorphism, $\phi.$
The sum of this term and $\dhoch(F)$ is the total variational
derivative of the comultiplication relative to a combined
infinitesimal automorphism and twisting. If this gives the
infinitesimal comultiplication then the combined effect
of the automorphism $\id +t\phi$ and twisting by $1 +t F$ is
to transform the infinitesimal comultiplication to zero.

The operator $\dqcoh :
\jitka02 \to \jitka03$ is
given by

\[
\dqcoh(F) =
(1\ot F)\cdot\Phi-\Phi\cdot(\de\ot\id)(F)+
(\id\ot\de)(F)\cdot\Phi-\Phi\cdot(F\ot1),
\]
for $F \in \jitka02 = \bigotimes^2V$.
This is the variational derivative of the (co)associativity
operator $\Phi$ relative to an infinitesimal twisting, $F$.

\begin{figure}
\setlength{\unitlength}{.05cm}
\begin{center}
\begin{picture}(200,120)(10,-5)

%prvni rada
\put(48,-2){\makebox(0,0){$\circ$}\makebox(0,0){$\times$}}
\put(78,-2){\makebox(0,0){$\circ$}}
\put(108,-2){\makebox(0,0){$\circ$}}
\put(138,-2){\makebox(0,0){$\circ$}}
\put(168,-2){\makebox(0,0){$\circ$}}
\put(198,-2){\makebox(0,0){$\circ$}}

%druha rada
\put(48,28){\makebox(0,0){$\circ$}\makebox(0,0){$\times$}}
\put(78,28){\makebox(0,0){$\circ$}}
\put(108,28){\makebox(0,0){$\circ$}}
\put(138,28){\makebox(0,0){$\circ$}}
\put(168,28){\makebox(0,0){$\circ$}}
\put(198,28){\makebox(0,0){$\circ$}}

%treti rada
\put(48,58){\makebox(0,0){$\circ$}\makebox(0,0){$\times$}}
\put(78,58){\makebox(0,0){$\circ$}}
\put(108,58){\makebox(0,0){$\circ$}}
\put(138,58){\makebox(0,0){$\circ$}}
\put(168,58){\makebox(0,0){$\circ$}}
\put(198,58){\makebox(0,0){$\circ$}}

%ctvrta rada
\put(48,88){\makebox(0,0){$\circ$}\makebox(0,0){$\times$}}
\put(78,88){\makebox(0,0){$\circ$}}
\put(108,88){\makebox(0,0){$\circ$}}
\put(138,88){\makebox(0,0){$\circ$}}
\put(168,88){\makebox(0,0){$\circ$}}
\put(198,88){\makebox(0,0){$\circ$}}

%pata rada
\put(48,118){\makebox(0,0){$\circ$}\makebox(0,0){$\times$}}
\put(78,118){\makebox(0,0){$\circ$}}
\put(108,118){\makebox(0,0){$\circ$}}
\put(138,118){\makebox(0,0){$\circ$}}
\put(168,118){\makebox(0,0){$\circ$}}
\put(198,118){\makebox(0,0){$\circ$}}

\put(108,-2){\makebox(0,0){$\bullet$}}
\put(138,-2){\makebox(0,0){$\bullet$}}
\put(168,-2){\makebox(0,0){$\bullet$}}
\put(78,28){\makebox(0,0){$\bullet$}}
\put(108,28){\makebox(0,0){$\bullet$}}
\put(138,28){\makebox(0,0){$\bullet$}}
\put(78,58){\makebox(0,0){$\bullet$}}
\put(108,58){\makebox(0,0){$\bullet$}}
\put(78,88){\makebox(0,0){$\bullet$}}

\put(78,28){\vector(2,-1){57}}
\put(78,88){\vector(2,-1){57}}
\put(65,28){$\phi$}
\put(78,58){\vector(2,-1){57}}
\put(65,58){$\chi$}
\put(65,88){$\Omega_a$}
\put(108,28){\vector(2,-1){57}}
\put(95,58){$\Omega_c$}
\put(108,58){\vector(2,-1){57}}
\put(125,22){$\Omega_q$}
\put(138,28){\vector(2,-1){57}}
\put(95,28){$D$}
\end{picture}

\caption{\label{fig11}
All the occurrences of the operator $d_1$ in deformation theory.}
\end{center}
\end{figure}
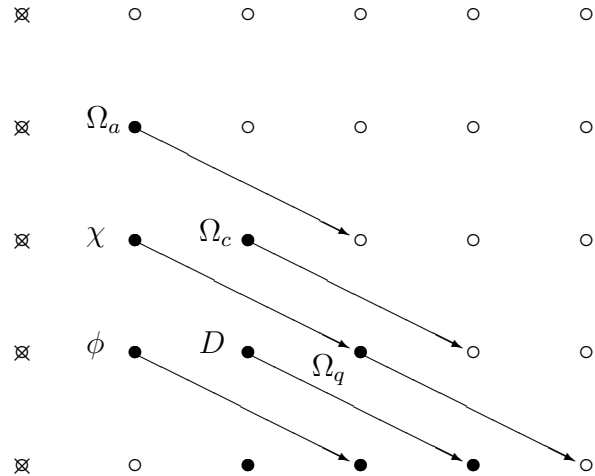

We shall give the explicit formula for three of the six occurrences
$d_1$ in deformation theory as illustrated in Figure~\ref{fig11}.

The variational derivative of the quasi-coassociativity relative
to an infinitesimal change in multiplication, $\chi,$ is given by
the map $d_1:\jitka21 \to \jitka13$,
\begin{eqnarray*}
d_1(\chi) &=&\sum_{(\bcc_1,\bcc_2,\bcc_3)\in A}\{\Sum
(\bcc_1\ot\bcc_2\ot\bcc_3)((\id\ot \D)\D(v)\bar\otimes\Phi
-\Phi\bar\otimes(\D\ot \id)\D(v))\\
&=&\sum_{(\bcc_1,\bcc_2,\bcc_3)\in A}\{\Sum
\BC_1(\D'(v),\Phi_1)\ot \BC_2(\D'(\D''(v)),\Phi_2)\ot
\BC_3(\D''(\D''(v)),\Phi_3)-
\\
&&\hphantom{\sum_{(\bcc_1,\bcc_2,\bcc_3)\in A}x}-
\Sum\BC_1(\Phi_1,\D'(\D'(v)))\ot \BC_2(\Phi_2,\D''(\D'(v))\ot
\BC_3(\Phi_3,\D''(v)\},
\end{eqnarray*}
for $\chi \in \jitka 21$, $v\in \oV$, $\Phi = \Sum
\Phi_1\ot\Phi_2\ot\Phi_3$, $\Delta = \Sum\D'\ot\D''$ and
\[
A :=
\{(\chi,\mu,\mu),(\mu,\chi,\mu), (\mu,\mu,\chi)\}.
\]
We hope that the meaning of this last notation is clear. For example,
if $(\bcc_1,\bcc_2,\bcc_3) =(\chi,\mu,\mu)$, then
\begin{eqnarray*}
\lefteqn{(\bcc_1\ot\bcc_2\ot\bcc_3)((\id\ot \D)\D(v)\bar\otimes\Phi)=}
\\
&&=\BC_1(\D'(v),\Phi_1)\ot \BC_2(\D'(\D''(v)),\Phi_2)\ot
\BC_3(\D''(\D''(v)),\Phi_3)
\\
&&=
\chi(\D'(v),\Phi_1)\ot \D'(\D''(v))\cdot\Phi_2\ot
\D''(\D''(v))\cdot\Phi_3.
\end{eqnarray*}
Equation (\ref{c}) is the total variational derivative
of quasi-coassociativity relative to all three components of the
infinitesimal deformation. Its vanishing is the condition for
preserving quasi-coassociativity under an infinitesimal deformation.

The map $d_1 : \jitka12 \to \jitka 04$ describes the variational
derivative of the pentagon identity relative to an infinitesimal
variation in comultiplication $D$,
\begin{eqnarray*}
d_1(D)&=& -(\id^2\ot \D)(\Phi)\cdot (D\ot
\id^2)(\Phi)+(1\ot\Phi)\cdot(\id \ot D\ot \id)(\Phi)\cdot(1\ot \Phi)-
\\
&&-(\id^2\ot D)(\Phi)\cdot (\D\ot\id^2)(\Phi),
\end{eqnarray*}
for $D \in \jitka12$.

\begin{figure}
\setlength{\unitlength}{.05cm}
\begin{center}
\begin{picture}(200,120)(15,-5)

%prvni rada
\put(48,-2){\makebox(0,0){$\circ$}\makebox(0,0){$\times$}}
\put(78,-2){\makebox(0,0){$\circ$}}
\put(108,-2){\makebox(0,0){$\circ$}}
\put(138,-2){\makebox(0,0){$\circ$}}
\put(168,-2){\makebox(0,0){$\circ$}}
\put(198,-2){\makebox(0,0){$\circ$}}

%druha rada
\put(48,28){\makebox(0,0){$\circ$}\makebox(0,0){$\times$}}
\put(78,28){\makebox(0,0){$\circ$}}
\put(108,28){\makebox(0,0){$\circ$}}
\put(138,28){\makebox(0,0){$\circ$}}
\put(168,28){\makebox(0,0){$\circ$}}
\put(198,28){\makebox(0,0){$\circ$}}

%treti rada
\put(48,58){\makebox(0,0){$\circ$}\makebox(0,0){$\times$}}
\put(78,58){\makebox(0,0){$\circ$}}
\put(108,58){\makebox(0,0){$\circ$}}
\put(138,58){\makebox(0,0){$\circ$}}
\put(168,58){\makebox(0,0){$\circ$}}
\put(198,58){\makebox(0,0){$\circ$}}

%ctvrta rada
\put(48,88){\makebox(0,0){$\circ$}\makebox(0,0){$\times$}}
\put(78,88){\makebox(0,0){$\circ$}}
\put(108,88){\makebox(0,0){$\circ$}}
\put(138,88){\makebox(0,0){$\circ$}}
\put(168,88){\makebox(0,0){$\circ$}}
\put(198,88){\makebox(0,0){$\circ$}}

%pata rada
\put(48,118){\makebox(0,0){$\circ$}\makebox(0,0){$\times$}}
\put(78,118){\makebox(0,0){$\circ$}}
\put(108,118){\makebox(0,0){$\circ$}}
\put(138,118){\makebox(0,0){$\circ$}}
\put(168,118){\makebox(0,0){$\circ$}}
\put(198,118){\makebox(0,0){$\circ$}}

\put(108,-2){\makebox(0,0){$\bullet$}}
\put(138,-2){\makebox(0,0){$\bullet$}}
\put(168,-2){\makebox(0,0){$\bullet$}}
\put(78,28){\makebox(0,0){$\bullet$}}
\put(108,28){\makebox(0,0){$\bullet$}}
\put(138,28){\makebox(0,0){$\bullet$}}
\put(78,58){\makebox(0,0){$\bullet$}}
\put(108,58){\makebox(0,0){$\bullet$}}
\put(78,88){\makebox(0,0){$\bullet$}}

\put(78,58){\vector(3,-2){88}}
\put(78,88){\vector(3,-2){88}}
\put(108,58){\vector(3,-2){88}}
\put(68,58){\makebox(0,0){$\chi$}}
\put(98,58){\makebox(0,0){$\Omega_c$}}
\put(68,88){\makebox(0,0){$\Omega_a$}}

\end{picture}
\end{center}

\caption{\label{fig12}
All the occurrences of the operator $d_2$ in deformation theory.}
\end{figure}
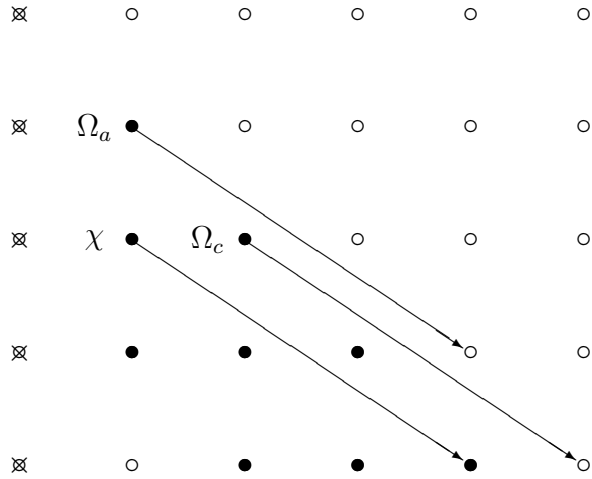

The variational derivative of the associativity operator
relative to an infinitesimal automorphism, $\phi,$ is given by
the operator $d_1
:\jitka11 \to \jitka03$,
\[
d_1(\phi)=
(\phi\ot \id^{\ot 2})(\Phi)+(\id\ot\phi\ot\id)(\Phi)+(\id^{\ot 2}\ot
\phi)(\Phi),
\]
for $\phi \in \jitka11$. Together with $\dqcoh(F)$ this is the
infinitesimal variation in $\Phi$ due to an infinitesimal twisting
and an infinitesimal automorphism.
In order to complete the discussion of~(\ref{d})
we need to compute
$d_2(\chi)$, see Figure~\ref{fig12}.

The variational derivative of the pentagon identity
relative to an infinitesimal variation in multiplication,
$\chi$, is described by the map $d_2:\jitka 21 \to \jitka 04$
\begin{eqnarray*}
d_2(\chi)&=&
\Sum_B
\{ -\Sum\BC_1(\Phi_1,\D'(\Phi_1))\ot \BC_2(\Phi_2,\D''(\Phi_1))\ot
\BC_3(\D'(\Phi_3),\Phi_2) \ot \BC_4(\D''(\Phi_3),\Phi_3)+
\\
&+&\Sum\BC_1(1, \Phi_1\cdot\Phi_1)\ot
\BC_2(\Phi_1,\D'(\Phi_2)\cdot\Phi_2)
\ot\BC_3(\Phi_2, \D''(\Phi_2)\cdot\Phi_3)\ot \BC_4(\Phi_3,\Phi_3)+
\\
&+&\Sum (1\otimes \Phi)\cdot[\BC_1(\Phi_1,\Phi_1)\ot
\BC_2(\D'(\Phi_2),\Phi_2)\ot
\BC_3(\D''(\Phi_2),\Phi_3)\ot \BC_4(\Phi_3,1)]\},
\end{eqnarray*}
for $\chi \in \jitka 21$ and $\D = \Sum \D' \ot \D''$, where the
``big'' summation is taken over
\[
(\bcc_1,\bcc_2,\bcc_3,\bcc_4)\in
B:= \{(\chi,\mu,\mu,\mu),(\mu,\chi,\mu,\mu),
(\mu,\mu,\chi,\mu),(\mu,\mu,\mu,\chi)\}.
\]
The elements of $A^{\otimes 4}$ whose components appear in the
previous expression correspond to the associativity operators
corresponding to the edges in the triangulation of the pentagon
$K_4$ illustrated in Figure~\ref{fig7} with the first,
second and third summands
corresponding to the top, middle and bottom triangles respectively.
Note that the sequence of transition operators read from
left to right is the opposite of their order with respect to the
ordering of the vertices. This is a simple consequence of the
usual rules for writing functional composition.
 The negative sign in the first summand comes from the fact that the
orientation given by the ordering of the vertices is the reverse of
the natural orientation in the boundary of the pentagon.

We see that~(\ref{d}) expresses the total variational derivative of
the pentagon identity relative to all three components of the
infinitesimal deformation. This finishes the proof that
$\hat H\sp2(A)$describes the infinitesimal deformations modulo
infinitesimal automorphisms and twistings.

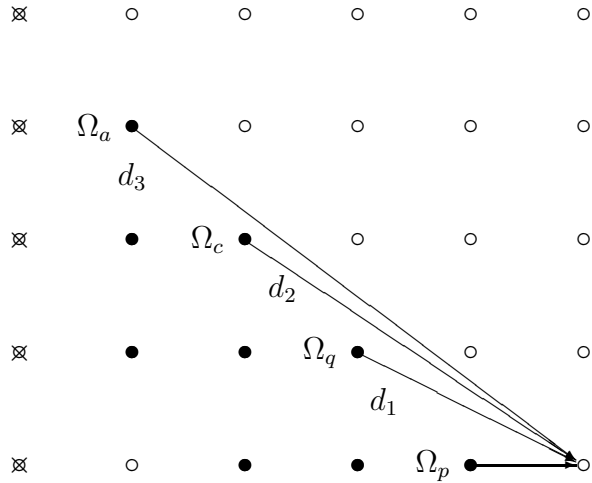
\begin{figure}
\setlength{\unitlength}{.05cm}
\begin{center}
\begin{picture}(200,120)(15,-5)
%prvni rada
\put(48,-2){\makebox(0,0){$\circ$}\makebox(0,0){$\times$}}
\put(78,-2){\makebox(0,0){$\circ$}}
\put(108,-2){\makebox(0,0){$\circ$}}
\put(138,-2){\makebox(0,0){$\circ$}}
\put(168,-2){\makebox(0,0){$\circ$}}
\put(198,-2){\makebox(0,0){$\circ$}}

%druha rada
\put(48,28){\makebox(0,0){$\circ$}\makebox(0,0){$\times$}}
\put(78,28){\makebox(0,0){$\circ$}}
\put(108,28){\makebox(0,0){$\circ$}}
\put(138,28){\makebox(0,0){$\circ$}}
\put(168,28){\makebox(0,0){$\circ$}}
\put(198,28){\makebox(0,0){$\circ$}}

%treti rada
\put(48,58){\makebox(0,0){$\circ$}\makebox(0,0){$\times$}}
\put(78,58){\makebox(0,0){$\circ$}}
\put(108,58){\makebox(0,0){$\circ$}}
\put(138,58){\makebox(0,0){$\circ$}}
\put(168,58){\makebox(0,0){$\circ$}}
\put(198,58){\makebox(0,0){$\circ$}}

%ctvrta rada
\put(48,88){\makebox(0,0){$\circ$}\makebox(0,0){$\times$}}
\put(78,88){\makebox(0,0){$\circ$}}
\put(108,88){\makebox(0,0){$\circ$}}
\put(138,88){\makebox(0,0){$\circ$}}
\put(168,88){\makebox(0,0){$\circ$}}
\put(198,88){\makebox(0,0){$\circ$}}

%pata rada
\put(48,118){\makebox(0,0){$\circ$}\makebox(0,0){$\times$}}
\put(78,118){\makebox(0,0){$\circ$}}
\put(108,118){\makebox(0,0){$\circ$}}
\put(138,118){\makebox(0,0){$\circ$}}
\put(168,118){\makebox(0,0){$\circ$}}
\put(198,118){\makebox(0,0){$\circ$}}

\put(108,-2){\makebox(0,0){$\bullet$}}
\put(138,-2){\makebox(0,0){$\bullet$}}
\put(168,-2){\makebox(0,0){$\bullet$}}
\put(78,28){\makebox(0,0){$\bullet$}}
\put(108,28){\makebox(0,0){$\bullet$}}
\put(138,28){\makebox(0,0){$\bullet$}}
\put(78,58){\makebox(0,0){$\bullet$}}
\put(108,58){\makebox(0,0){$\bullet$}}
\put(78,88){\makebox(0,0){$\bullet$}}

\put(78, 88){\vector(4,-3){118}}
\put(68,88){\makebox(0,0){$\Omega_a$}}
\put(108,58){\vector(3,-2){88}}
\put(98,58){\makebox(0,0){$\Omega_c$}}
\put(78,75){\makebox(0,0){$d_3$}}
\put(138,28){\vector(2,-1){58}}
\put(128,28){\makebox(0,0){$\Omega_q$}}
\put(118,45){\makebox(0,0){$d_2$}}
\put(168,-2){\vector(1,0){28}}
\put(158,-2){\makebox(0,0){$\Omega_p$}}
\put(145,15){\makebox(0,0){$d_1$}}

\end{picture}
\caption{\label{fig13}
The one occurrence of $d_3$ in deformation cohomology.}
\end{center}
\end{figure}

The proof that $\hat H^3 (A)$ describes the
obstructions to the integrability of an infinitesimal deformation,
involves the first and only appearance of the operator $d_3$. It acts
on
precisely one component of the obstruction class, $\Omega_a$, the
obstruction to associativity appearing in position $(1,3)$.
For the full
proof that the obstruction class, consisting of four components
$\Omega_a$, $\Omega_c$, $\Omega_q$ and $\Omega_p$ as in
Figure~\ref{fig13}, is a cocycle
we may either use the classical
approach of~\cite{shnider-sternberg:preprint}, or the deviation
calculus of~\cite{MS}. This straightforward
computation would stretch the length of the
paper beyond any reasonable limit and is therefore omitted.
\qed

In the deformation theory, $d_3(\Omega_a)$ is one of the
terms in the coboundary of the obstruction class, see
Figure~\ref{fig13}.
We include the explicit formula even though it is very
complicated since it is the first really non-trivial
example of the general construction. We will write $d_3(\Omega_a)$ as
\[
d_3(\Omega_a) = \ix244 + \ix243+\ix 242+\ix422 + \ix332,
\]
where
\def\oo{\overline{\!\otimes\!}}
\def\ot{\!\!\otimes\!\!}
\def\mmm{\Sum\!(\BC_1\!\!\otimes\cdots\otimes\!\!\BC_5)}
{\small
\begin{eqnarray*}
\ix244&=&
-\mmm(\XV\oo\XIII\oo\XII)\\
&&+\mmm(\VIII\oo(\IV\!\!\cdot\!\!\VII)\oo\XII)\\
&&+\VIII\!\!\cdot\!\!\mmm(\IV\oo\VII\oo\XII),
\end{eqnarray*}
\begin{eqnarray*}
\lefteqn{\ix243=}&&
\\
&&\!\!\!\!\!+\IXX\!\!\cdot\!\!\mmm(\XVI\oo\XXII\oo\IX)
\\
&&\!\!\!\!\!-\IXX\!\!\cdot\!\!\mmm
((1\ot(\id\ot\Delta\ot\id)(\ph))
\oo(\id\ot(\id\ot\Delta)
\Delta\ot\id)(\Phi)\!\!\cdot\!\!((\id\sp2\ot\Delta)
(\Phi)\ot1)\oo\IX)
\\
&&\!\!\!\!\!-\IXX\!\!\cdot\!\!\XVIII\!\!\cdot\!\!\mmm(\XX\oo\X\oo\IX)
\end{eqnarray*}
\begin{eqnarray*}
\lefteqn{\ix242=}&&
\\
&&\!\!\!\!\!-\VIII\!\!\cdot\!\!\mmm(\IV\oo\III\oo\VI)
\\
&&\!\!\!\!\!+\VIII\!\!\cdot\!\!\mmm(\V\oo\II\!\!\cdot\!\!\I\oo\VI)
\\
&&\!\!\!\!\!+\VIII\!\!\cdot\!\!\V\!\!\cdot\!\!\mmm(\II\oo\I\oo\VI)
\end{eqnarray*}
\begin{eqnarray*}
\lefteqn{\ix422=}&&\\
&&\!\!\!\!\!-\mmm(\VIII\oo\V\oo\II\!\!\cdot\!\!\I\!\!\cdot\!\!\VI)
\\
&&\!\!\!\!\!+\IXX\!\!\cdot\!\!\mmm(\XVIII\oo\XVII\oo\II
\!\!\cdot\!\!\I\!\!\cdot\!\!\VI)
\\
&&\!\!\!\!\!+\mmm(\IXX\oo\XVIII\!\!\cdot\!\!\XVII\oo
\II\!\!\cdot\!\!\I\!\!\cdot\!\!\VI)
\end{eqnarray*}
\begin{eqnarray*}
\lefteqn{\ix332=}&&
\\
&&\!\!\!\!\!-\IXX\!\!\cdot\!\!\XVIII\!\!\cdot\!\!\mmm
(\XVII\oo\II\oo\I\!\!\cdot\!\!\VI)
\\
&&\!\!\!\!\!+\IXX\!\!\cdot\!\!\XVIII\!\!\cdot\!\!\mmm
(\XX\oo\XI\oo\I\!\!\cdot\!\!\VI)
\end{eqnarray*}
}
and, finally
{\small
\begin{eqnarray*}
\ix333&=&-\mmm(\IXX\oo\XVI\oo\XXII\!\!\cdot\!\!\IX)
\\
&+&\mmm(\XV\oo\XXI\oo\XXII\!\!\cdot\!\!\IX)
\end{eqnarray*}
}
\def\oo{\overline{\otimes}}
\def\ot{\otimes}
Here the summation is taken over all
$(\bcc_1,\bcc_2,\bcc_3,\bcc_4,\bcc_5)\in B$, where
\[
B:= \{(\Omega_a,\mu,\mu,\mu,\mu),(\mu,\Omega_a,\mu,\mu,\mu),
(\mu,\mu,\Omega_a,\mu,\mu),(\mu,\mu,\mu,\Omega_a,\mu),(\mu,\mu,
\mu,\mu,\Omega_a)\}.
\]

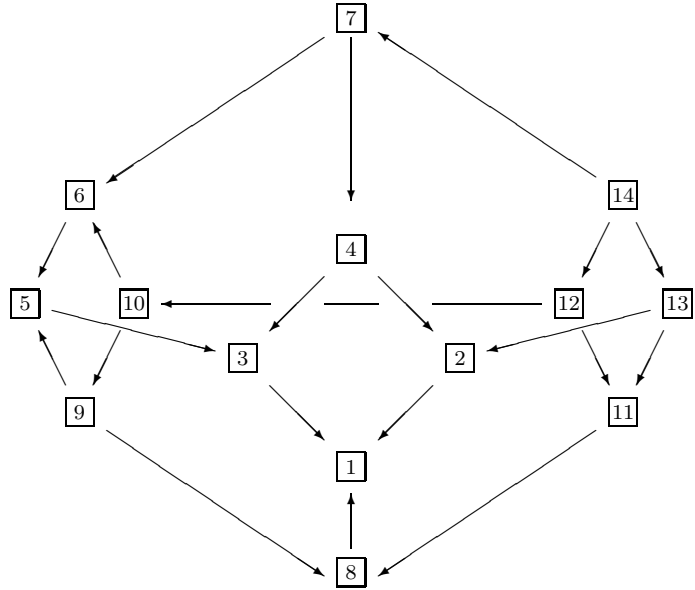
\begin{figure}
\setlength{\unitlength}{1.75em}
\begin{center}
\begin{picture}(12,9.5)(0,-.5)
\put(1.75,3.75){\ctv{10}}
\put(0.75,1.75){\ctv9}
\put(-0.25,3.75){\ctv5}
\put(0.75,5.75){\ctv6}
\put(9.75,3.75){\ctv{12}}
\put(5.75,-1.2){\ctv8}
\put(3.75,2.75){\ctv3}
\put(5.75,9){\ctv7}
\put(10.75,1.75){\ctv{11}}
\put(5.75,.75){\ctv1}
\put(5.75,4.75){\ctv4}
\put(10.75,5.75){\ctv{14}}
\put(11.75,3.75){\ctv{13}}
\put(7.75,2.75){\ctv2}

\put(1.75,4.5){\vector(-1,2){0.5}}
\put(1.75,3.5){\vector(-1,-2){0.5}}
\put(0.75,5.5){\vector(-1,-2){0.5}}
\put(0.75,2.5){\vector(-1,2){0.5}}
\put(5.5,4.5){\vector(-1,-1){1}}
\put(6.5,4.5){\vector(1,-1){1}}
\put(4.5,2.5){\vector(1,-1){1}}
\put(7.5,2.5){\vector(-1,-1){1}}
\put(10.75,5.5){\vector(-1,-2){0.5}}
\put(11.25,5.5){\vector(1,-2){0.5}}
\put(10.25,3.5){\vector(1,-2){.5}}
\put(11.75,3.5){\vector(-1,-2){.5}}
\put(.5,3.875){\vector(4,-1){3}}
\put(11.5,3.875){\vector(-4,-1){3}}
\put(4.5,4){\vector(-1,0){2}}
\put(5.5,4){\line(1,0){1}}
\put(7.5,4){\line(1,0){2}}
\put(10.5,6.33){\vector(-3,2){4}}
\put(10.5,1.66){\vector(-3,-2){4}}
\put(1.5,1.66){\vector(3,-2){4}}
\put(5.5,8.9){\vector(-3,-2){4}}
\put(6,8.9){\vector(0,-1){3}}
\put(6,-.5){\vector(0,1){1}}
\end{picture}
\end{center}
\caption{\label{fig14}
$K_5$ as a cell in three dimensions.}
\end{figure}

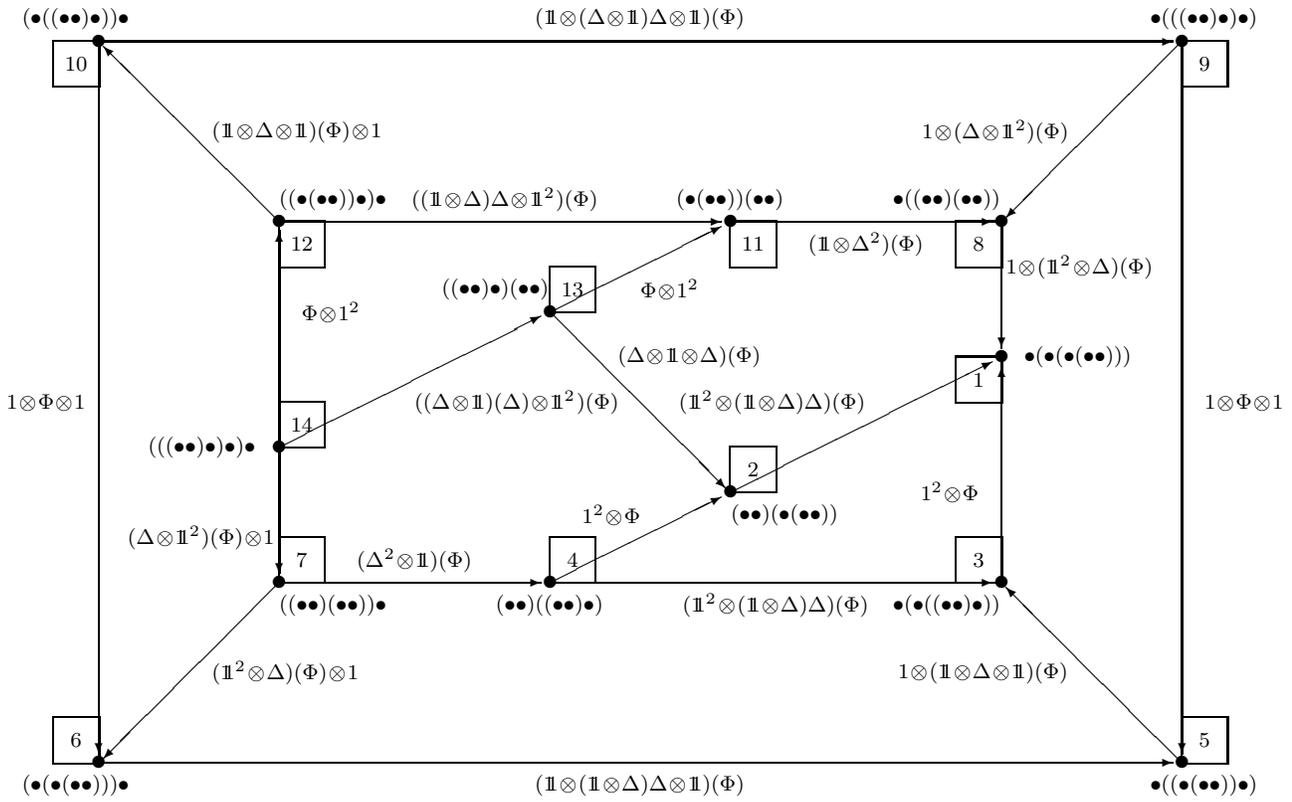
\begin{figure}
\setlength{\unitlength}{1.2cm}
\begin{center}
\begin{picture}(12,8.5)(-6,-4.5)

\put(-6,-4){\vectorjedna}
\put(-6,4){\vectorjedna}
\put(-6,4){\vectordva}
\put(6,4){\vectordva}
\put(-4,-2){\vectortri}
\put(6,4){\vectortri}
\put(6,-4){\vectorctiri}
\put(-4,2){\vectorctiri}
\put(-4,2){\vectorpet}
\put(-1,-2){\vectorpet}
\put(-4,-2){\vectorsest}
\put(1,2){\vectorsest}
\put(-4,-.5){\vectorsedm}
\put(4,-2){\vectorsedm}
\put(-4,-.5){\vectorosm}
\put(4,2){\vectorosm}
\put(-4,-.5){\vectordevet}
\put(1,-1){\vectordevet}
\put(-1,1){\vectordeset}
\put(-1,-2){\vectordeset}
\put(-1,1){\vectorjedenact}

\put(-6,-4){\kulka}
\put(6,-4){\kulka}
\put(-6,4){\kulka}
\put(6,4){\kulka}
\put(-4,-2){\kulka}
\put(4,-2){\kulka}
\put(-4,2){\kulka}
\put(4,2){\kulka}
\put(-4,-.5){\kulka}
\put(4,.5){\kulka}
\put(-1,1){\kulka}
\put(1,-1){\kulka}
\put(-1,-2){\kulka}
\put(1,2){\kulka}

\def\b{\bullet}
\put(-6.25,-4.25){\ccc{(\b(\b(\b\b)))\b}}
\put(-6.5,-4){\ctv6}
\put(6.25,-4.25){\ccc{\b((\b(\b\b))\b)}}
\put(6,-4){\ctv5}
\put(6.25,4.25){\ccc{\b(((\b\b)\b)\b)}}
\put(6,3.5){\ctv9}
\put(-6.25,4.25){\ccc{(\b((\b\b)\b))\b}}
\put(-6.5,3.5){\ctv{10}}
\put(-4,-2.25){\lll{((\b\b)(\b\b))\b}}
\put(-4,-2){\ctv7}
\put(4,-2.25){\rrr{\b(\b((\b\b)\b))}}
\put(3.5,-2){\ctv3}
\put(4,2.25){\rrr{\b((\b\b)(\b\b))}}
\put(3.5,1.5){\ctv8}
\put(-4,2.25){\lll{((\b(\b\b))\b)\b}}
\put(-4,1.5){\ctv{12}}
\put(-1,-2.25){\ccc{(\b\b)((\b\b)\b)}}
\put(-1,-2){\ctv4}
\put(1,2.25){\ccc{(\b(\b\b))(\b\b)}}
\put(1,1.5){\ctv{11}}
\put(-4.25,-.5){\rrr{(((\b\b)\b)\b)\b}}
\put(-4,-.5){\ctv{14}}
\put(4.25,.5){\lll{\b(\b(\b(\b\b)))}}
\put(3.5,0){\ctv1}
\put(1,-1.25){\lll{(\b\b)(\b(\b\b))}}
\put(1,-1){\ctv2}
\put(-1,1.25){\rrr{((\b\b)\b)(\b\b)}}
\put(-1,1){\ctv{13}}
\def\id{1\!\! 1}
\def\OO{\!\otimes\!}
\def\de{\Delta}
\put(0,-4.25){\ccc{(\id\OO(\id\OO\de)\de\OO\id)(\Phi)}}
\put(0,4.25){\ccc{(\id\OO(\de\OO\id)\de\OO\id)(\Phi)}}

\put(-4.75,-3){\lll{(\id^2\OO\de)(\Phi)\OO1}}
\put(4.75,3){\rrr{1\OO(\de\OO\id^2)(\Phi)}}
\put(4.75,-3){\rrr{1\OO(\id\OO\de\OO\id)(\Phi)}}
\put(-4.75,3){\lll{(\id\OO\de\OO\id)(\Phi)\OO1}}

\put(1.5,-2.25){\ccc{(\id^2\OO(\id\OO\de)\de)(\Phi)}}
\put(-1.5,2.25){\ccc{((\id\OO\de)\de\OO\id^2)(\Phi)}}

\put(-2.5,-1.75){\ccc{(\de^2\OO\id)(\Phi)}}
\put(2.5,1.75){\ccc{(\id\OO\de^2)(\Phi)}}

\put(-6.15,0){\rrr{1\OO\Phi\OO1}}
\put(6.25,0){\lll{1\OO\Phi\OO1}}

\put(0,-1.25){\rrr{1^2\OO\Phi}}
\put(0,1.25){\lll{\Phi\OO1^2}}

\put(-.25,.5){\lll{(\de\OO\id\OO\de)(\Phi)}}

\put(-2.5,0){\lll{((\de\OO\id)(\de)\OO\id^2)(\Phi)}}
\put(2.5,0){\rrr{(\id^2\OO(\id\OO\de)\de)(\Phi)}}

\put(3.75,-1){\rrr{1^2\OO\Phi}}
\put(-3.75,1){\lll{\Phi\OO1^2}}

\put(-4.05,-1.5){\rrr{(\de\OO\id^2)(\Phi)\OO1}}
\put(4.05,1.5){\lll{1\OO(\id^2\OO\de)(\Phi)}}
\end{picture}
\caption{\label{fig15}
The $2$-skeleton of
$K_5$ under stereographic projection from a pole at extreme left.}
\end{center}
\end{figure}

On the one hand these terms can be understood in terms of the
edges of the simplices in $\T 5$, our canonical triangulation
of $K_5$. Recall that the
triangulation is induced by the cone over the
triangulations of $(K_a\times K_b)_c$
for $a+b=6$ and $2\leq c\leq b$.
The bracketed expression on the left side of these equations
indicates the part of the boundary of $K_5$ whose
cone determines the components on the right.
Figures~\ref{fig14} and~\ref{fig15} illustrate the structure of $K_5$,
and
we list explicitly the
$3$-simplices in $\tilde \T 5$ which uses the decreasing
ordering the vertices to be consistent with the order in which
the transition operators appear.

\begin{eqnarray*}
\mbox{Cone}((K_2\times K_4)_4, \xi_5) &:&
[1,2,13,14],[1,8,13,14],[8,11,13,14]\\
\mbox{Cone}((K_2\times K_4)_3, \xi_5) &:&
 [3,4,7,14],[3,5,7,14],[5,6,7,14] \\
\mbox{Cone}((K_2\times K_4)_2, \xi_5) &:&
[8,11,12,14],[8,9,12,14],[9,10,12,14]\\
\mbox{Cone}((K_4\times K_2)_2, \xi_5)
 &:& [1,8, 9,14],[3,5,9,14], [1,3,9,14]\\
\mbox{Cone}((K_3\times K_3)_2, \xi_5)&:& [5,9,10,14],[5,6,10,14]\\
\mbox{Cone}((K_3\times K_3)_3, \xi_5) &:& [1,3,4,14],[1,2,4,14]
\end{eqnarray*}

On the other hand we can understand the terms as arising from
the substitution of $\Omega_a$ wherever
the associativity of multiplication is assumed
in the original proof of $\dqcoh(\Omega_p)=0$, the cocycle condition
for {\em restricted} deformations where multiplication and
comultiplication are not deformed, as given in
\cite{shnider-sternberg:preprint} or
using the deviation calculus of \cite{MS}.
We can distinguish the summands in terms of the part of
$\dqcoh(\Omega_p)$ in which we want to use associativity.
The first set of three summands, $M[(K_2\times K_4)_4]$, in our
list comes from the part of $\dqcoh(\Omega_p)$
involving $(\id^3\otimes\Delta)(\Omega_p)$,
the second set of three terms comes from the
part involving $(\id^2\otimes \Delta\otimes \id)(\Omega_p)$,
the third set comes from $(\id\otimes \Delta \otimes \id^2)(\Omega_p)$
etc.

Before ending this section we describe an application to
the deformation theory of the Drinfel'd's quasi-Hopf quantization
$A_h=(U({\cal G})[[h]],\MU, \Delta, \Phi_h)$ of the
universal enveloping algebra of a simple Lie algebra ${\cal G}$
over the rational field, $\bk=\bf Q$. Here $A_h$ is as
in~\cite[Chapter~11]{shnider-sternberg:book},
i.e.~$\cdot$ and $\Delta$ are undeformed and
$\Phi_h= 1+$ a polynomial in $h^2$. Let us prove first the following
lemma.

\begin{lemma}
\label{Kveta}
The second cohomology group $\hat H^2(A_0)$ of the undeformed
algebra $A_0$ is trivial. The third cohomology group $\hat H^3(A_0)$
has dimension one and is generated by the infinitesimal of the
deformation $A_h$.
\end{lemma}

\noindent
{\bf Proof.}
As we have already observed, for the undeformed algebra
$A_0=(U({\cal G}),\MU, \Delta, \Phi_0=1)$ the cohomology
$\hat H^n(A_0)$ is the same as the cohomology of the extended
bicomplex constructed by M.~Gerstenhaber and D.~Schack for the
bialgebra cohomology, see
also~\cite[Chapter~11]{shnider-sternberg:book} for details. An easy
spectral sequence argument together with the fact that there are no
invariants in $\land^2{\cal G}$ and that invariants in $\land^3{\cal
G}$
form an one-dimensional subspace spanned by the infinitesimal of
$A_h$ (see again~\cite[Chapter~11]{shnider-sternberg:book}) give the
desired result.
\qed

\begin{proposition}
$\hat H^2(A_h)=0$ and $\hat H^3(A_h)\cong \hat H^3(A_0)
\otimes {\bf Q}[[h]]$.
\end{proposition}

\noindent
{\bf Proof.}
The first part is an easy consequence of the semicontinuity principle
for the cohomology of a deformed algebra~\cite[par.~7]{ADT88} and the
triviality of $\hat H^2(A_0)$ (Lemma~\ref{Kveta}).

To compute $\hat H^3(A_h)$ we can use the spectral
sequence associated to the filtration
$$F^p(\hat C(A_h))=\bigoplus_{i\geq p}E^{i,j}(A_h).$$
The first term of the spectral sequence is
$$E_1^{i,j}=\mbox{Hoch}^j(U({\cal G}),U({\cal G})^{\otimes i})\otimes
{\bf Q}[[h]]=
\mbox{H}^j_{CE}({\cal G})\otimes (U({\cal G})^{\otimes i})^{\cal G}
\otimes {\bf Q}[[h]],$$
where $\mbox{Hoch}(U({\cal G}),M)$ is the Hochschild cohomology
of $U({\cal G})$ with coefficients in $M$, H$_{CE}({\cal G})$
is the Chevalley-Eilenberg cohomology with trivial coefficients,
and $M^{\cal G}$ is the space of ${\cal G}$ invariants.
Since H$^j_{CE}({\cal G})=0$ for $j=1,2$, and we must have
$i\geq 1$, the
component of $E_1$ in dimension $i+j-1\leq 2$ is
$(U({\cal G})^{\otimes i}))^{\cal G}[[h]]$. The coboundary
operator on $E_1,$ is $d_1=\dqcoh$.

All the higher coboundary operators, $d_n$, in the spectral sequence
are zero in the relevant range. Therefore, to conclude the proof of
the proposition, it is enough to show that
$H^*(E_1, d_1)$ is a free rank one module over ${\bf Q}[[h]]$.

\begin{lemma}{\label{elem}}
The cohomology $H^3(E_1,d_1)$, is
a free ${\bf Q}[[h]]$ module of rank one generated
by $\psi_h=\frac1h\frac{d}{dh}\Phi_h$.
\end{lemma}
\noindent
{\bf Proof.} The coboundary operator $\dqcoh$ has an expansion in
even powers of $h$,
$$\dqcoh=d_h=\sum d_{2n} h^{2n}$$
with $d_0=\dcoh$ the standard Cartier coboundary.
As we saw above the $\dqcoh$ is, up to conjugation,
 the variational
derivative of the pentagon identity for $\Phi_h$. Therefore
differentiating the pentagon identity using the chain rule
for derivatives gives $d_h\frac{d}{dh}\Phi_h=0$.
Since $\Phi_h$ is a formal power series in $h^2$ we can divide
by $h$ and conclude that $\psi_h$ is a $d_h$
cocycle. We claim that any $d_h$ 3-cocycle
is cohomologous to a ${\bf Q}[[h]]$ multiple of $\psi_h$.
Let $\sigma_h$ be such a
3-cocycle. We will prove that there exist a formal power series
$c_h$ and a 2-cochain $\rho_h$ such that
$ \sigma_h= c_h \psi_h + d_h\rho_h.$
The proof is by induction. Consider the
following truncations at $h^{n+1}$: let $d^n$ be the truncation
of $d_h$, $\psi^n$, the truncation
of $\psi_h$ , and $\sigma^n$, the truncation of
$\sigma_h$. Assume that there exist partial series
 $c^n=c_0 + \cdots + c_n h^n$ and
$\rho^n=\rho_0 + \cdots +\rho_n h^n$ such that
$$ \sigma^n= c^n \psi^n + d^n \rho^n\quad\mbox{mod}\quad h^{n+1}$$
Define $\xi_{n+1}$ by
$$\xi_{n+1}h^{n+1}=\sigma^{n+1}- (c^n \psi^{n+1}
+ d^{n+1} \rho^n)\quad\mbox{mod}\quad h^{n+2}.$$
Then
$$d_0\xi_{n+1}h^{n+1}=d^{n+1}(\sigma^{n+1}- c^n \psi^{n+1}
- d^{n+1} \rho^n)=0\quad\mbox{mod}\quad h^{n+2}$$
since $d^{n+1}\sigma^{n+1}=d^{n+1}\psi^{n+1}=
d^{n+1}d^{n+1}\rho^n=
0\quad\mbox{mod}\quad h^{n+2}.$ Thus $\xi_{n+1}$ is a
$d_0$ cocycle. Now $\phi=\psi_0$ generates the ${\cal G}$
invariant $d_0$ cohomology
so there exist a constant $c_{n+1}$ and a 2-cochain
$\rho_{n+1}$ such that
$$\xi_{n+1}=c_{n+1}\psi_0+d_0\rho_{n+1}.$$
Define $\rho^{n+1}=\rho^n +\rho_{n+1}h^{n+1}$ and
$c^{n+1}=c^n+ c_{n+1}h^{n+1}.$ Then
$$ \sigma^{n+1}= c^{n+1} \psi^{n+1} +
d^{n+1} \rho^{n+1}\quad\mbox{mod}\quad h^{n+2},$$
completing the induction.

This shows that the cohomology group under consideration is generated
by
$\psi_h$. Next we must show that it is torsion-free as a ${\bf
Q}[[h]]$
module. Suppose that there exists an exponent $n$ such that
\begin{equation}
\label{cobound}
h^n\psi_h=d_h\xi_h
\end{equation}
and consider the minimal such exponent, $n'$. We
know that $n'>0$ since $\psi_0=\phi$ is not a $d_0$ coboundary.
Assume~(\ref{cobound}) holds for some $n'>0$, then we can use the fact
that
$H^2(A, d_0)=0$ to find a $\chi_h\in C^1$ such that $\xi_h=
d_h\chi_h$ mod $h$. Therefore
$$d_h(\xi_h- d_h\chi_h)=d_h\xi_h=h^{n'}\psi_h.$$
The term in parentheses is divisible by $h$ so we can reduce the
exponent $n'$ by one, in
contradiction to minimality. Since the cohomology is generated by
$\psi_h$ this shows that it is a torsion-free
module over ${\bf Q}[[h]]$ and since the latter is a local ring,
any torsion-free module is free.
\qed

 The interpretation of Lemma~\ref{elem} relates to the problem of
``lifting'' cohomology classes to the deformation.
We proved in Lemma~\ref{Kveta}, $\hat H^3(A_0)$ is generated by
the infinitesimal of a deformation. This implies that $\hat H^3(A_0)$
is
generated by liftable cocycles, as the infinitesimal of a
deformations is always liftable.
On the other hand, the triviality of $\hat H^2(A_0)$ implies that
there are no nontrivial cocycles which lift to a
coboundary~\cite[Proposition~5.4]{fox:JPAA93}. The theorem then
follows from the analysis~\cite[page~56]{ADT88}.
 For the terminology and related analysis we refer the reader
to~\cite[par.~7]{ADT88} and~\cite{fox:JPAA93}, in particular,
[4, Corollary 6.3]. These papers work
with the Hochschild cohomology, but their results obviously apply
to our situation as well.

Finally we have the following theorem.
\begin{theorem}
The Drinfel'd deformation $A_h$ is not a jump deformation, that is,
the
`generic' two-parameter deformation $A_{h(1+u)}$ is not equivalent to
the deformation $A_h[[u]]$, which is trivial as a deformation in the
parameter $u$.
\end{theorem}

\section{Proofs of the technical sublemmas}

We now prove the two technical sublemmas from section 4.

\noindent{\bf Proof of sublemma 4.9:}
Suppose we have some $(j,\pi)\in S''_{a\Psi}$,
i.e.~ some $l$, $1\leq l\leq
k$, such that $\pi(l)> \pi(l-1)$ and $0< j = \pi(l)+l-2$.
We have, by
definition,
\[
(\id^{\odot(r-1)}\odot \db^j \odot \id^{\odot(n-r)})M^r_\pi(\hat a) =
\BB{v_k}{v_0}(\Sum \Mu(\psi_1)\ots \Mu(\psi_{r-1})\ot \psi_r\ot
\Mu(\psi_{r+1})\ots \Mu(\psi_n))
\]
where
\[
\psi_i =\Sum \hhh {a}{v_0}i0 \ots \hhh{a}{v_{l-1}}i{\pi(l)-1} \cdot
\Psi_{{v_{l-1}},{v_{l}}}^i \ots \hhh{a}{v_k}i{q+1},\ 1\leq i\leq n,
\]
while
\[
(\id^{\odot(r-1)}\odot \db^j \odot \id^{\odot(n-r)})
M^r_{\alpha(\pi)}(\hat a) =
\BB{v_k}{v_0}(\Sum \Mu(\psi'_1)\ots
\Mu(\psi'_{r-1})\ot \psi'_r\ot
\Mu(\psi'_{r+1})\ots\Mu(\psi'_n))
\]
with
\[
\psi'_i = \Sum\hhh{a}{v_0}i0 \ots \Psi_{{v_{l-1}},{v_l}}^i \cdot \hhh
{a}{v_l}i{\alpha(\pi)(l)}
\ots \hhh {a}{v_k}i{q+1},\ 1\leq i\leq n.
\]
But $\alpha(\pi)(l) = \pi(l)-1$ by definition and
\[
\Sum\Bigotimes_{i=1}^n\hhh{a}{v_{l-1}}i{\pi(l)-1} \cdot
\Psi_{{v_{l-1}},{v_{l}}}^i =\Sum\Bigotimes_{i=1}^n
\Psi_{{v_{l-1}},{v_l}}^i \cdots \hhh
{a}{v_l}i{\pi(l)-1}
\]
by\REF{angrevar}, and\REF{A}\ follows ``modulo the signs''. We must
show that $\sgn(\pi,\sigma) =
-\sgn(\alpha(\pi),\sigma)$ which is, by the definition of
$\sgn(-,-)$, the same as $(-1)^{o(\pi)}= -(-1)^{o(\alpha(\pi))}$,
which follows from $o(\alpha(\pi))=o(\pi)-1$. Thus\REF{A}\ is proven.

Let $(j,\pi)\in T$, i.e.~ $0\leq j\leq q$ and $\pi \in P(q-1,k)$. Then
\[
M^r_\pi[\sigma](\db^j(\hat a)) =
\BB{v_k}{v_0}(\Sum \Mu(\omega_1)\ots \Mu(\omega_{r-1})\ot \omega_r\ot
\Mu(\omega_{r+1})\ots \Mu(\omega_n)),
\]
where
\[
\Omega_i =\Sum \hhh a{v_0}i0\ots \Psi^i_{v_{l-1},v_l} \ots
\hoj{(a_j\cdot a_{j+1})}{v_l}i \ots\Psi^i_{v_{l},v_{l+1}} \ots \hhh
a{v_k}i{q+1}, \ 1\leq i\leq n,
\]
while
\[
(\id^{\odot(r-1)}\odot \db^{\beta(j)} \odot
\id^{\odot(n-r)})M^r_{\beta(\pi)}(\hat a) =
\BB{v_k}{v_0}(\Sum \Mu(\omega'_1)\ots \Mu(\omega'_{r-1})
\ot \omega'_r\ot
\Mu(\omega'_{r+1})\ots \Mu(\omega'_n)),
\]
with
\[
\omega'_i =\Sum \hhh a{v_0}i0\ots \Psi^i_{v_{l-1},v_l} \ots
\hhh a{v_l}ij\cdot
\hhh a{v_l}i{j+1} \ots\Psi^i_{v_{l},v_{l+1}} \ots \hhh
a{v_k}i{q+1}, \ 1\leq i\leq n,
\]
and~\REF{B}\ ``modulo the signs'' follows from the fact that
$\sum\otimes_{i=1}^n\hoj{(a_j\cdot a_{j+1})}{v_l}i =
\sum\otimes_{i=1}^n\hhh a{v_l}ij\cdot
\hhh a{v_l}i{j+1}$ (the multiplicativity of $\Delta$).
It remains to
show that
\[
(-1)^{n+\beta(j)}\cdot \sgn(\beta(\pi),\sigma) =
(-1)^{k+n+j}\cdot \sgn(\pi,\sigma).
\]
The desired relation reduces to congruence modulo 2
\[
\beta(j)+o(\beta(\pi))\equiv j+k+ o(\pi)
\]
which follows from $o(\beta(\pi))=o(\pi)+k-l$ and $\beta(j)=j+l.$
 The equation\REF{B}\ is proven.

Let $\pi \in P(q,l-1)$. By definition,
\[
M^r_\pi[d_S^0(\sigma)] =
\BB{v_k}{v_0}(\Sum \Mu(\nu_1)\ots \Mu(\nu_{r-1})\ot \nu_r\ot
\Mu(\nu_{r+1})\ots \Mu(\nu_n)),
\]
where
\[
\nu_i = \Sum \hhh a{v_1}i0\ots \hhh a{v_1}i{\pi(1)-1}\ot
\Psi_{v_1,v_2}^i \ots
\hhh a{v_k}i{q+1},\ 1\leq i\leq n,
\]
while
\[
(\id^{\odot(r-1)}\odot \db^{0} \odot
\id^{\odot(n-r)})M^r_{\gamma(\pi)}(\hat a) =
\BB{v_k}{v_0}(\Sum \Mu(\nu'_1)\ots \Mu(\nu'_{r-1})\ot \nu'_r\ot
\Mu(\nu'_{r+1})\ots \Mu(\nu'_n)),
\]
with
\[
\nu'_i = \Sum \hhh a{v_0}i0\cdot \Psi^i_{v_0,v_1} \ots
\hhh a{v_1}i{\pi(1)-1} \Psi_{v_1,v_2}^i \ots
\hhh a{v_k}i{q+1},\ 1\leq i\leq n.
\]
{}From the equation $\sum\bigotimes_{i=1}^n\hhh a{v_0}i0\cdot
\Psi^i_{v_0,v_1}=\sum \bigotimes_{i=1}^n \Psi^i_{v_0,v_1}\cdot \hhh
a{v_1}i0$
(see\REF{angrevar}) we get that
\begin{eqnarray*}
\lefteqn{
\Sum\Mu(\nu'_1)\ots \Mu(\nu'_{r-1})\ot \nu'_r\ot
\Mu(\nu'_{r+1})\ots \Mu(\nu'_n)=}
\\
&&=\PP {v_0}{v_1} \b( \Sum \Mu(\nu_1)\ots \Mu(\nu_{r-1})\ot \nu_r\ot
\Mu(\nu_{r+1})\ots \Mu(\nu_n))
\end{eqnarray*}
and\REF{C}\ ``modulo the signs'' follows from Lemma~\ref{susenka}. We
must also verify that $\sgn(\pi,d^0_S(\sigma))\! = \!
(-1)^{n-1} \sgn(\gamma(\pi),\sigma).$
The verification reduces to checking that
$(-1)^{o(\pi)+(k-1)n}\! =\!(-1)^{o(\gamma(\pi)+kn+n-1}$ which
follows from $o(\pi)=o(\gamma(\pi))+1$.
This finishes the proof of\REF{C}.

The proof of\REF{D}\ is a ``mirror image'' of the
proof of\REF{C}, except that here the desired relation between the
signs of the partitions is $ (-1)^{k+o(\pi)+(k-1)n}
=(-1)^{n+k+q-1+o(\delta(\pi)+kn}.$\qed

\noindent{\bf Proof of sublemma 4.11:}
First, observe that $M^r_\pi[\sigma \dia_{t,\phi} \tau]
(\hat a)$ can be written as
\[
M^l_\pi[\sigma \dia_{t,\phi} \tau](\hat a) =
\BB {(v_r,w_s)_t}{(v_0,w_0)_t}(\Sum \Mu(\xi_1)\ots
\Mu(\xi_{l-1})\ot \xi_l
\ot \Mu(\xi_{l+1})\ots \Mu(\xi_{a+b-1}),
\label{Ml}\]
where, for $1\leq d \leq a+b-1$,
\begin{eqnarray*}
\xi_d& =& \Sum a_0^{((v_{\phi'(0)},w_{\phi''(0)})_t,d)} \ots
a_{\pi(1)-1}^{((v_{\phi'(0)},w_{\phi''(0)})_t,d)} \ot
\Psi^d_{(v_{\phi'(0)},w_{\phi''(0)})_t,(v_{\phi'(1)},
w_{\phi''(1)})_t}
\ot \cdots
\\
&& \cdots \otimes \Psi^d_{(v_{\phi'(r+s-1)},
w_{\phi''(r+s-1)})_t,
(v_{\phi'(r+s)},w_{\phi''(r+s)})_t}
\ot a_{\pi(r+s)}^{((v_{\phi'(r+s)},w_{\phi''(r+s)})_t,d)}\ots
a_{q+1}^{((v_{\phi'(r+s)},w_{\phi''(r+s)})_t,d)}
\end{eqnarray*}

Let us introduce the following terminology to distinguish those
edges in $\sigma \dia_{t,\phi} \tau$ along which the vertex from
$\sigma$ changes and those in which the vertex from $\tau$ changes.
We say that a shuffle $\phi \in
\sh rs$ has a jump of the 1st type
(resp.~of the 2nd type) at some point $h$,
$1\leq h \leq r+s$, if $h \in \{i_1< \cdots < i_r\}$
(resp.~$h \in \{j_1< \cdots < j_s\}$) or,
equivalently, if $\phi'(h) =
\phi'(h-1)+1$ and $\phi''(h) = \phi''(h-1)$
(resp.~$\phi'(h) =
\phi'(h-1)$ and $\phi''(h) = \phi''(h-1)+1$).

It follows directly from the
definitions that if $\phi$ has a jump of the 1st
type at $h$, then
\begin{eqnarray*}
\Psi_{(v_{\phi'(h-1)},w_{\phi''(h-1)})_t,(v_{\phi'(h)},
w_{\phi''(h)})_t}
&=& \Psi_{(v_{\phi'(h-1)},w_{\phi''(h-1)})_t,
(v_{\phi'(h-1)+1},w_{\phi''(h-1)})_t}=\\
&=& 1^{\ot (t-1)} \otimes\PP{v_{\phi'(h-1)}}{v_{\phi'(h)}}
\ot 1^{\ot (b-t)},
\end{eqnarray*}
while, in the case of a jump of the 2nd type,
\begin{eqnarray*}
\Psi_{(v_{\phi'(h-1)},w_{\phi''(h-1)})_t,(v_{\phi'(h)},
w_{\phi''(h)})_t}
&=& \Psi_{(v_{\phi'(h-1)},w_{\phi''(h-1)})_t,
(v_{\phi'(h-1)},w_{\phi''(h-1)+1})_t}=\\
&=&
(\id^{\ot (1-t)}\ot \deltacurl{v_{\phi'(h-1)}}\ot
\id^{\ot (b-t)})
 \PP{w_{\phi''(h-1)}}{w_{\phi''(h)}}.
\end{eqnarray*}
So, we see that, for $d \not\in [t,t+a-1]$, we have
\[
\Psi^d_{(v_{\phi'(h-1)},w_{\phi''(h-1)})_t,
(v_{\phi'(h)},w_{\phi''(h)})_t}=
\left\{
\begin{array}{ll}
1,& \mbox{at the jump of the 1st type,}
\\
\Psi^d_{w_{\phi''(h-1)},w_{\phi''(h)}},
& \mbox{jump of the 2nd type, $1\leq d\leq t-1
$,}
\\
\Psi^{d-a+1}_{w_{\phi''(h-1)},w_{\phi''(h)}},
& \mbox{jump of the 2nd type,
$t\leq d-a\leq b-1$.}
\end{array}
\right.
\]
We get easily from the definition that
\[
\deltacurl{(v_{\phi'(h)},w_{\phi''(h)})_t} =
(\id^{\ot (t-1)}\ot \deltacurl {v_{\phi'(h)}}\ot
\id^{\ot(b-t)})\deltacurl{w_{\phi''(h)}},
\]
consequently,
\[
a^{((v_{\phi'(h)},w_{\phi''(h)})_t,d)} =
\left\{
\begin{array}{ll}
a^{(w_{\phi''(h)},d)},& \mbox{for $1\leq d\leq t-1$, and}
\\
a^{(w_{\phi''(h)},d-a+1)},& \mbox{for $t\leq d-a\leq b-1$.}
\end{array}
\right.
\]
Because each $\phi$ has at least one jump of the 1st type, we see
that $\xi_l$, which should be considered as an element of
the normalized bar resolution $\BV{-q-r-s}$, is of the form
\[
\xi_l = \Sum (\Rada \alpha0{i-1}|1|\Rada \alpha{i+1}{q+r+s+1}),
\]
for some $1\leq i \leq q+r+s$, hence it is zero.
This also proves the first
part of the sublemma.

On the other hand, we have
\[
M^t_{\pi''}[\tau] (\hat a) =
\BB {w_s}{w_0}(\Sum \Mu(\zeta_1)\ots \Mu(\zeta_{t-1})\ot \zeta_t
\ot \Mu(\zeta_{t+1})\ots \Mu(\zeta_{b})),
\]
with
\[
\zeta_e = \Sum\hhh a{w_0}e0 \ots \hhh a{w_0}e{\pi''(1)-1}\ot
\Psi^e_{w_0,w_1}\ots \Psi^e_{w_{s-1},w_s}\ot
\hhh a{w_s}e{\pi''(s)}\ots\hhh a{w_s}e{q+1},\ 1\leq e\leq b.
\]
Define then $(b_0|b_1|\cdots |b_{q+s+1}):= \zeta_t$. By
Lemma~\ref{EPS}, for
$t\leq l\leq t+a-1$,
\begin{eqnarray*}
\lefteqn{(\id^{\odot( t-1)}\odot M_{\pi'}^{l-t+1}[\sigma]
\odot \id^{\odot( b-t)})M^t_{\pi''}[\tau] (\hat a) =}
\\
&=&\BB {(v_r,w_s)_t}{(v_0,w_0)_t}
(\Sum \Mu(\zeta_1)\ots \Mu(\zeta_{t-1})\ot
\Mu(\eta_1)\ot\cdots
\\
&&\hskip15mm\cdots\ot \Mu(\eta_{l-t})\ot \eta_{l-t+1}
\ot \Mu(\eta_{l-t+2})\ots \Mu(\eta_{a})
\ot \Mu(\zeta_{t+1})\ots \Mu(\zeta_{b})),
\end{eqnarray*}
where
\[
\eta_f = \Sum\hhh b{v_0}f0 \ots \hhh b{v_0}f{\pi'(1)-1}\ot
\Psi^f_{v_0,v_1}\ots \Psi^f_{v_{r-1},v_r}\ot
\hhh b{v_r}f{\pi'(r)}\ots\hhh b{v_r}f{s+q+1}, 1\leq f\leq a.
\]

We finish the proof of the sublemma by showing that
\begin{eqnarray}
\label{Opulinka}
\lefteqn{\Sum \Mu(\zeta_1)\ots \Mu(\zeta_{t-1})\ot
\Mu(\eta_1)\ot\cdots}
\\
\nonumber
&&\cdots\ot \Mu(\eta_{l-t})\ot \eta_{l-t+1}
\ot \Mu(\eta_{l-t+2})\ots \Mu(\eta_{a})
\ot \Mu(\zeta_{t+1})\ots \Mu(\zeta_{b})
\end{eqnarray}
is the same as
\begin{eqnarray}
\label{Zebrulinka}
\Sum \Mu(\xi_1)\ots \Mu(\xi_{l-1})\ot \xi_l
\ot \Mu(\xi_{l+1})\ots \Mu(\xi_{a+b-1}).
\end{eqnarray}
Observe first that the terms at the $d$-th place for $d\not\in
[t,t+a-1]$ agree. Indeed, for such $d$ put
\[
g(d) :=
\left\{
\begin{array}{ll}
d,& \mbox{for $1\leq d \leq t-1$, and}
\\
d-a+1,& \mbox{for $t\leq d-a \leq b-1$}.
\end{array}
\right.
\]
It is clear from the above considerations that
\[
\Mu(\xi_d) =
\Mu(\Sum a_0^{(w_{\phi''(0)},g(d))}\ots
a_{\pi(1)-1}^{(w_{\phi''(0)},g(d))}
\ot X_1 \ots X_{r+s} \ot a_{\pi(r+s)}^{(w_{\phi''(r+s)},g(d))}\ots
a_{q+1}^{(w_{\phi''(r+s)},g(d))}
\]
with
\[
X_h :=
\left\{
\begin{array}{ll}
1,& \mbox{at the jump of the 1st type, and}
\\
\Psi^{g(d)}_{w_{\phi''(h-1)},w_{\phi''(h)}},
&\mbox{at the jump of the 2nd type.}
\end{array}
\right.
\]
Since the multiplication $\cdot$ on $V$ is supposed to be unital, we
may simply ``forget'' all the $X_h$'s at the
jumps of the first type. The
claim that $\Mu(\xi_d) = \Mu(\zeta_{g(d)})$ is now obvious.
To show the
equality of the terms at the $d$-th place for $d\in
[t,t+a-1]$ is even easier, as we see
immediately that for such $d$,
$\xi_d = \eta_{d-t+1}$.
This finishes the proof of the sublemma ``modulo the signs''. It
remains to show that
\[
\sgn(\pi,\sigma\dia_{t,\phi}\tau)\cdot\sgn(\phi)=
(-1)^{(a+1)s+(b+1)r}\cdot
\sgn(\pi',\sigma)\cdot \sgn(\pi'',\tau)
\]
which is, by the definition of $\sgn(-,-)$, the same as congruence
modulo 2
\begin{eqnarray*}
o(\pi)&+&(a+b-1)(r+s)+\sum \hat i_a\\
&\equiv&(a+1)s +(b+1)r +o(\pi')+ar +o(\pi')
+bs.
\end{eqnarray*}
All the terms involving $a,b,r,s,$ cancel leaving
$o(\pi)+\sum \hat i_a=o(\pi')+o(\pi''),$ which follows immediately from
the construction of $\pi$ from $\pi'$ and $\pi''$ as described
before the statement of Sublemma~\ref{slune}.
\qed

This completes the proofs of the technical sublemmas and
thus we have established conclusively the
existence of the permanent paranormal
object described in Theorem~\ref{existence_of_homomorphism}.
See~\cite[p.~200]{B} for the terminology.

\vskip3mm
\catcode`\@=11
\noindent
M.~M.: Mathematical Institute of the Academy, \v Zitn\'a 25, 115 67
Praha 1, Czech Republic,\hfill\break\noindent
\hphantom{M.~M.:} email: {\bf markl@earn.cvut.cz}\hfill\break\noindent
\hphantom{M.~M.:} Current address (until March 1994): Math-UNC,
Chapel Hill, NC 27599-3250, USA

\noindent
S.~S.: Department of Mathematics, University of Bar Ilan, Israel,
\hfill\break\noindent
\hphantom{S.~S.:} email: {\bf shnider@bimacs.cs.biu.ac.il}
\end{document}